\documentstyle[12pt,epsfig]{article}
\input{epsf}

\newcommand{\be}{\begin{eqnarray}}
\newcommand{\ee}{\end{eqnarray}}
\newcommand{\bea}{\begin{eqnarray}}
\newcommand{\eea}{\end{eqnarray}}

\newcommand{\beq}{\begin{equation}}
\newcommand{\eeq}{\end{equation}}
\newcommand{\nn}{\nonumber}

\def\ga{\mathrel{\mathpalette\fun >}}
\def\fun#1#2{\lower3.6pt\vbox{\baselineskip0pt\lineskip.9pt
\ialign{$\mathsurround=0pt#1\hfil##\hfil$\crcr#2\crcr\sim\crcr}}}

\begin{document}

\title{Searching for the quark--diquark systematics of baryons
 composed by  light quarks $q=u,d$}

\author{A.V. Anisovich,
        V.V. Anisovich\footnote{anisovic@thd.pnpi.spb.ru}, M.A. Matveev,\\
        V.A. Nikonov,
        A.V. Sarantsev and T.O. Vulfs}

\maketitle

\begin{abstract}
Supposing  quark--diquark structure of baryons, we look for systematics of
baryons composed of light quarks ($q=u,d$).
We systematize baryons using the notion of two diquarks:
(i) axial--vector state, $D^{1}_{1}$, with the spin $S_D=1$ and isospin
 $I_D=1$  and (ii) scalar one,
 $D^{0}_{0}$, with the spin $S_D=0$ and isospin
 $I_D=0$. We consider several schemes for the composed baryons:
(1) with different  diquark masses, $M_{D^0_0} \ne M_{D^1_1}$, (2) with
$M_{D^0_0} = M_{D^1_1}$ and overlapping $qD^0_0$ and $qD^1_1$
states (resonances), (3) with/without $SU(6)$ constraints for low-lying
states (with quark--diquark orbital momenta $L=0$). In the high-mass
region the model  predicts several  baryon resonances at $M\sim
2.0-2.9$ GeV. Moreover, the model gives us the double pole structure
 (i.e. two poles with the same Re$M$ but  different Im$M$)  in many
 amplitudes at masses $M\ga 2.0$ GeV.
  We see also that for description of low-lying baryons (with
 $L=0$), the $SU(6)$ constraint is needed.

\end{abstract}

\section{Introduction}

The experiment gives us much lesser number of highly excited baryons
than the model with three constituent quarks predicts. One of the
plausible explanation is that the excited baryons  do not prefer
to be formed as  three-body systems of spatially separated colored
quarks. Instead,  similarly to mesons, they are  two-body systems
of quark and diquark:
\beq \label{I100}
 q_\alpha D^{\alpha} =
q_\alpha\bigg[\varepsilon^{\alpha\beta\gamma}q_\beta q_\gamma \bigg].
\eeq
Here $\varepsilon^{\alpha\beta\gamma}$ is the three-dimensional totally
 antisymmetrical tensor which works  in   the
 color space. Below we omit color indices, imposing  the
 symmetry anzatz for the spin--flavor--coordinate variables of wave
 functions.

It is an old idea that a $qq$-system inside the baryon can be
regarded as a specific object -- diquark. Thus, interactions with a
baryon may be considered as interactions with quark, $q$, and
two-quark system, $(qq)$: such a hypothesis was used in \cite{Y_5vva65}
for the description of the high-energy hadron--hadron collisions. In
\cite{ida,Y_5licht,ono}, baryons were described as quark--diquark
systems. In hard processes on nucleons (or nuclei), the coherent
$qq$ state (composite diquark) can be responsible for interactions
in the region of large Bjorken-$x$ values, at $x \sim 2/3$; deep
inelastic scatterings were considered in the framework of such an
approach in \cite{Y_5vva75,sch,fec,kawabe,fred}. More detailed
considerations of the diquark and the applications to different
processes may be found in \cite{Y_5diquark1,Y_5goeke,Y_5diquark2}.

Here we suppose that  excited baryons
are quark--diquark systems.  It means that  in the space
of three colors $({\bf c_3})$ the excited baryons, similarly to
excited mesons, are $\bigg({\bf \bar c}_3(D^0_0){\bf c_3}(q)\bigg)$ or
$\bigg({\bf \bar c}_3(D^1_1){\bf c_3}(q)\bigg)$
systems.

The two-particle system has considerably less degrees of freedom
than three-particle one and, consequently, much less
 excited states. At the same time, the comparison of experimental data
 with model calculations \cite{Izgur-3q,Gloz,Petry} demonstrates that
 the number of  predicted three-quark states
is much larger than the number of observed ones. The aim of
this paper is to analyze
mechanisms which may reduce the number of predicted
excited states. Generally, it is the main motivation for the developing
of quark--diquark models, see discussion in \cite{book2,book3}.

Now let us have a look at what type of states  appears in $qD^0_0$ and
$qD^1_1$ systems.
The $qD^0_0$ systems with total spin $S=1/2$ and
isospin $I=1/2$ contain the following baryon states
$J^P$ at different orbital momenta $L$ (we restrict ourselves by $L\leq
6$):
\begin{equation} \label{I2} \begin{tabular}{lccccccc}
  qD$^0_0$:&&&&&&\\
 $L=0:$  & $\frac {1}{2}^+$ & & & &&\\
 $L=2:$ &&$\frac {3}{2}^+$,& $\frac {5}{2}^+$&& &\\
 $L=4:$ &&&&$\frac {7}{2}^+$,& $\frac {9}{2}^+$&\\
 $L=6:$ &&&&&&$\frac {11}{2}^+$, &$\frac {13}{2}^+$
       \\
       \hline
 $L=1:$ & $\frac {1}{2}^-$, &$\frac {3}{2}^-$ & & &&\\
 $L=3:$ &&& $\frac {5}{2}^-$,&$\frac {7}{2}^-$& &\\
 $L=5:$ &&&&& $\frac {9}{2}^-$,&$\frac {11}{2}^-$&\\
\end{tabular}
\end{equation}
The $qD^1_1$ systems have quark--diquark total spins $S=1/2,\,3/2$ and
isospins $I=1/2$ (nucleons) and $I=3/2$ ($\Delta$ isobars), thus
creating the following baryon states $J^P$ at orbital momenta $L\leq
6$:
\begin{equation}
\label{I3}
%\begin{tabular}{lccccccccc}
\begin{tabular}{lccccccccc}
  qD$^1_1$:&&&&&&&&\\
 $L=0$, $S=\frac{1}{2}:$  & $\frac {1}{2}^+$ & & & &&&\\
 $L=0$, $S=\frac{3}{2}:$  && $\frac {3}{2}^+$ & & & &&\\
 $L=2$, $S=\frac{1}{2}:$ &&$\frac {3}{2}^+$,& $\frac {5}{2}^+$&& &&\\
 $L=2$, $S=\frac{3}{2}:$ &$\frac {1}{2}^+$&$\frac {3}{2}^+$,& $\frac {5}{2}^+$
 &$\frac {7}{2}^+$&& &\\
  $L=4$, $S=\frac{1}{2}:$ &&&&$\frac {7}{2}^+$,& $\frac {9}{2}^+$&&\\
   $L=4$, $S=\frac{3}{2}:$ &&&$\frac {5}{2}^+$,&$\frac {7}{2}^+$,
   & $\frac {9}{2}^+$&$\frac {11}{2}^+$,&\\
 $L=6$, $S=\frac{1}{2}:$ &&&&&&$\frac {11}{2}^+$, &$\frac {13}{2}^+$
       &\\
$L=6$, $S=\frac{3}{2}:$ &&&&&$\frac {9}{2}^+$,&$\frac {11}{2}^+$, &$\frac {13}{2}^+$&
  $\frac {15}{2}^+$
  \\
 \hline
 $L=1$, $S=\frac{1}{2}:$ & $\frac {1}{2}^-$, &$\frac {3}{2}^-$ & & &&&&\\
 $L=1$, $S=\frac{3}{2}:$ & $\frac {1}{2}^-$, &$\frac {3}{2}^-$, &$\frac {5}{2}^-$, & &&&&\\
  $L=3$, $S=\frac{1}{2}:$ &&& $\frac {5}{2}^-$,&$\frac {7}{2}^-$& &&&\\
  $L=3$, $S=\frac{3}{2}:$ && $\frac {3}{2}^-$,& $\frac {5}{2}^-$,&$\frac {7}{2}^-$&
   $\frac {9}{2}^-$, &&&\\
 $L=5$, $S=\frac{1}{2}:$ &&&&& $\frac {9}{2}^-$,&$\frac {11}{2}^-$&&&\\
  $L=5$, $S=\frac{3}{2}:$ &&&&$\frac {7}{2}^-$,& $\frac {9}{2}^-$,&$\frac {11}{2}^-$&
  $\frac {13}{2}^-$&&\\
 \end{tabular}
\end{equation}
Symmetry properties, such as those of the $SU(6)$, lead to certain
constraints in the realization of these states.

In Eqs. (\ref{I2}) and (\ref{I3}), the basic states are included only.
Actually,
 every state in (\ref{I2}) and (\ref{I3}) is also  characterized by its
 radial quantum number $n=1,2,3,\ldots$. So, in (\ref{I2}) and
 (\ref{I3}) every state labelled by $J^P$ represents a set of
 baryons:
 \beq \label{I4}
  J^P \to (n,J^P),\quad n=1,2,3,\ldots
 \eeq
 The states with different $L$ and $S$ but with the same $(n,J^P)$ can mix
 with each other.
 However, the meson systematics tell us that $L$ may be considered
 as a good quantum number for $q\bar q $ systems. Below, in the
 consideration of quark--diquark models, we use the same hypothesis and
 characterize $qD^0_0$ and $qD^1_1$ systems by the orbital momentum
 $L$. We also consider total spin $S$ in $qD^1_1$ systems as another
 conserved quantum number, though we realize that it should be regarded
 as a rough approximation only.

\begin{figure}[h]
%Fig. 1
\centerline{\epsfig{file=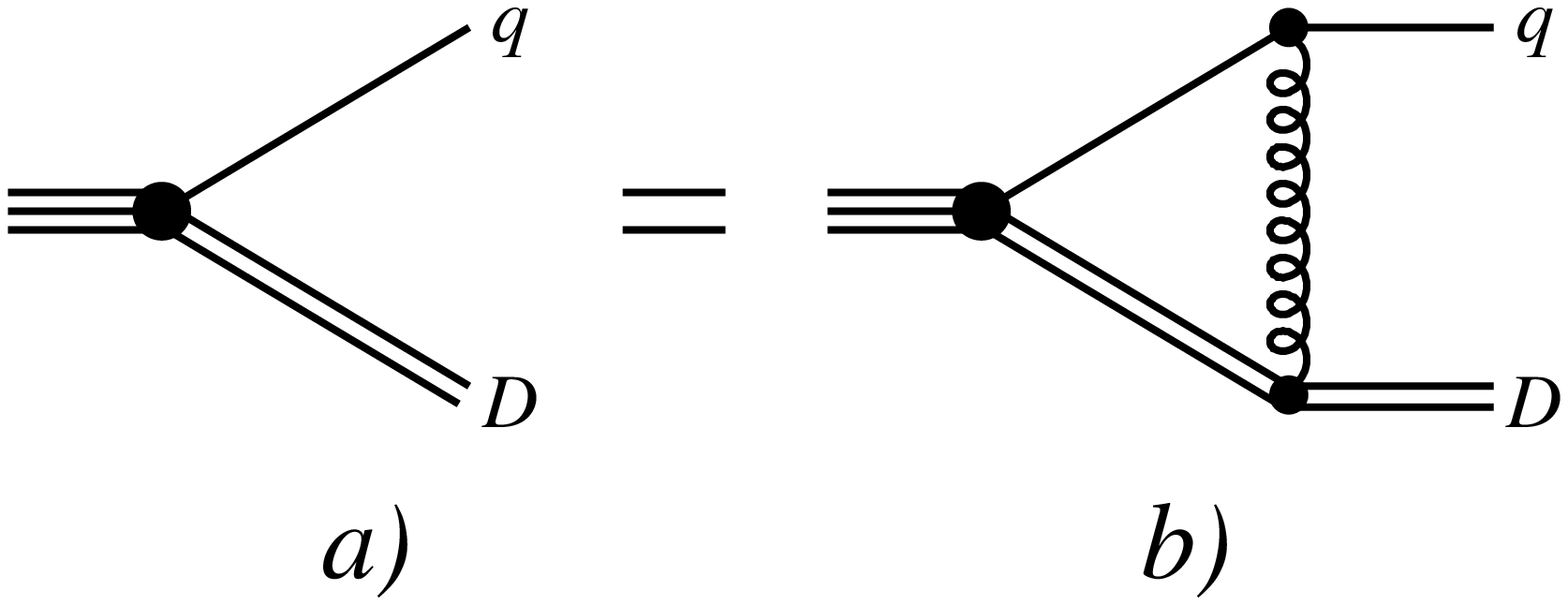,width=75mm}
            }
\centerline{\epsfig{file=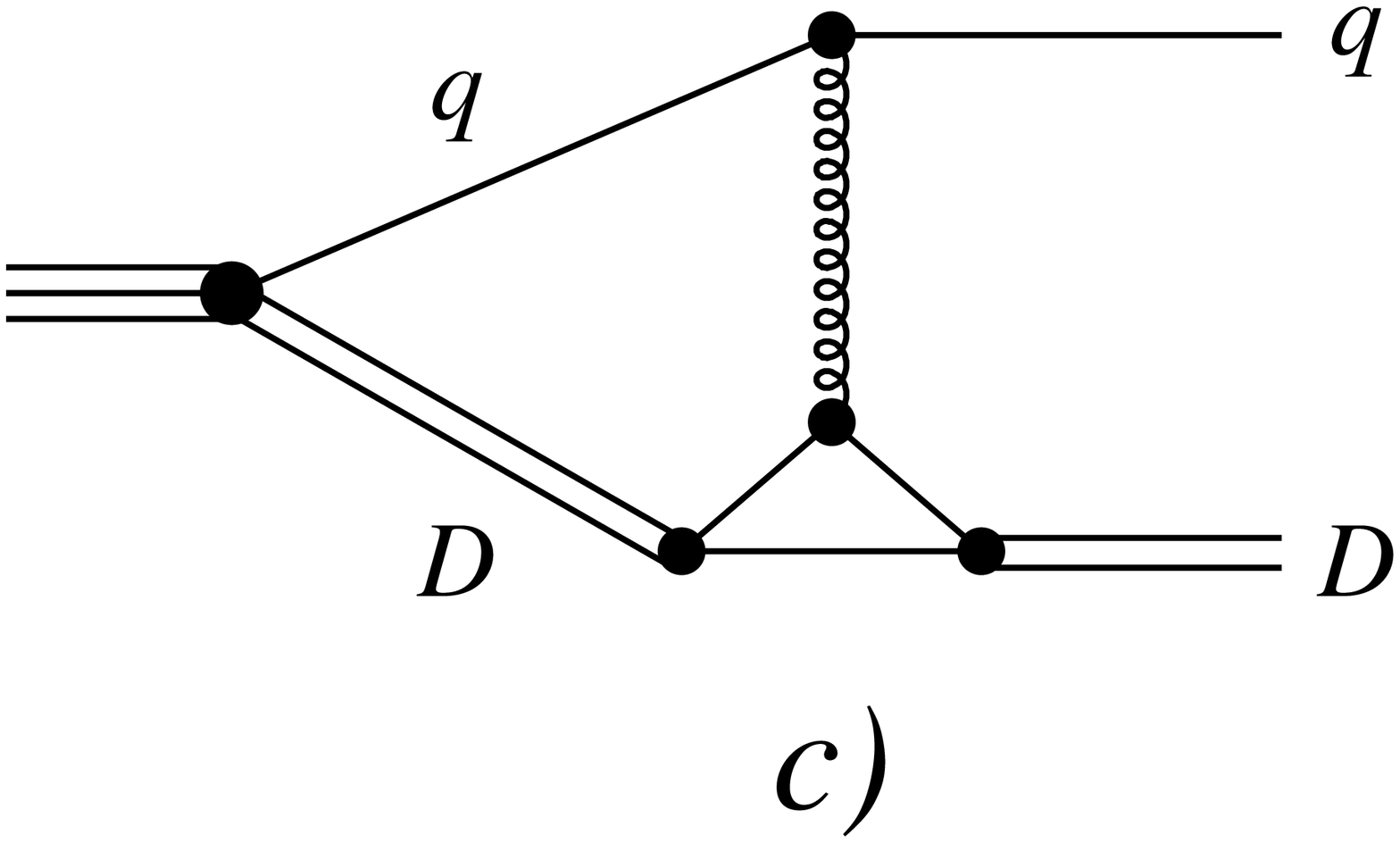,width=45mm}
            \epsfig{file=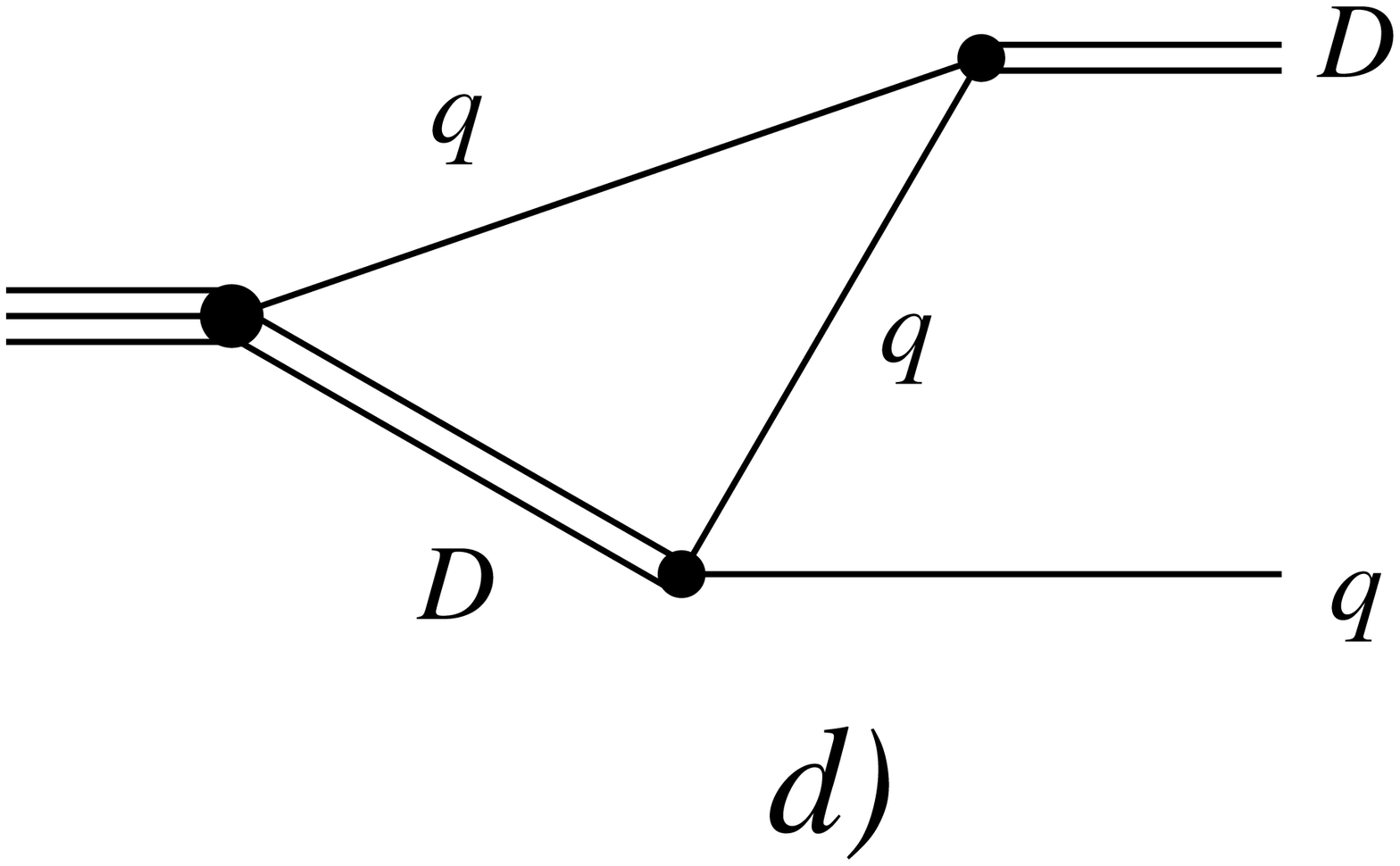,width=45mm}}
\caption{a,b) Equation for quark--diquark system (the flavor-neutral
interaction denoted by helix-type line). c,d) Processes
 considered as negligibly small in the quark--diquark model for
highly excited states. }\label{Int}
\end{figure}

 Now let us present in a more detail the arguments in favor of a
 possible realization of the quark--diquark structure of highly excited
 baryons. We use as a guide the spectral integral (or Bethe--Salpeter
 equation) for understanding quark--diquark systems
 considered here -- the equation is shown schematically in Figs.
 \ref{Int}a,b. If the interaction (a helix-type line in  Fig.
 \ref{Int}b) is flavor-neutral (gluonic or confinement singularity
 exchange), diquarks retain their quantum numbers,  $qD^0_0\to qD^0_0$
 and $qD^1_1\to qD^1_1$, and the states $qD^1_1$ and $qD^0_0$  do not
 mix. In the equation shown in Figs. \ref{Int}a,b, it was supposed that
 three-quark intermediate states are absent. It means that, first, the
 diquarks should be effectively point-like (diquark form factors lead
 to $qqq$ state, see Fig. \ref{Int}c) and, second, the quark-exchange
 processes, Fig. \ref{Int}d, are  suppressed (these processes include
  three-quark states). Both requirements can be fulfilled, if
 the diquark size is much less than the baryon one,
 $R_{diquark}\ll  R_{baryon}$, that may happen for highly
 excited states. Regretfully, we do not know
 for which states $R_{diquark}\ll R_{baryon}$, thus we
 consider here several variants.

 The paper is organized as follows.

 In section 2, we consider wave functions for quark--diquark systems in
 the nonrelativistic approximation (the relativization of vertices
 $B\to  qqq$ is not difficult, it can be found, for example, in
 \cite{book3,AMN}). In this section, we also demonstrate the way to
 transform the quark--diquark wave function into three-quark
 $SU(6)$-symmetrical one.

 In sections 3, 4 and 5  we consider different
 variants of the classification of baryon states.

 First, on the basis of the absence of
 $N_{\frac 32^+}^{L=0}(\sim 1200)$ and $\Delta_{\frac 32^+}^{L=0}(\sim 950)$,
 predicted by the quark--diquark model in its general form
 (section 4), we justify  the $(L=0)$ states to obey the $SU(6)$
 symmetry rules (section 3).

 In section 4 the model with $qD^0_0$ and $qD^1_1$
 systems is considered in general form and the overall predictions are
 given.
 With an exception for $(L=0)$ states, we suggest in section 4 the
 setting of quark--diquark baryons which are in a qualitative
 agreement with data. Still, some uncertainties exist in the
 $\Delta_{\frac 52^-}$ sector owing to certain contradictions in data.
 To be illustrative, we present in this section the $(J,M^2)$ and
 $(n,M^2)$ plots.

 In section 5 we consider the model with
 $qD^0_0$ resonances  overlapping with those of  $qD^1_1$, with
 the same $J,L,S=1/2$. It reduces the number of
 easily visible bumps, though not substantially.

 We see that even in
 the quark--diquark  model the  number of resonances is noticeably
 larger than  presently observed in the experiment. Also, the model
 predicts a set of
 overlapping resonances, resulting in a hide
 of some of them in visible bumps.
 It is a common prediction inherent in all
 considered schemes, being therefore a challenge for the experiment.

In Conclusion, we summarize the problems which appear in the
consideration of the quark--diquark scheme.

\section{Baryons as quark--diquark systems  }

Here, to be illustrative,  we consider wave functions of
quark--diquark systems, $qD^0_0$ and $qD^1_1$,  in the
non-relativistic limit. Relativistic generalization of the $B\to 3q$
vertices may be found, for example, in Chapter 7 of \cite{book3}.

\subsection{$S$-wave diquarks and baryons}

Recall that we have two $S$-wave diquarks with color numbers
${\bf\bar c}=3$: scalar diquark $ D^{0}_{0}$ and axial--vector  one,
$D^{1I_Z}_{1S_Z}$. The diquark spin--flavor wave functions
with $I_D=1,\, S_D=1$ and $I_D=0,\,S_D=0$ read as follows:
\bea \label{D-1}
&&
D^{11}_{11}(ij)= u^\uparrow (i)u^\uparrow (j),\nn \\
 &&D^{11}_{10}(ij)= \frac{1}{\sqrt 2}
 \bigg(u^\uparrow (i)u^\downarrow (j)+ u^\downarrow (i)u^\uparrow
(i)\bigg),\nn \\
&&D^{10}_{11}(ij)= \frac{1}{\sqrt 2}\bigg(u^\uparrow
(i)d^\uparrow (j)+ d^\uparrow (i)u^\uparrow (j)\bigg),\nn\\
&&D^{10}_{10}(ij)= \frac{1}{ 2}\bigg(u^\uparrow (i)d^\downarrow (j)+
u^\downarrow (i)d^\uparrow (j)+d^\uparrow (i)u^\downarrow (j)+
d^\downarrow (i)u^\uparrow (j)\bigg),\nn \\
&& D^{0}_{0}(ij)= \frac{1}{ 2}\bigg(u^\uparrow (i)d^\downarrow (j)-
u^\downarrow (i)d^\uparrow (j)-d^\uparrow (i)u^\downarrow (j)+
d^\downarrow (i)u^\uparrow (j)\bigg).
\eea

\subsection{Wave functions of  quark--diquark systems with $L=0$}

In general case, we have the following sets of baryon states:
\bea
\label{D-2}
&& (qD_1^1)_{J^{\pm},L,S(=1/2,\;3/2)\, ,n}\ , \nn \\
&& (qD_0^0)_{J^{\pm},L,S(=1/2), n}.
\eea
Recall that positive and negative parities $P=\pm$ are determined by
the orbital momentum $L$ between quark and diquark: $(+)$ and $(-)$
for even and odd  $L$. The total spin of the quark--diquark states runs
$S=1/2,\;3/2$. The states with the same $(I,J^P)$ may have different
radial excitation numbers $n$.

Here we consider the wave functions of  quark--diquark systems with
$L=0$, namely,
 $\Delta^{++}_{3/2}$, $\Delta^{+}_{3/2}$
 and $p$, as well as $\Delta^{++}_{1/2}$, $N^{+}_{3/2}$ and
 corresponding radial excitations. These examples
give us a guide for  writing other wave functions of the quark--diquark
states composed by light quarks ($u,d$).

\subsubsection{The $\Delta$ isobar: quark--diquark wave function
for arbitrary $n$ and its transformation into the $SU(6)$ wave function }

The wave function, up to the normalizing coefficient, for
$\Psi(\Delta_{\uparrow \uparrow \uparrow }^{++})$ with arbitrary $n$
 reads
\bea \label{qds-2}
&&
    u^\uparrow (1)D^{11}_{11}(23)\Phi_1(1;23)
  + u^\uparrow (2)D^{11}_{11}(13)\Phi_1(2;13)
  + u^\uparrow (3)D^{11}_{11}(12)\Phi_1(3;12)
\nn \\
=&&
    u^\uparrow (1)u^\uparrow (2)u^\uparrow (3)\Phi_1(1;23)
  + u^\uparrow (2)u^\uparrow (1)u^\uparrow (3)\Phi_1(2;13)
  + u^\uparrow (3)u^\uparrow (1)u^\uparrow (2)\Phi_1(3;12)
\nn \\
&& \equiv  \Psi(\Delta_{\uparrow \uparrow \uparrow }^{++}) .
\eea
Here the  indices (1,2,3) label the  momenta (or coordinates) of
quarks.
 The wave functions of quarks are symmetrical in the spin, coordinate
and flavor   spaces. Momentum/coordinate component of the wave
function is normalized in a standard way:
$\int |\Psi(\Delta_{\uparrow \uparrow \uparrow
}^{++})|^2d\Phi_3=1$,
where $d\Phi_3$ is the three-particle phase space.

Let us consider $n=1$. If the momentum/coordinate wave function of
the basic state is
symmetrical,
\beq \label{qds-SU6}
 \Phi_1(1;23)= \Phi_1(2;13)= \Phi_1(3;12)\equiv
 \varphi^{(sym)}_1(1,2,3) ,
 \eeq
  we have the $SU(6)$ symmetry for
$\Delta_{\uparrow\uparrow\uparrow}^{+}$:
\bea \label{qds-3}
\Psi_{SU(6)}(\Delta_{\uparrow \uparrow \uparrow }^{++})& = &
\Big [
    u^\uparrow (1)u^\uparrow (2)u^\uparrow (3)
  + u^\uparrow (2)u^\uparrow (1)u^\uparrow (3)
  + u^\uparrow (3)u^\uparrow (1)u^\uparrow (2)\Big ]\varphi_1(1,2,3)\nn
  \\ &=& u^\uparrow(1) u^\uparrow (2)u^\uparrow (3)\;3\varphi^{(sym)}_1(1,2,3)
  \equiv
\{u^\uparrow u^\uparrow u^\uparrow\}  \varphi_1(1,2,3).
\eea
 Here and below we omit the index $^{(sym)}$, i.e.
$\varphi^{(sym)}_1(1,2,3)\to \varphi_1(1,2,3)$.

The $\Delta_{\uparrow  }^{++}$ wave function with arbitrary $n$ reads
\bea \label{qds-4}
\Psi(\Delta_{\uparrow }^{++})&=&
 C^{\frac 32\frac 32}_{1\,1\;\;\frac 12\frac 12}\,
 C^{\frac 32\,\frac 12}_{1\,0\;\;\frac 12\,\frac 12}
\Big[  u^\uparrow (1)D^{11}_{10}(23)\Phi_1(1;23)
      + u^\uparrow (2)D^{11}_{10}(31)\Phi_1(2;31)
\Big. \nn  \\ &&\Big.\hspace{25mm}
  + u^\uparrow (3)D^{11}_{10}(12)\Phi_1(3;12)\Big]
  \nn \\
&+&C^{\frac32\frac32}_{1\,1\;\;\frac12\frac12}\,
C^{\frac32\frac12}_{1\,1\;\;\frac12\,-\frac12}
\Big[ u^\downarrow (1)D^{11}_{11}(23)\Phi_1(1;23)
  +    u^\downarrow (2)D^{11}_{11}(31)\Phi_1(2;31)
\Big. \nn  \\ &&\Big.\hspace{25mm}
  +    u^\downarrow (3)D^{11}_{11}(12)\Phi_1(3;12)\Big ]\ .
  \eea
If at $n=1$, as previously,
the momentum/coordinate wave function is symmetrical, see Eq.
(\ref{qds-SU6}),
  we deal with the $SU(6)$ symmetry for  basic
$\Delta_{\uparrow}^{+}$:
\bea \label{qds-6}
 \Psi_{SU(6)}(\Delta_{\uparrow  }^{++})
 &=& \frac{1}{\sqrt 3}\Big [
    u^\downarrow (1)u^\uparrow (2)u^\uparrow   (3)
  + u^\uparrow   (1)u^\uparrow   (2)u^\downarrow (3)
  + u^\uparrow   (1)u^\downarrow (2)u^\uparrow (3)\Big ]
  \nn \\
 &\times& \varphi_1(1,2,3).
\eea
In a more compact form, it reads
\bea \label{qds-7}
\Psi_{SU(6)}(\Delta_{\uparrow  }^{++})
&=&
 \frac{1}{\sqrt 3}\Big [
    u^\downarrow u^\uparrow   u^\uparrow
  + u^\uparrow   u^\uparrow   u^\downarrow
  + u^\uparrow   u^\downarrow u^\uparrow \Big ]\varphi_1(1,2,3)
 \nn \\
 &\equiv&
   \{ u^\downarrow u^\uparrow u^\uparrow\}\varphi_1(1,2,3).
\eea

The $\Delta_\uparrow^{+}$ wave function, $\Psi(\Delta_{\uparrow  }^{+})$,
is proportional to
\bea \label{iii-4}
&&
C^{\frac32\,\frac12}_{1\,0\;\;\frac12\,\frac12}\,
 C^{\frac32\,\frac12}_{1\,0\;\;\frac12\,\frac12}
\Big[
    u^\uparrow (1)D^{10}_{10}(23)\Phi_1(1;23)
  + u^\uparrow (2)D^{10}_{10}(31)\Phi_1(2;31)
\Big. \nn  \\ &&\Big.\hspace{25mm}
  + u^\uparrow (3)D^{10}_{10}(12)\Phi_1(3;12)\Big]
\nn  \\
&&+C^{\frac32\,\frac12}_{1\,0\;\;\frac 12\,\frac12}\,
 C^{\frac32\,\frac12}_{1\,1\;\;\frac12\,-\frac12}
\Big[
    u^\downarrow (1)D^{10}_{11}(23)\Phi_1(1;23)
  + u^\downarrow (2)D^{10}_{11}(31)\Phi_1(2;31)
\Big. \nn  \\ &&\Big.\hspace{25mm}
  + u^\downarrow (3)D^{10}_{11}(12)\Phi_1(3;12)\Big ]
\nn \\
&&+C^{\frac32\,\frac12}_{1\,1\;\;\frac12\,-\frac12}\,
  C^{\frac32\,\frac12}_{1\,0\;\;\frac12\,\frac12}
\Big[
    d^\uparrow (1)D^{11}_{10}(23)\Phi_1(1;23)
  + d^\uparrow (2)D^{11}_{10}(31)\Phi_1(2;31)+
\Big.
\nn  \\
&&\Big.\hspace{25mm}
  + d^\uparrow (3)D^{11}_{10}(12)\Phi_1(3;12)\Big ]
\nn  \\
&&+C^{\frac32\,\frac12}_{1\,1\;\;\frac12\,-\frac12}\,
C^{\frac32\,\frac12}_{1\,1\;\;\frac12\,-\frac12}
\Big[
    d^\downarrow (1)D^{11}_{11}(23)\Phi_1(1;23)
  + d^\downarrow (2)D^{11}_{11}(31)\Phi_1(2;31)
\Big.
\nn  \\
&&\Big.\hspace{25mm}
  + d^\downarrow (3)D^{11}_{11}(12)\Phi_1(3;12)
  \Big],
  \eea
  so we have:
  \bea \label{iii-9}
&&\Psi(\Delta_{\uparrow  }^{+})
\nn \\
&&
=\frac23
\Big(
    u^\uparrow (1)D^{10}_{10}(23)\Phi_1(1;23)
  + u^\uparrow (2)D^{10}_{10}(31)\Phi_1(2;31)
  + u^\uparrow (3)D^{10}_{10}(12)\Phi_1(3;12)\Big)
\nn  \\
&&
  +\frac{\sqrt 2}{3}
\Big(
    u^\downarrow (1)D^{10}_{11}(23)\Phi_1(1;23)
  + u^\downarrow (2)D^{10}_{11}(31)\Phi_1(2;31)
  + u^\downarrow (3)D^{10}_{11}(12)\Phi_1(3;12)\Big )
\nn \\
&&
  +\frac{\sqrt 2}{3}
\Big(
    d^\uparrow (1)D^{11}_{10}(23)\Phi_1(1;23)
  + d^\uparrow (2)D^{11}_{10}(31)\Phi_1(2;31)
  + d^\uparrow (3)D^{11}_{10}(12)\Phi_1(3;12)\Big )
\nn  \\
&&
  +\frac 13
\Big(
d^\downarrow     (1)D^{11}_{11}(23)\Phi_1(1;23)
  + d^\downarrow (2)D^{11}_{11}(31)\Phi_1(2;31)
  + d^\downarrow (3)D^{11}_{11}(12)\Phi_1(3;12)\Big).
\nn \\ &&
\eea
The $SU(6)$-symmetrical wave function reads
\bea \label{iii-10}
\Psi_{SU(6)}(\Delta_{\uparrow  }^{+})=\Big ( \sqrt{\frac{1}{3}}
\{u^\uparrow u^\uparrow d^\downarrow\}+ \sqrt{\frac{2}{3}}
\{u^\uparrow u^\downarrow d^\uparrow\}\Big)\varphi_1(1,2,3),
\eea
where
\beq \label{iii-11}
\{q_iq_jq_\ell\}=\frac{1}{\sqrt 6}
\{q_iq_jq_\ell+q_iq_\ell q_j
+q_jq_iq_\ell+q_jq_\ell q_i
 +q_\ell q_iq_j+q_\ell q_jq_i \}
\eeq
for $q_j\neq q_i\neq q_\ell$.

Above, to simplify the presentation, we  transformed the
wave functions to the $SU(6)$-symmetry ones for  $n=1$. One
can present certain examples with an easy generalization
to $n>1$. Assuming for $n=1$ that
\beq
\label{example1} \varphi_1(1,2,3)\equiv
\varphi_1^{(n=1)}(1,2,3)=A^{(1)}_1\exp(-b^{(1)}s),
\eeq
where $s$ is the
total energy squared $s=(k_1+k_2+k_3)^2$,  one may have for $n=2$:
\beq \label{example2}
\varphi_1^{(n=2)}(1,2,3)=A^{(1)}_2\exp(-b^{(1)}s)(s-B^{(1)}_2).
\eeq
Here $B^{(1)}_2$ is chosen to introduce a node into the $(n=2)$ wave
 function. Likewise, we can write down wave functions for higher $n$. It
is not difficult to construct models  with wave functions of the type
(\ref{example1}), (\ref{example2}) -- the variants of  corresponding
models are duscussed in section 2.4.

\subsubsection{The state with $I=3/2$, $J=1/2$ at $L=0$ and $S=1/2$,
$\Delta_{J^P=\frac 12 ^+}(L=0,S=1/2)$}

Let us present the wave function for
$\Delta^\uparrow _{J^P=\frac 12 ^+}(L=0,S=1/2)$.
It reads for $I_Z=\frac 32$ and $S_Z=\frac 12$ as follows:
\bea \label{qds-19}
&&
 C^{\frac 32\frac 32}_{1\,1\;\;\frac 12\frac 12}\,
 C^{\frac 12\,\frac 12}_{1\,0\;\;\frac 12\,\frac 12}
\Big[  u^\uparrow (1)D^{11}_{10}(23)\Phi_1(1;23)
      + u^\uparrow (2)D^{11}_{10}(31)\Phi_1(2;31)+
\nn \\
&&\qquad\qquad\qquad\qquad
      + u^\uparrow (3)D^{11}_{10}(12)\Phi_1(3;12)\Big]
\nn  \\
+&&
C^{\frac32\frac32}_{1\,1\;\;\frac12\frac12}\,
C^{\frac 12\frac12}_{1\,1\;\;\frac12\,-\frac12}
\Big[ u^\downarrow (1)D^{11}_{11}(23)\Phi_1(1;23)
  +    u^\downarrow (2)D^{11}_{11}(31)\Phi_1(2;31)
  \nn \\
&&\qquad\qquad\qquad\qquad
  +    u^\downarrow (3)D^{11}_{11}(12)\Phi_1(3;12)\Big ]
\nn \\
&& =
 -\frac{1}{\sqrt 3}
\Big[  u^\uparrow (1)
\frac{u^\uparrow (2)u^\downarrow (3)+u^\downarrow (2)u^\uparrow
(3)}{\sqrt 2}\Phi_1(1;23)+(1\rightleftharpoons 2)
                         +(1\rightleftharpoons 3)  \Big] \nn \\
&& \quad
+
\sqrt{\frac 23}
\Big[ u^\downarrow (1)u^\uparrow (2)u^\uparrow (3)\Phi_1(1;23)
+(1\rightleftharpoons 2)
                         +(1\rightleftharpoons 3)  \Big]\, .
\eea
In the $SU(6)$ limit, the wave functions for
$\Delta^\uparrow _{J^P=\frac 12 ^+}(L=0,S=3/2)$, which depend on $s$
only, are equal to zero:
\bea \label{qds-20} &&
 -\frac{1}{\sqrt 3}
\Big[  u^\uparrow (1)
\frac{u^\uparrow (2)u^\downarrow (3)+u^\downarrow (2)u^\uparrow
(3)}{\sqrt 2}+(1\rightleftharpoons 2)
                         +(1\rightleftharpoons 3)
                         \Big]\varphi_1(s) \nn \\
&&
+
\sqrt{\frac 23}
\Big[ u^\downarrow (1)u^\uparrow (2)u^\uparrow (3)
+(1\rightleftharpoons 2)
  +(1\rightleftharpoons 3)
 \Big]\varphi_1(s)=0\, .
\eea
Radial excitation wave functions in the $SU(6)$ limit, if
they depend on $s$ only (for example, see Eq.
(\ref{example2})), are also equal to zero.
So, in the $SU(6)$ limit we have for $L=0$ the
 state $\Delta_{J^P=\frac 32 ^+}$ only.

\subsubsection{The nucleon $N_{1/2}$: quark--diquark wave function
for arbitrary $n$ and its transformation into the $SU(6)$ one }

The  $S$-wave functions for $N_{1/2}^{+\,\uparrow}(qD^0_0)$ state with
 arbitrary $n$ reads
\bea \label{iv-12}
&&\Psi_{J=1/2}^{+\,\uparrow}(qD^0_0)= \Big [
 u^\uparrow (1)D^0_0(23)\Phi_0(1;23)+u^\uparrow (2)D^0_0(31)\Phi_0(2;31)
 \nn \\
&&\qquad\qquad\qquad\qquad
+u^\uparrow (3)D^0_0(12)\Phi_0(3;12)\Big].
\eea
For the symmetrical momentum/coordinate wave function,
\beq \label{qd0-SU6}
\Phi_0(1;23)= \Phi_0(2;31)= \Phi_0(3;12)\equiv \varphi_0(1,2,3) ,
\eeq
we have:
\bea \label{iii-13}
\Psi_{SU(6)}^{+\,\uparrow}(qD^0_0)=\Big (
\sqrt{\frac{2}{3}}\{u^\uparrow u^\uparrow d^\downarrow\}
-\sqrt{\frac{1}{3}}
\{u^\uparrow u^\downarrow d^\uparrow\}\Big)\varphi_0(1,2,3).
\eea

Likewise, we can construct a nucleon as $qD^1_1$ system --
 the wave function of $N_{J=1/2}^{+\,\uparrow}(qD^1_1)$ is written
at arbitrary $n$ as
\bea
\label{v-14}
&&\Psi_{J=1/2}^{+\,\uparrow}(qD^1_1)
=\frac 13\Big(
    u^\uparrow (1)D^{10}_{10}(23)\Phi_1(1;23)
  + u^\uparrow (2)D^{10}_{10}(31)\Phi_1(2;31)
  \nn \\
&&\qquad\qquad\qquad\qquad
  + u^\uparrow (3)D^{10}_{10}(12)\Phi_1(3;12)\Big)+
\nn  \\
&+&(-\frac{\sqrt 2}{3})\Big(
    u^\downarrow (1)D^{10}_{11}(23)\Phi_1(1;23)
  + u^\downarrow (2)D^{10}_{11}(31)\Phi_1(2;31)
  \nn \\
&&\qquad\qquad\qquad\qquad
  + u^\downarrow (3)D^{10}_{11}(12)\Phi_1(3;12)\Big )
\nn \\
&+&(-\frac{\sqrt 2}{3})\Big(
    d^\uparrow (1)D^{11}_{10}(23)\Phi_1(1;23)
  + d^\uparrow (2)D^{11}_{10}(31)\Phi_1(2;31)
  \nn \\
&&\qquad\qquad\qquad\qquad
  + d^\uparrow (3)D^{11}_{10}(12)\Phi_1(3;12)\Big )
\nn  \\
&+&\frac{ 2}{3}\Big(
    d^\downarrow (1)D^{11}_{11}(23)\Phi_1(1;23)
  + d^\downarrow (2)D^{11}_{11}(31)\Phi_1(2;31)
  \nn \\
&&\qquad\qquad\qquad\qquad
  + d^\downarrow (3)D^{11}_{11}(21)\Phi_1(3;12)\Big)\, .
\eea
In the limit of  Eq. (\ref{qds-SU6}), which means the $SU(6)$ symmetry
for $qD^1_1$ states, we have
\bea \label{v-15}
&&\Psi_{SU(6)}^{+\,\uparrow}(qD^1_1)= \Big(
\sqrt{\frac{2}{3}}\{u^\uparrow u^\uparrow d^\downarrow\}
-\sqrt{\frac{1}{3}}
\{u^\uparrow u^\downarrow d^\uparrow\}\Big)\varphi_1(1,2,3) .
\eea
One can see that, if
 \beq
 \label{v-15a}
\varphi_0(1,2,3)\neq \varphi_1(1,2,3),
 \eeq
 we have two different nucleon states corresponding to two different
 diquarks, $D^0_0$ and $D^1_1$.

If we require
 \beq
 \label{v-15b}
\varphi_0(1,2,3)= \varphi_1(1,2,3),
 \eeq
it makes possible to have one level  only,
not two, that means the $SU(6)$ symmetry.

Recall that in the $SU(6)$ limit the nucleon can be
presented as a mixture of  both
diquarks: to be illustrative, we rewrite the spin--flavour part of the
proton wave function as follows:
\bea \label{v-15aa}
&&\sqrt{\frac{2}{3}}\{u^\uparrow u^\uparrow d^\downarrow\}
-\sqrt{\frac{1}{3}}
\{u^\uparrow u^\downarrow d^\uparrow\}
=
\frac{1}{\sqrt{2}}u^\uparrow (1) D^0_0(23)
+\frac{1}{3\sqrt{2}}u^\uparrow (1) D^{10}_{10}(23)
\nn \\
&&
-\frac{1}{3}d^\uparrow (1) D^{11}_{10}(23)
-\frac{1}{3}u^\downarrow (1) D^{10}_{11}(23)
+\frac{\sqrt{2}}{3}d^\downarrow (1) D^{11}_{11}(23) \ .
\eea
So, the nucleon in the $SU(6)$ limit is a mixture of $qD^0_0$ and
$qD^1_1$ states in equal proportion.

\subsubsection{The state with $I=1/2$, $J=3/2$, $L=0$ and $S=3/2$,
$N_{J^P=\frac 32 ^+}(L=0,S=3/2)$}

Let us write down the wave function for
$N^\uparrow _{J^P=\frac 32 ^+}(L=0,S=3/2)$ with
$I_Z=\frac 12$ and $S_Z=\frac 32$:
\bea \label{qds-26} &&
 C^{\frac 12\frac 12}_{1\,0\;\;\frac 12\frac 12}\,
 C^{\frac 32\,\frac 32}_{1\,1\;\;\frac 12\,\frac 12}
\Big[  u^\uparrow (1)D^{10}_{11}(23)\Phi_1(1;23)
      + u^\uparrow (2)D^{10}_{11}(31)\Phi_1(2;31)
      \nn \\
&&\qquad\qquad\qquad\qquad
      + u^\uparrow (3)D^{10}_{11}(12)\Phi_1(3;12)\Big]
\nn  \\
&+&
C^{\frac 12\frac 12}_{1\,1\;\;\frac12 -\frac12}\,
C^{\frac 32\frac 32}_{1\,1\;\;\frac12\,\frac12}
\Big[ d^\uparrow (1)D^{11}_{11}(23)\Phi_1(1;23)
  +    d^\uparrow (2)D^{11}_{11}(31)\Phi_1(2;31)
  \nn \\
&&\qquad\qquad\qquad\qquad
  +    d^\uparrow (3)D^{11}_{11}(12)\Phi_1(3;12)\Big ]
\nn \\
&=&
 -\frac{1}{\sqrt 3}
\Big[  u^\uparrow (1)
\frac{u^\uparrow (2)d^\uparrow (3)+d^\uparrow (2)u^\uparrow
(3)}{\sqrt 2}\Phi_1(1;23)+(1\rightleftharpoons 2)
                         +(1\rightleftharpoons 3)  \Big] \nn \\
&&
+
\sqrt{\frac 23}
\Big[ d^\uparrow (1)u^\uparrow (2)u^\uparrow (3)\Phi_1(1;23)
+(1\rightleftharpoons 2)
                         +(1\rightleftharpoons 3)  \Big]\, .
\eea
In the $SU(6)$ limit, under the constraint of Eq. (\ref{qds-SU6}), the
wave function for the $\Delta^\uparrow _{J^P=\frac 12 ^+}(L=0,S=3/2)$
is equal to zero:
\bea \label{qds-27} &&
 -\frac{1}{\sqrt 3}
\Big[  u^\uparrow (1)
\frac{u^\uparrow (2)d^\uparrow (3)+d^\uparrow (2)u^\uparrow
(3)}{\sqrt 2}+(1\rightleftharpoons 2)
                         +(1\rightleftharpoons 3)
                         \Big]\varphi_1(1,2,3) \nn \\
&&
+
\sqrt{\frac 23}
\Big[ d^\uparrow (1)u^\uparrow (2)u^\uparrow (3)
+(1\rightleftharpoons 2)
  +(1\rightleftharpoons 3)
 \Big]\varphi_1(1,2,3)=0\, .
\eea
Radial excitation wave functions in the $SU(6)$ limit, if
they depend on $s$ only (see Eq.
(\ref{example2}) for example),  are equal to zero too.

So, in the $SU(6)$ limit the nucleon state with $L=0$ and
$J^P=\frac 32 ^+$ does not exist.

\subsection{Wave functions of quark--diquark systems with $L\neq 0$
  } \label{Section-WF-L123}

Let us consider, first, the $\Delta$ isobar at $I_Z=3/2$ with fixed
$J,J_Z$,  total spin $S$ and orbital momentum $L$.
The wave function for this state at arbitrary $n$ reads
\bea \label{L-28}
&&
\sum\limits_{S_Z,m_z}
 C^{J\,J_Z}_{L\,J_Z-S_Z\;\;S\,S_Z}
 C^{S\,S_Z}_{1\,S_Z-m_z\;\;\frac 12\,m_z}
 C^{\frac 32\frac 32}_{1\,1\;\;\frac 12\frac 12}
 \Big(  u^{m_z} (1)D^{11}_{1\, S_Z-m_z}(23)
\nn \\
&&\qquad\qquad\times
|\vec k_{1\, cm}|^LY_{L}^{J_Z-S_Z}(\theta_1,\phi_1)\Phi_1^{(L)}(1;23)
+(1\rightleftharpoons 2)
  +(1\rightleftharpoons 3)
      \Big)\, .
  \eea
Here $|\vec k_{1\, cm}|$  and $(\theta_1,\phi_1)$ are the momenta
and momentum angles of the first quark in the c.m. system.

For other $I_Z$, one should include into  wave function the
summation over isotopic states, that means the following substitution in
(\ref{L-28}):
\beq \label{L-29}
 C^{\frac 32\frac 32}_{1\,1\;\;\frac 12\frac 12}
  u^{m_z} (1)D^{11}_{1\, S_Z-m_z}(23)\to
\sum\limits_{j_z}
 C^{\frac 32 I_Z}_{1\,I_Z-j_z\;\;\frac 12 \, j_z}
  q^{m_z}_{j_z} (1)D^{1\, I_Z-j_z}_{1\, S_Z-m_z}(23) \ .
  \eeq
One can see that wave functions of neither(\ref{L-28}) nor
(\ref{L-29}) give us zeros, when $\Phi_1^{(L)}(1;23)$ depends on $s$
only. Indeed, in this limit we have
\bea \label{L-30}
&&
\sum\limits_{S_Z,m_z}
 C^{J\,J_Z}_{L\,J_Z-S_Z\;\;S\,S_Z}
 C^{S\,S_Z}_{1\,S_Z-m_z\;\;\frac 12\,m_z}
\sum\limits_{j_z}
 C^{\frac 32 I_Z}_{1\,I_Z-j_z\;\;\frac 12 \, j_z}
\nn \\
&&\times\Big(  q^{m_z}_{j_z} (1)D^{1\, I_Z-j_z}_{1\, S_Z-m_z}(23)
|\vec k_{1\, cm}|^LY_{L}^{J_Z-S_Z}(\theta_1,\phi_1)
+(1\rightleftharpoons 2)
  +(1\rightleftharpoons 3)
      \Big)\phi_1^{(L)}(s)\, .\qquad
  \eea
The factor $|\vec k_{1\, cm}|^LY_{L}^{J_Z-S_Z}(\theta_1,\phi_1)$ and
analogous ones in
$(1\rightleftharpoons 2)$ and $(1\rightleftharpoons 3)$ prevent the
cancelation of different terms in big parentheses of Eq. (\ref{L-30}),
which are present in case of $L=0$, see Eq. (\ref{qds-20}).

For nucleon states $(I=1/2)$ we write:
\bea \label{L-31}
&&
\sum\limits_{S_Z,m_z}
 C^{J\,J_Z}_{L\,J_Z-S_Z\;\;S\,S_Z}
 C^{S\,S_Z}_{1\,S_Z-m_z\;\;\frac 12\,m_z}
\sum\limits_{j_z}
 C^{\frac 12 I_Z}_{1\,I_Z-j_z\;\;\frac 12 \, j_z}
 \Big(  q^{m_z}_{j_z} (1)D^{1\, I_Z-j_z}_{1\, S_Z-m_z}(23)
\nn \\
&&\qquad\qquad\times
|\vec k_{1\, cm}|^LY_{L}^{J_Z-S_Z}(\theta_1,\phi_1)
\Phi_{1}^{(L)}(1;23) +(1\rightleftharpoons 2)
  +(1\rightleftharpoons 3)
      \Big)\, .\qquad
  \eea

The $SU(6)$ limit, as previously, is
realized at $\Phi_{1}^{(L)}(i;j\ell)\to\varphi_1^{(L)}(s)$. Then one has
instead of (\ref{L-31}):
\bea \label{L-31a}
&&
\sum\limits_{S_Z,m_z}
 C^{J\,J_Z}_{L\,J_Z-S_Z\;\;S\,S_Z}
 C^{S\,S_Z}_{1\,S_Z-m_z\;\;\frac 12\,m_z}
\sum\limits_{j_z}
 C^{\frac 12 I_Z}_{1\,I_Z-j_z\;\;\frac 12 \, j_z}
\nn \\
&&\times\Big(  q^{m_z}_{j_z} (1)D^{1\, I_Z-j_z}_{1\, S_Z-m_z}(23)
|\vec k_{1\, cm}|^LY_{L}^{J_Z-S_Z}(\theta_1,\phi_1)
 +(1\rightleftharpoons 2)
  +(1\rightleftharpoons 3)
      \Big)\varphi_{1}^{(L)}(s)\, .\qquad
  \eea
For $qD^0_0$ states the wave function in the general case reads
\bea \label{L-31b}
&&
\sum\limits_{m_z}
 C^{J\,J_Z}_{L\,J_Z-m_z\;\;\frac 12\,m_z}
\Big(  q^{m_z}_{I_z} (1)D^{0}_{0}(23)
|\vec k_{1\, cm}|^LY_{L}^{J_Z-m_z}(\theta_1,\phi_1)\Phi_{0}^{(L)}(1;23)
\nn \\
&&\qquad\quad\qquad\qquad\qquad
+(1\rightleftharpoons 2)
  +(1\rightleftharpoons 3)
      \Big)\ . \qquad
  \eea
In the $SU(6)$ limit we have:
\bea \label{L-31c}
&&
\sum\limits_{m_z}
 C^{J\,J_Z}_{L\,J_Z-m_z\;\;\frac 12\,m_z}
\Big(  q^{m_z}_{I_z} (1)D^{0}_{0}(23)
|\vec k_{1\, cm}|^LY_{L}^{J_Z-m_z}(\theta_1,\phi_1)
\nn \\
 &&\qquad\quad\qquad\qquad\qquad
 +(1\rightleftharpoons 2)
  +(1\rightleftharpoons 3)
      \Big)\varphi_0^{(L)}(s)\, .\qquad
  \eea
  Baryons are characterized by $I$ and $J^P$ -- the states with different
  $S$ and $L$ can mix. To select independent states, one may
  orthogonalize wave functions with the same isospin and $J^P$. The
  orthogonalization depends on the structure of  momentum/coordinate
  parts $\Phi_{1}^{(L)}(i;j\ell)$. But in  case of the $SU(6)$ limit
  the momentum/coordinate wave functions transform in a common
  factor $\Phi_{1}^{(L)}(i;j\ell),\; \Phi_{0}^{(L)}(i;j\ell)\to
  \varphi_{SU(6)}^{(L)}(s)$, and one can orthogonalize the spin
  factors. Namely, we can present the $SU(6)$ wave function as follows:
\beq \label{L-31d}
  \Psi^{(A)}_{J^P}=Q^{(A)}_{J^P}\varphi_{SU(6)}^{(A)}(s) ,
\eeq
where $Q^{(A)}_{J^P}$ is the spin operator and
$A=I,II,III,...$ refer to different $(S,L)$. The orthogonal set of operators
$Q^{(A)}_{J^P}$ is constructed in a standard way:
\bea \label{L-31e}
&& Q^{(\perp I)}_{J^P}\equiv Q^{(\perp A)}_{J^P}\, ,
\nn \\
&&
Q^{(\perp II)}_{J^P}= Q^{( II)}_{J^P}-Q^{(\perp I)}_{J^P}
\frac{\Big(Q^{(\perp I)+}_{J^P}Q^{( II)}_{J^P} \Big)}
{\Big(Q^{(\perp I)+}_{J^P}Q^{(\perp I)}_{J^P} \Big)}\, ,
\nn \\
&&
Q^{(\perp III)}_{J^P}= Q^{(III)}_{J^P}-
Q^{(\perp I)}_{J^P}
\frac{\Big(Q^{(\perp I)+}_{J^P}Q^{( III)}_{J^P} \Big)}
{\Big(Q^{(\perp I)+}_{J^P}Q^{(\perp I)}_{J^P} \Big)}-
Q^{(\perp II)}_{J^P}
\frac{\Big(Q^{(\perp II)+}_{J^P}Q^{( III)}_{J^P} \Big)}
{\Big(Q^{(\perp II)+}_{J^P}Q^{(\perp II)}_{J^P} \Big)},
  \eea
and so on. The convolution of operator
$\Big(Q^{(A)+}_{J^P}Q^{(B)}_{J^P} \Big)$ includes both the summation over
quark spins and integration over quark momenta.

\begin{figure}[h]
%Fig. 2
\centerline{\epsfig{file=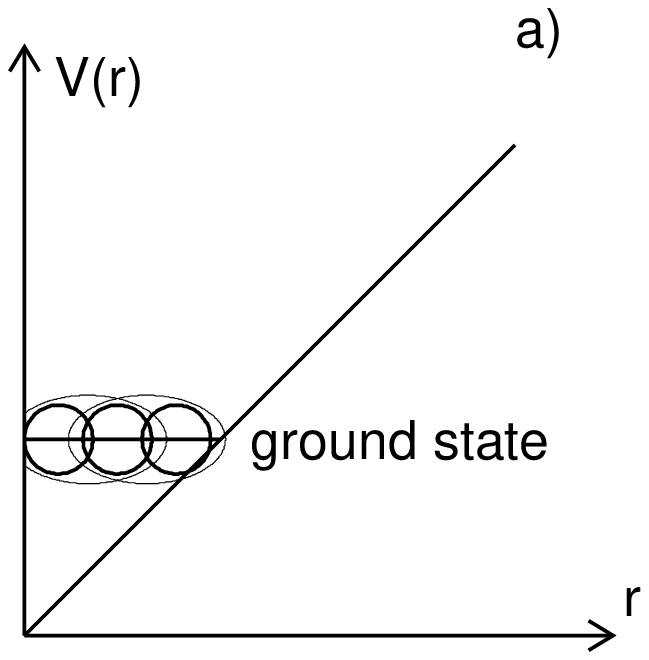,width=40mm}
            \epsfig{file=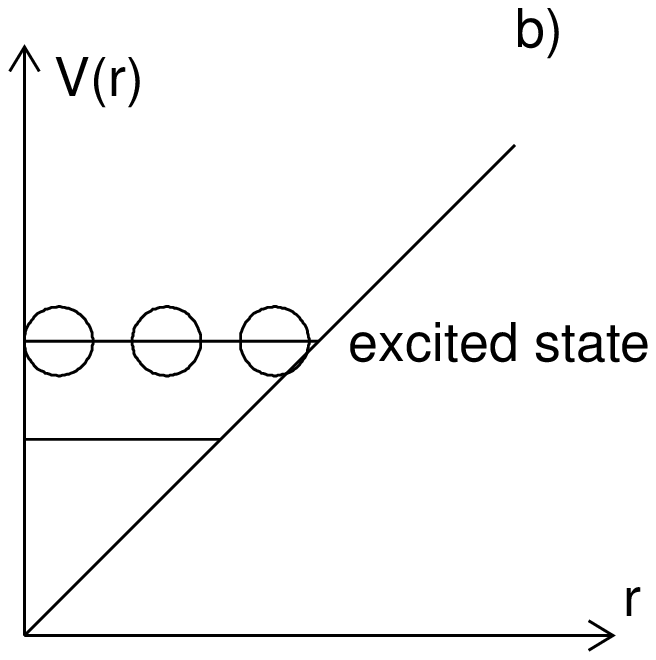,width=40mm}
            \epsfig{file=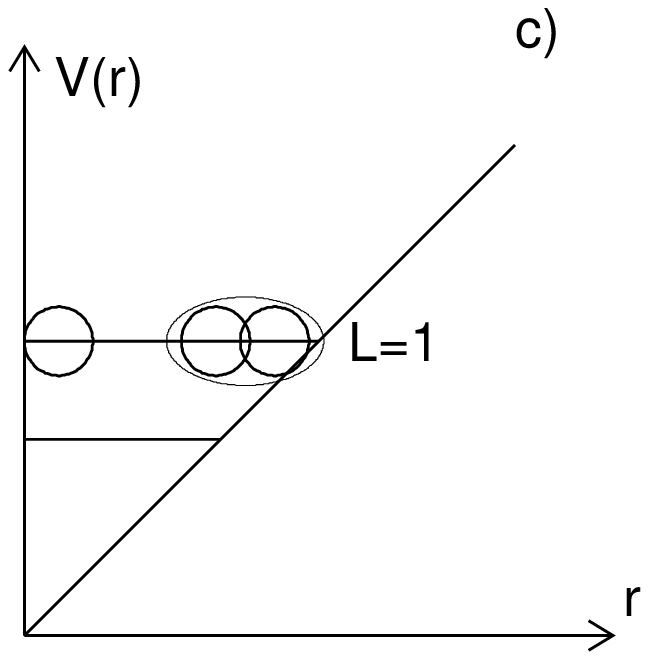,width=40mm}}
\caption{Illustration of
the quark--diquark structure of baryon levels.
a) Ground state  with a complete mixing of  constituent
quarks: the bound quarks and diquarks, being compressed states, provide
us the  three-quark $SU(6)$-symmetry structure.
b) Conventional picture for an excited state in the standard three-quark
model with three spatially separated quarks. c) Example of the excited
state in the quark--diquark model: the quark--diquark state with $L=1$,
 quark and diquark being
 spatially separated. }\label{1}
\end{figure}

\subsection{Quarks and diquarks in baryons}

Exploring the notion of constituent quark and composite diquark, we
propose several schemes for the structure of  low-lying baryons.
To be illustrative, let us turn to  Fig. \ref{1}.

Using potential picture, the
standard scheme of the three-quark baryon is shown in Figs. \ref{1}a
and \ref{1}b.
On the lowest level, there are three $S$-wave quarks ---
it is a compact system (the radii of constituent
quarks are of the order of $\sim 0.2$ fm, while the nucleon size is
$\sim 0.8$ fm \cite{book2,ALR}).
So, for the ground state, which  is a  system
with the overlapping different quark pairs
(Fig. \ref{1}a), the hypothesis about
 the $SU(6)$ classification seems rather reliable.

As concern the excited states, the
quarks of the
 standard quark model (see the example in Fig. \ref{1}b),  are in the
 average located at comparatively large distances from each other.
 Such a
three-quark composite system is characterized by pair excitations --
the number of pair excitations may be large, thus resulting in a quick
increase of the number of excited baryons.

The quark--diquark structure of levels, supposed in our consideration
for $L>0$, is demonstrated in Fig. \ref{1}c. For excited states, we
assume the following quark--diquark picture: two quarks are at
comparatively small distances, being a diquark state, and the third
quark is separated from this diquark. The number of quark--diquark
excitations is noticeably less than the number of excitations in the
three-quark system.

   \subsubsection{ Diquark composite systems and mass distributions of diquarks}

 Constituent quarks and  diquarks are effective particles.
 We assume that propagators of the diquark composite systems
  can be well described using K\"allen--Lehman representation
  \cite{K-L}. For scalar and axial--vector diquarks, the propagators
  read
    \bea \label{A-1}
 && \Pi^{(D^0_0)}(p)=
  \int\limits_{m^2_{min}}^\infty dm^2_D
  \frac{\rho_{D^0_0}(m^2_D)}{m^2_D-p^2-i0}  ,\nn \\
&&
  \Pi_{\mu\nu}^{(D^1_1)}(p)=-g_{\mu\nu}^{\perp p}
  \int\limits_{m^2_{min}}^\infty
  dm^2_D \frac{\rho_{D^1_1}(m^2_D)}{m^2_D-p^2-i0},
  \eea
  where the mass  distributions $\rho_{D^0_0}(m^2_D)$ and
  $\rho_{D^1_1}(m^2_D)$
 are characterized by the compactness of the scalar and axial--vector
 diquarks. The use of mass propagator
  (\ref{A-1}) is definitely needed in the calculation of subtle effects in
  baryonic  reactions.
  However, in a rough approximation
  one may treat diquarks, similarly to constituent quarks, as
  effective particles:
  \bea \label{A-2}
  &&
  \rho_{D^0_0}(m^2_D) \to \delta(m^2_D-M^2_{D^0_0}),  \nn\\
  &&
  \rho_{D^1_1}(m^2_D) \to \delta(m^2_D-M^2_{D^1_1}).
 \eea
 We expect the diquark mass to be in the region of the 600-900 MeV
 \cite{AGS}.

\begin{figure}[h]
%Fig. 3
\centerline{\epsfig{file=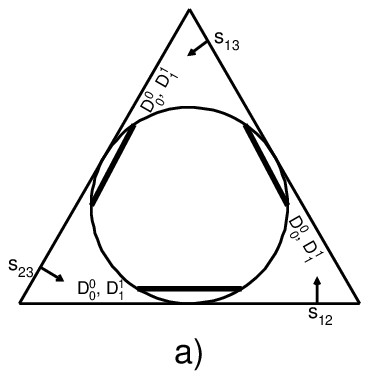,width=70mm}
            \epsfig{file=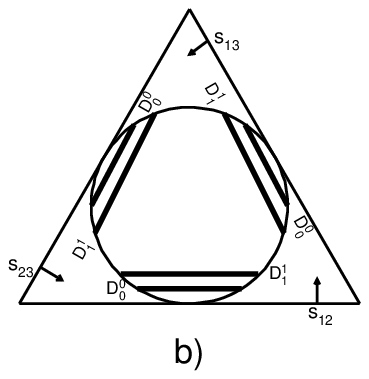,width=70mm}}
\caption{  Dalitz plots for three-quark systems at a)
$M^2_{D^0_0}=M^2_{D^1_1}$ and b) $M^2_{D^0_0}\neq
M^2_{D^1_1}$}\label{dalitz} \end{figure}

 Mass distributions for three-quark systems in the approximation of
 (\ref{A-2}) at fixed $s=s_{12}+s_{13}+s_{23}-3m^2$ can be shown on the
 Dalitz-plot. In Fig. \ref{dalitz}, we show the Dalitz-plots for
 the approach of short-range  diquarks -- conventionally, below we
 use (\ref{A-2}).
  In Figs. \ref{dalitz}a and \ref{dalitz}b the cases
  $M^2_{D^0_0}=M^2_{D^1_1}$ and $M^2_{D^0_0}\neq M^2_{D^1_1}$
 are demonstrated, correspondingly.

\subsubsection{Normalization condition for wave functions of
quark--diquark systems}

The model with spatially separated quark and diquark
results in a specific orthogonality/normalization condition. The matter
is that interference terms with different diquarks provide a small
contribution. For example,
\bea \label{L-32}
&&\int d\Phi_3|\vec k_{1\,
cm}|^LY_{L}^{J_Z-S_Z}(\theta_1,\phi_1) \Phi_{1}^{(L)}(1;23)
\nn  \\
&&\times \Big (
|\vec k_{2\, cm}|^{L}Y_{L}^{J_Z-S_Z}(\theta_2,\phi_2)
\Phi_{1}^{(L)}(2;13)\Big)^+\simeq 0\, .
  \eea
 Below we neglect such interference terms.

Therefore, the normalization condition,
\be \label{L-32a}
\int d\Phi_3
\Big|\Psi_{J,J_Z}^{(L,S)}(1,2,3)\Big|^2=1\, ,
\ee
So, for the $qD^1_1$ systems, we re-write
(\ref{L-32a}) as follows:
\bea \label{L-33}
&&\int d\Phi_3\Big|\sum\limits_{S_Z,m_z}
 C^{J\,J_Z}_{L\,J_Z-S_Z\;\;S\,S_Z}
 C^{S\,S_Z}_{1\,S_Z-m_z\;\;\frac 12\,m_z}
\sum\limits_{j_z}
 C^{I, I_Z}_{1\,I_Z-j_z\;\;\frac 12 \, j_z}
 q^{m_z}_{j_z} (1)D^{1\, I_Z-j_z}_{1\, S_Z-m_z}(23)
\nn \\
&&\times
|\vec k_{1\, cm}|^LY_{L}^{J_Z-S_Z}(\theta_1,\phi_1)
\Phi_{1}^{(L)}(1;23) \Big |^2
+\int d\Phi_3\Big|1\rightleftharpoons 2\Big|^2
+\int d\Phi_3\Big|1\rightleftharpoons 3\Big|^2 =1,\qquad
\eea
while for  $qD^0_0$  we have
\bea \label{L-33a}
&&\int d\Phi_3\Big|\sum\limits_{m_z}
 C^{J\,J_Z}_{L\,J_Z-m_z\;\;1/2\,m_z}
q^{m_z}_{j_z} ( 1)D^{0}_{0}(23)
|\vec k_{1\, cm}|^LY_{L}^{J_Z-m_z}(\theta_1,\phi_1)
\Phi_{1}^{(L)}(1;23) \Big |^2
\nn \\
&&+\int d\Phi_3\Big| 1\rightleftharpoons 2\Big|^2
  +\int d\Phi_3\Big|1\rightleftharpoons 3\Big|^2 =1  .
\eea
Here we suppose that $L$ and $S$ are good quantum numbers. If no,
 one should take into account the mixing in each term of
(\ref{L-33}). Let us emphasize that under hypothesis (\ref{L-33}),
(\ref{L-33a}) the mixing of terms with different diquarks is forbidden.

\section{The setting of  states with $L=0$  and the  $SU(6)$ symmetry }

This section is devoted to the basic $L=0$ states and
their radial excitations. But, first, let us overlook the situation
with the observed baryons -- some of them need
comments.

\subsubsection{Baryon spectra for the excited
states}

The masses of the well-established states (3 or 4 stars in the
Particle Data Group classification \cite{PDG}) are taken as a mean
value over the interval given by PDG, with errors covering this
interval. But the states established not so definitely require special
 discussion.

We have introduced two $S_{11}$ states in the region of 1900 and
2200 MeV, which are classified by PDG as $S_{11}(2090)$. Indeed, the
observation of a state with mass
$2180\pm 80$ MeV by Cutkosky \cite{Cutkosky:1980rh} can be hardly
compatible with observations
 \cite{Manley:1992yb,Vrana:1999nt,Plotzke:1998ua} of an $S_{11}$ state
 with the mass around 1900 MeV.

The same procedure has been applied to the states $D_{13}$ around 2000
MeV. Here the first state is located in the region of 1880 MeV and was
observed in the analyses
\cite{Cutkosky:1980rh,Manley:1992yb,Bell:1983dm}. This
state is also well compatible with the analysis of photoproduction
reactions \cite{Sarantsev:2005tg}. The second state is located in
the region 2040 MeV: its mass has been obtained as an average value
over the results of
\cite{Cutkosky:1980rh,Vrana:1999nt,Hohler:1973ww}.

The $P_{11}(1880)$ state has been observed by Manley
\cite{Manley:1992yb} as well as in the analyses of the photoproduction
data with open strangeness \cite{Sarantsev:2005tg,Anisovich:2007bq}: we
consider this state as well established. Thus, for the state
$P_{11}(2100)$ we have taken the mass as an average value over
all the measurements quoted by PDG:
\cite{Cutkosky:1980rh,Vrana:1999nt,Plotzke:1998ua,Hohler:1973ww,Ablikim:2006mb,Batinic:1995kr}.

We also consider $D_{15}(2070)$ as an established state. It has been
observed in the $\eta$ photoproduction data \cite{Crede:2003ax},
although we understand that a confirmation of this state by other
data is needed.  Furthermore, we have taken for $D_{15}(2200)$ the
average value, using
\cite{Cutkosky:1980rh,Hohler:1973ww,Batinic:1995kr} analyses which give
compatible results.

As to the $\Delta$ sector, we see that the $\frac 52^-$ state
observed in the analysis of the GWU group \cite{Arndt:2006bf} with
mass 2233 MeV and quoted as $D_{35}(1930)$ can be hardly compatible
with other observations, which give the results in the
region of 1930 MeV. Moreover, the GWU result is compatible with the
analysis of Manley \cite{Manley:1992yb}, which is quoted by PDG as
$D_{35}(2350)$, though it gives the mass $2171\pm 18$  MeV. Thus, we
introduce the $\Delta(\frac 52^-)$ state with the mass 2210 MeV and
the error which covers both these results. Then, the mass of the
$D_{35}(2350)$ state is taken as an average value over the results of
\cite{Cutkosky:1980rh,Vrana:1999nt,Hohler:1973ww}.

We also consider the $D_{33}(1940)$ state, which has one star by PDG
classification, as an established one. It is seen very clearly
in the analysis of the $\gamma p\to \pi^0\eta p$ data
\cite{Horn:2007pp,Horn:2008qv}.

One of the most interesting observations concerns
$\Delta(\frac 52^+)$ states. The analyses of Manley
\cite{Manley:1992yb} and Vrana \cite{Vrana:1999nt}
give a state in the region 1740 MeV with
compatible widths. However, this state was confirmed neither by
$\pi N$ elastic nor by photoproduction data. This state is listed by PDG
as $F_{35}(2000)$ together with the observation \cite{Cutkosky:1980rh}
of a state in the elastic $\pi N$ scattering at 2200 MeV. Here we
consider these results as a possible indication to two states: one at
1740 MeV and another at 2200 MeV.

\subsection{The setting of  (L=0) states}

We consider $N_{J^P=\frac 12^+}$ and
 $\Delta_{J^P=\frac 32^+}$ states in two variants: \\
 ({\bf 1})
  $M^2_{D^0_0}=M^2_{D^1_1}$ (see Fig. \ref{dalitz}a),
  the $SU(6)$ symmetry being imposed,
  and \\
 ({\bf 2}) $M^2_{D^0_0}\neq M^2_{D^1_1}$ (see Fig. \ref{dalitz}b) with
the broken $SU(6)$ symmetry constraints.

\subsubsection{The $SU(6)$ symmetry for the nucleon $N_{\frac 12^+}(940)$,
 isobar  $\Delta_{\frac 32 ^+}(1238)$ and their radial
excitations }

In this way
we assume  $M^2_{D^0_0}=M^2_{D^1_1}$ (see Fig. \ref{dalitz}a) and
suppose the $SU(6)$ symmetry for the lowest baryons with $L=0$. It gives
us two ground states, the nucleon $N_{\frac 12^+}(940)$  and the isobar
$\Delta_{\frac 32 ^+}(1238)$ as well as their radial excitations, see
section 2:
\begin{equation}\label{L01}
{\renewcommand{\arraystretch}{0,5}
\begin{tabular}{l|ll}
 $L=0$         &$S=\frac12$, $N(\frac12^+)$&
$S=\frac32$, $\Delta(\frac 32^+) $ \\
      $n=1$    &$ 938\pm 2  $&$1232\pm 4$  \\
      $n=2$    &$1440\pm 40 $&$1635\pm 75$ \\
      $n=3$    &$1710\pm 30 $&$\sim 1920  $\\
      $n=4$    &$1900\pm 100$&$\sim 2190  $  \\
\end{tabular}
}
\end{equation}
Note that the mass-squared splitting of the nucleon
radial excitation  states, $\delta_n M^2(N_{\frac 12^+})$, is of the
order of $1.05\pm 0.15$ GeV$^2$.
 This value is close to that observed in
 meson sector \cite{book3,syst}:
\bea
\label{18spl}
&&M^2[N_{\frac 12^+}(1440)]-M^2[N_{\frac 12^+}(940)]
\nn \\ \simeq
&&M^2[N_{\frac 12^+}(1710)]-M^2[N_{\frac 12^+}(1440)]\equiv
\delta_n M^2(N_{\frac 12^+})\simeq 1.0 \,{\rm GeV}^2.
\eea
 The state with $n=4$ cannot be unambiguously determined.
 Namely, in the region of $1880$ MeV a resonance structure is
 seen, which may be either nucleon radial excitation $(n=4)$ or
  $(S=3/2,L=2,J^P=1/2^+)$ state.
  Also it is possible
  that in the region $\sim 1900$
  there are two poles, not one.
It means that one pole dives into the
 complex-$M$ plane and
  is not observed yet.

One can see that the mass-squared splitting of the
$\Delta_{\frac 32^+}$ isobars,
$\delta_n M^2(\Delta_{\frac 32^+})$, coincides with
that of a  nucleon, $\delta_n M^2(N_{\frac 12^+})$, with a good
accuracy:
\beq \label{17b}
\delta_n M^2(\Delta_{\frac 32^+})=1.07\pm
0.05 \ . \eeq
Let us emphasize that two states, $ \Delta_{\frac32^+}(1600)$
and $ \Delta_{\frac 32^+}(1920)$, are considered here as
radial excitations of $\Delta_{\frac 32^+}(1232)$ with $n=2$ and $n=3$.
However, the resonances  $\Delta_{\frac 32^+}(1600)$ and
$\Delta_{\frac 32^+}(1920)$ can be reliably classified as  $S=1/2,L=2$ and
$S=3/2,L=2$ states, with $n=1$ (see Section 4). Actually, it means that
around $\sim 1600$ MeV  one may expect the double-pole structure, while
the three-pole structure may be at $\sim 1920$ MeV.

\subsubsection{The setting of (L=0) states with broken
  $SU(6)$ symmetry, $M_{D^0_0}\ne M_{D^1_1}$ }

 Here, we consider alternative scheme supposing  diquarks
 $D^0_0$ and $D^1_1$ to have different
 masses, thus being different effective particles -- arguments in favor
 of different  effective masses of $D^0_0$ and $D^1_1$ may be found
 in \cite{Y_5vva75,AGS}.

 In the scheme with two different diquarks, $D^0_0$ and $D^1_1$,
 we have two basic nucleons with corresponding sets
 of  radial excitations.

 The first nucleonic set corresponds to the $qD^0_0$ states,
 the second one describes the $qD^1_1$ states:
\begin{equation}\label{L02}
{\renewcommand{\arraystretch}{0,5}
\begin{tabular}{l|lll}
 $L=0$         &$S=\frac12$, $N(\frac12^+)$&$S=\frac32$, $N(\frac12^+)$&
$S=\frac32$, $\Delta (\frac 32^+) $ \\
      $n=1$    &$ 938\pm 2  $&$1440\pm 40$&$1232\pm 4$  \\
      $n=2$    &$1440\pm 40 $&$1710\pm 30$&$1635\pm 75$ \\
      $n=3$    &$1710\pm 30 $&$2100\pm 100$&$\sim 1920  $\\
      $n=4$    &$2100\pm 100$&               &$\sim 2190  $  \\
\end{tabular}
}
\end{equation}
This scheme requires overlapping states (double-pole structure of the
partial amplitude) in the regions of $M\sim$1400 MeV, 1700 MeV, 1900
MeV. The double-pole structure may be considered as a signature of the
model with two different diquarks, $D^0_0$ and $D^1_1$.

The ($L=0$) set of isobar states coincides with that defined in the
($M^2_{D^0_0}=M^2_{D^1_1}$) scheme.

\section{The setting of baryons with $L>0$ as $(q D^1_1,\,q D^0_0)$  states}

Considering excited states, we analyze several variants,
assuming $M_{D^0_0}\neq M_{D^1_1}$ for the $L\geq 1$ states and the
$SU(6)$ constraints for $L=0$ ones.

\begin{figure}[h]
%Fig.4
\centerline{\epsfig{file=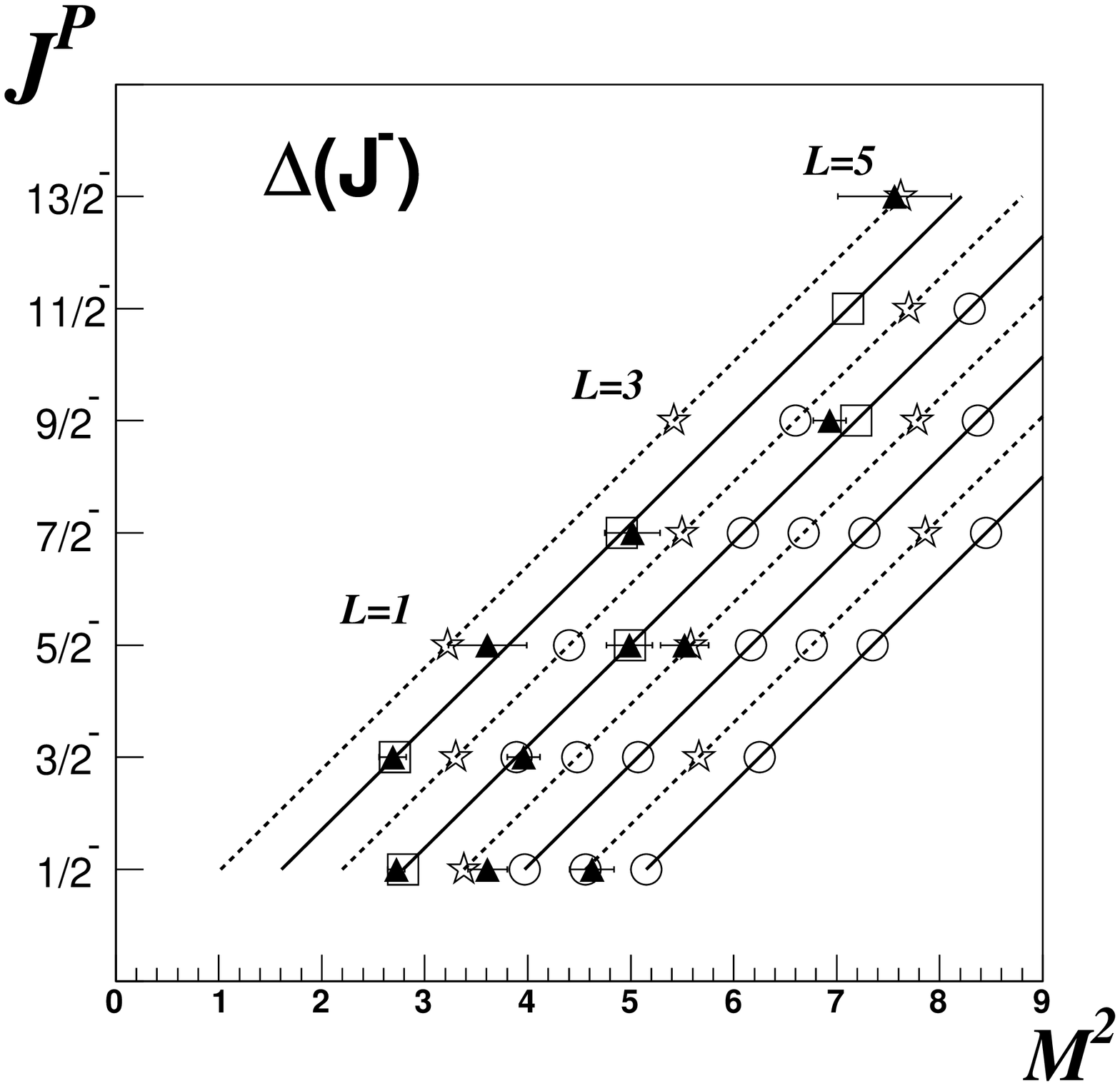,width=70mm}
            \epsfig{file=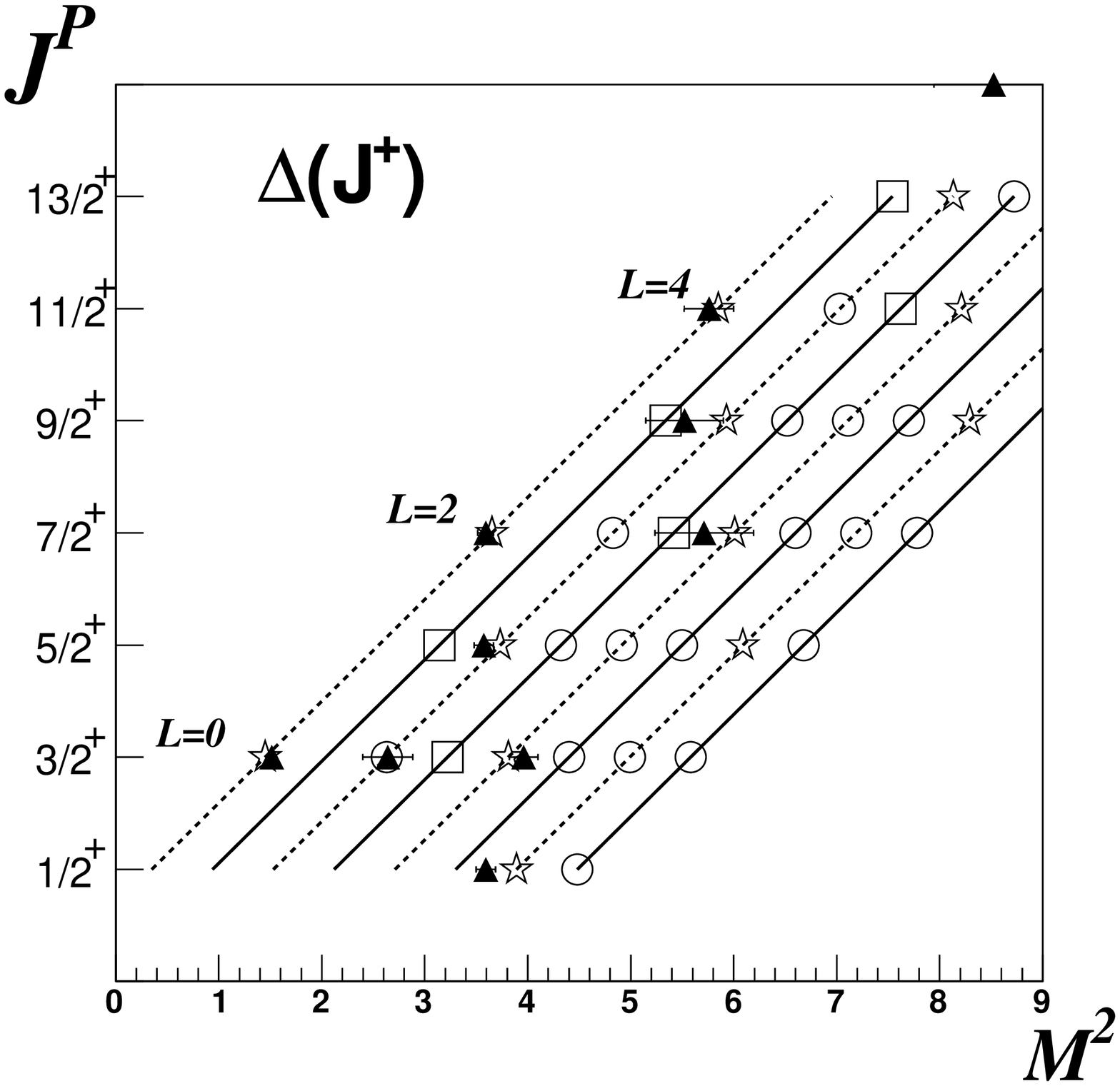,width=70mm}}
\centerline{\epsfig{file=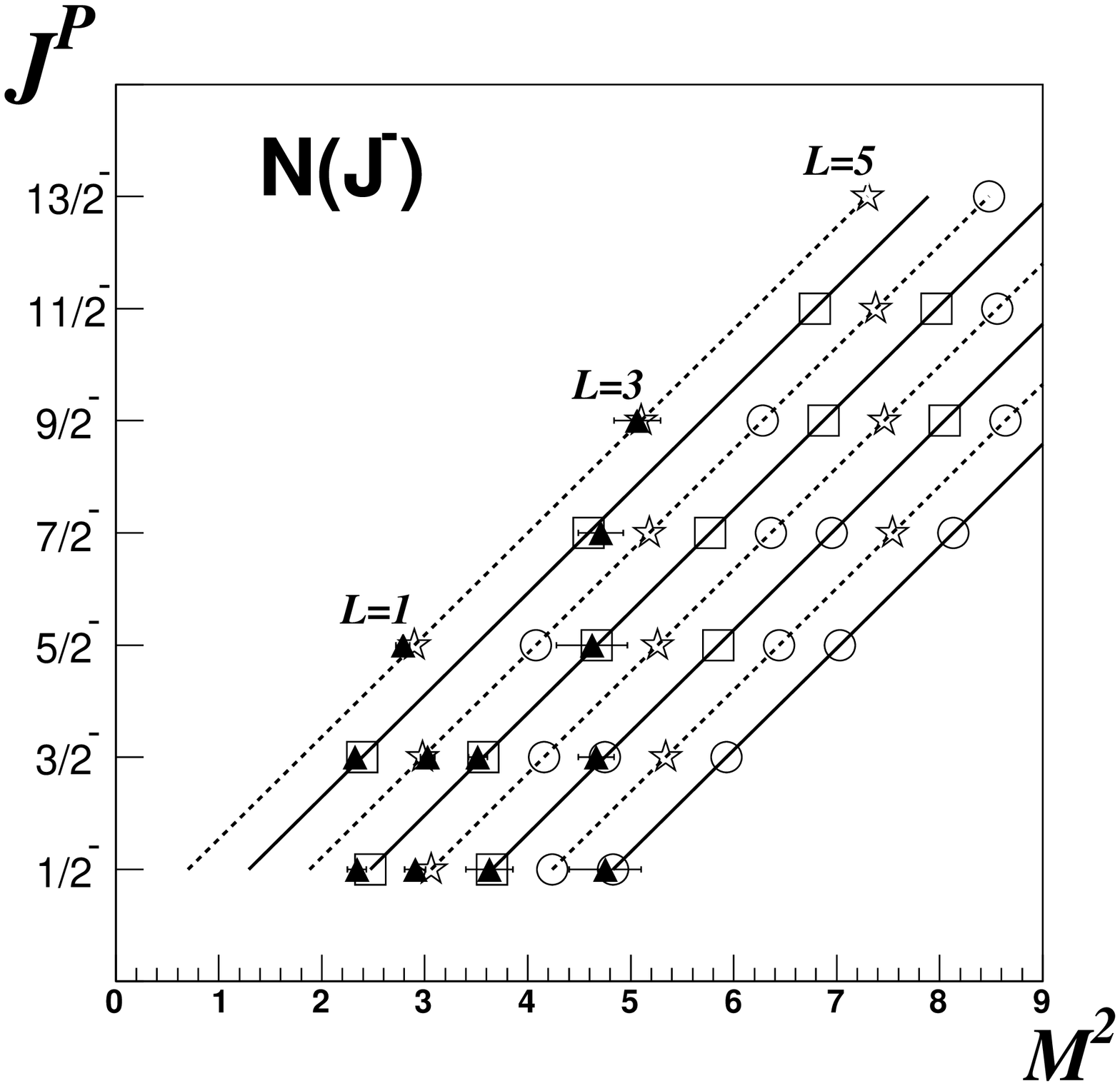,width=70mm}
            \epsfig{file=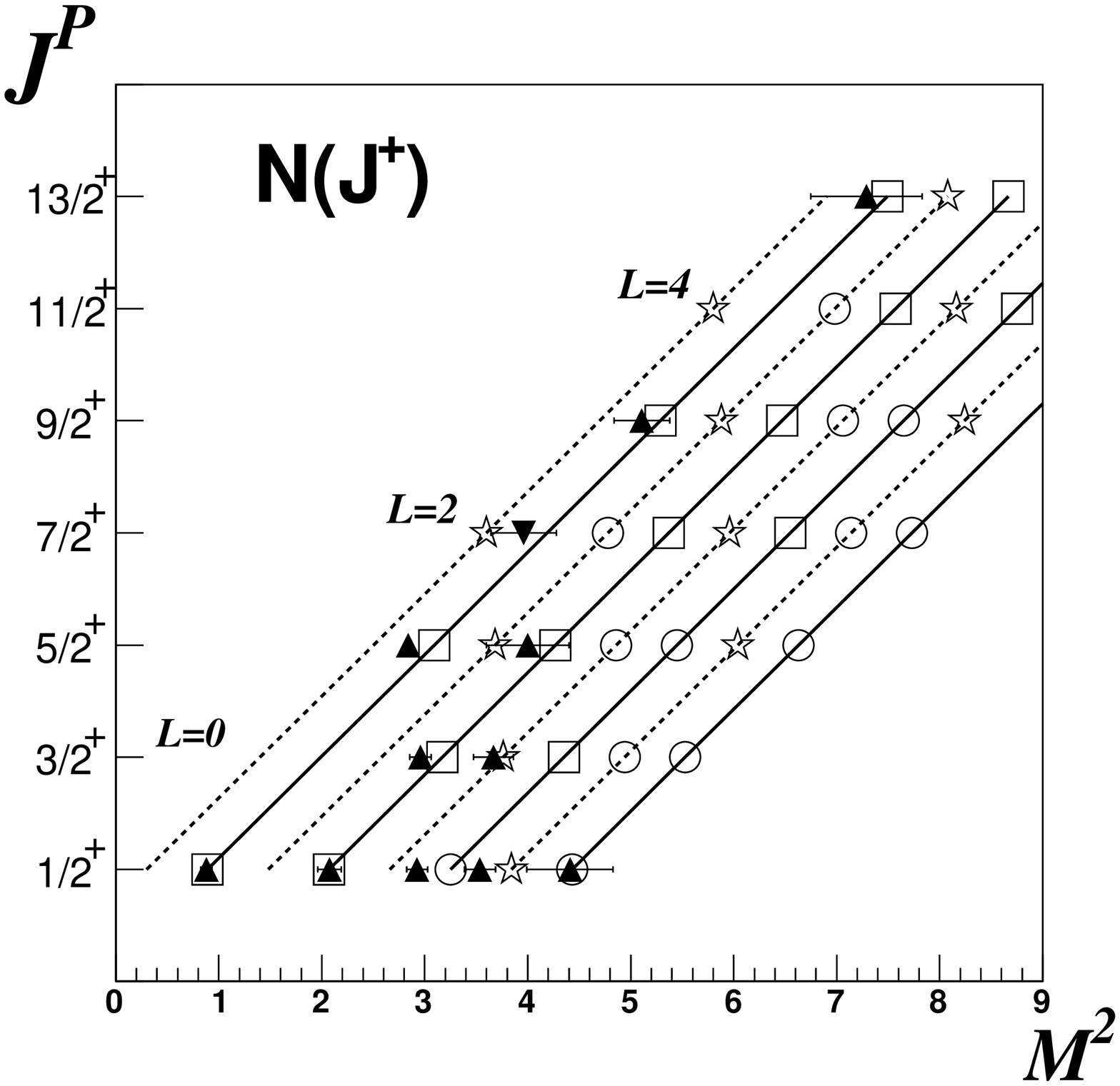,width=70mm}}
\caption{ $(J^P,M^2)$ planes for baryons at
$M_{D^0_0} \neq M_{D^1_1}$ with the $SU(6)$ constraints for $L=0$ states.
 Solid and dashed lines are the trajectories for the states with
 $S=1/2$ and $S=3/2$.
Squares: ground states ($n=1$) with $S=1/2$; stars and rombs: ground states
($n=1$) with $S=3/2$;
circles: radially exited states ($n>1$) with $S=1/2,\quad 3/2$ \label{J-M-G}}
\end{figure}

 In Fig. \ref{J-M-G} we demonstrate the setting of baryons on      the
 $(J^P,M^2)$ planes.
 We see reasonably good description of data, although the scheme
 requires
some additional states as well as  double pole structures in many
cases.

The deciphering of  baryon setting shown  in Fig. \ref{J-M-G} is given
in  (\ref{GMG-1}),  (\ref{GMG-2}), (\ref{GMG-3}), (\ref{GMG-4}) --
 the mass values (in MeV units) are taken from
\cite{PDG,eb,an,al}.

 \newpage

 Let us comment the trajectories in Fig. \ref {J-M-G}.  The states
 belonging to the same $J^P$ trajectories have $\delta J^\pm=2^\pm$ and
 $M^2_{(J+2)^\pm}-M^2_{J^\pm}\simeq 2$ GeV$^2$. Clear examples give us
 $\Delta(\frac 32^+)$ trajectory (the states
 $\Delta_{\frac  32^+}(1231)$, $\Delta_{\frac 72^+}(1895)$,
 $\Delta_{\frac {11}{2}^+}(2400)$, $\Delta_{\frac {13}{2}^+}(2920)$ )
 and $N(\frac  52^-)$ trajectory (the states $N_{\frac 52^-}(1670)$,
 $N_{\frac  92^-}(2250)$, $N_{\frac {13}{2}^-}(2270)$). At the same
 time, in Fig.  \ref {J-M-G} we see
  the lines with $\delta J^\pm=1^\pm$ and
 $M^2_{(J+2)^\pm}-M^2_{J^\pm}\simeq 1$ GeV$^2$: actually, such a line
 represents two overlapping trajectories.

\begin{figure}[h]
%Fig. 5
\centerline{\epsfig{file=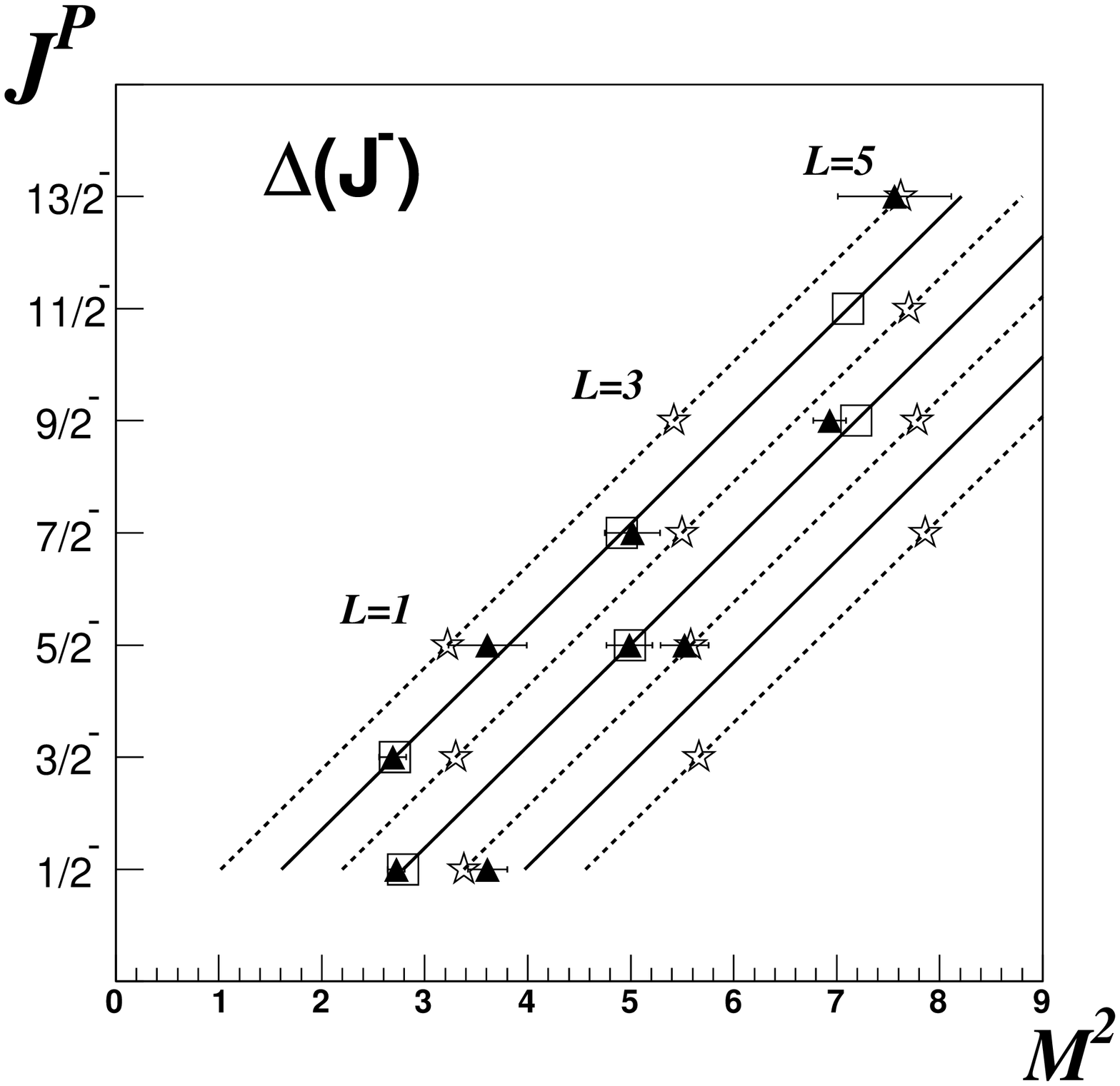,width=70mm}
            \epsfig{file=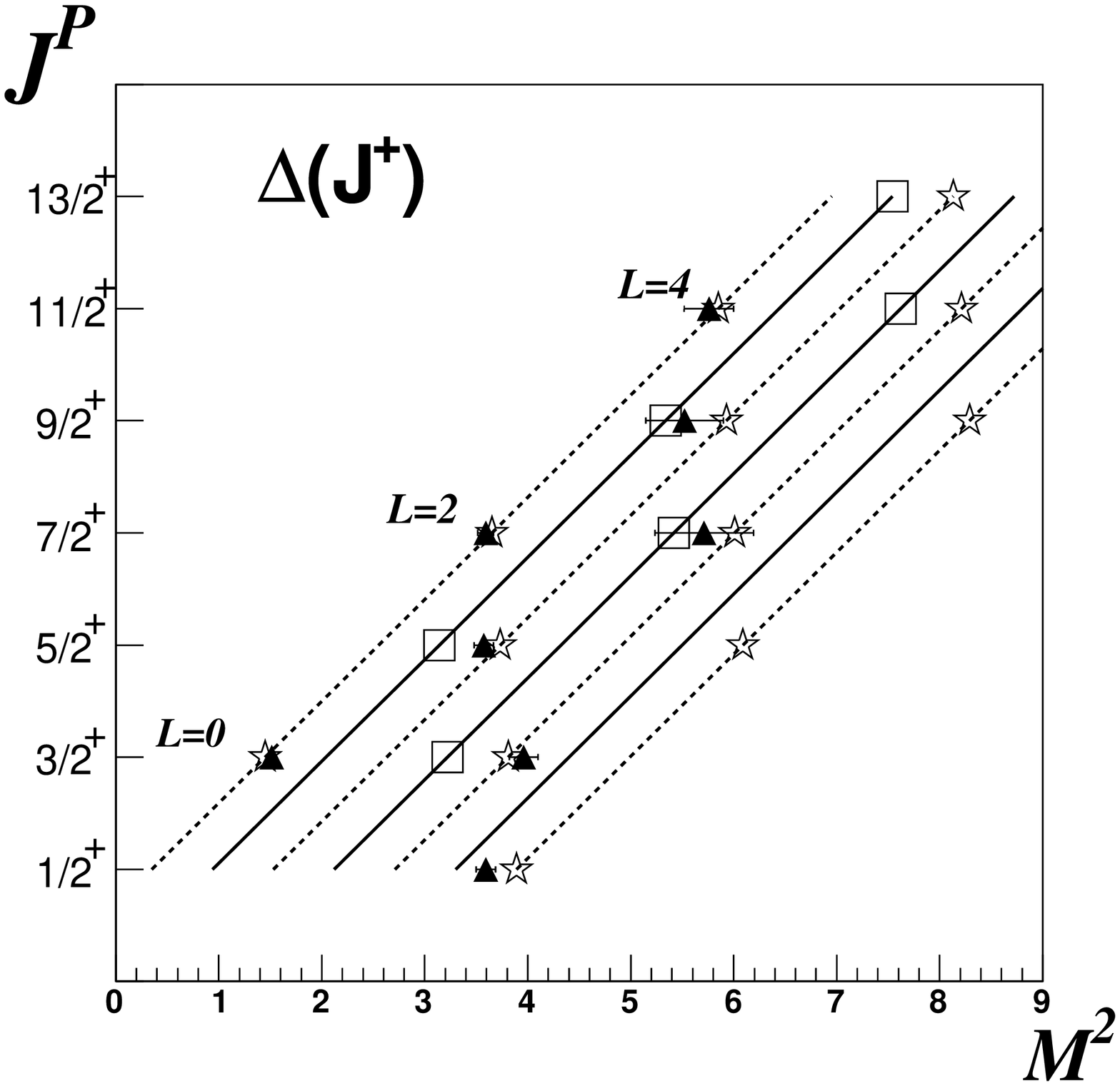,width=70mm}}
\centerline{\epsfig{file=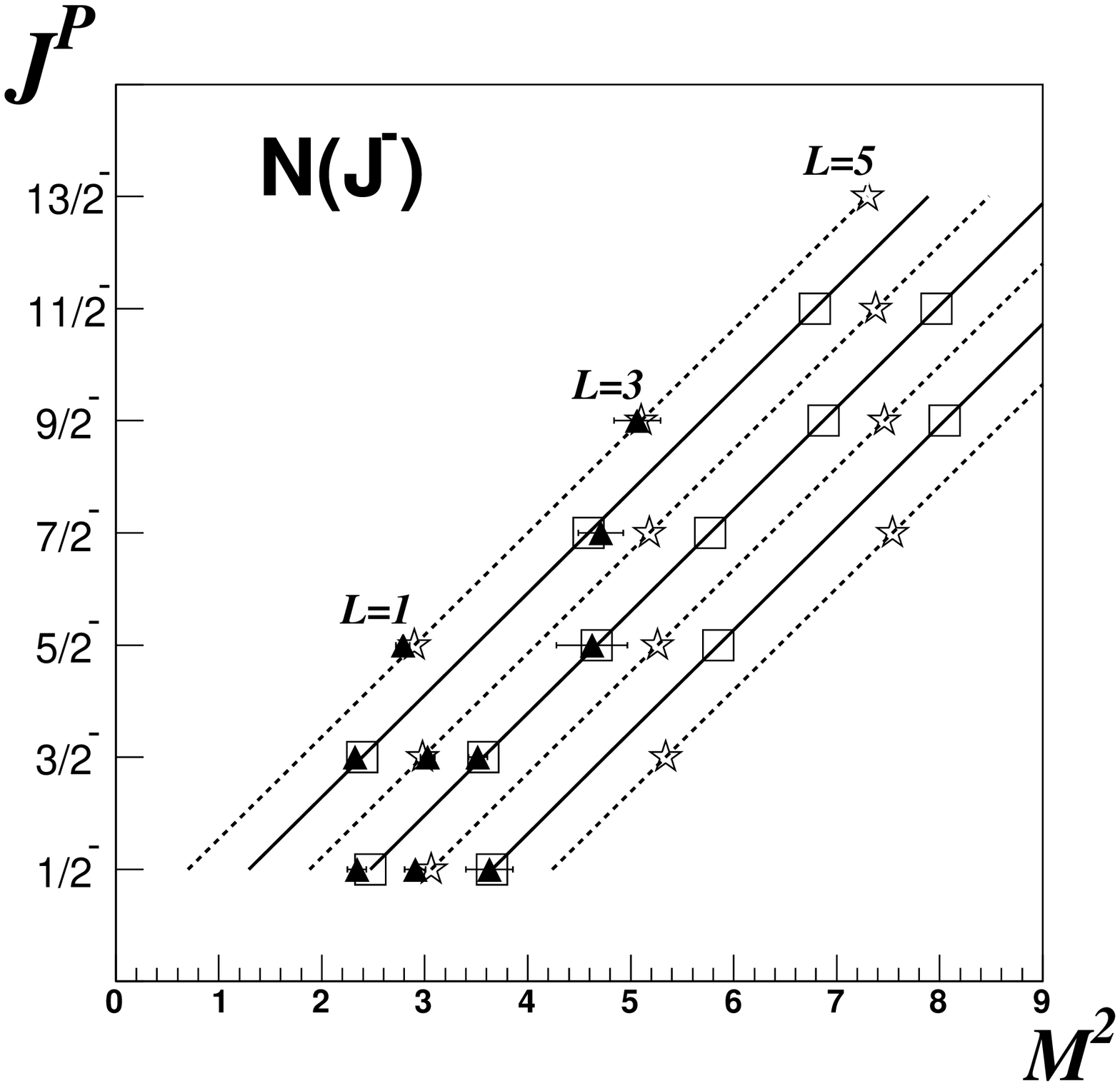,width=70mm}
            \epsfig{file=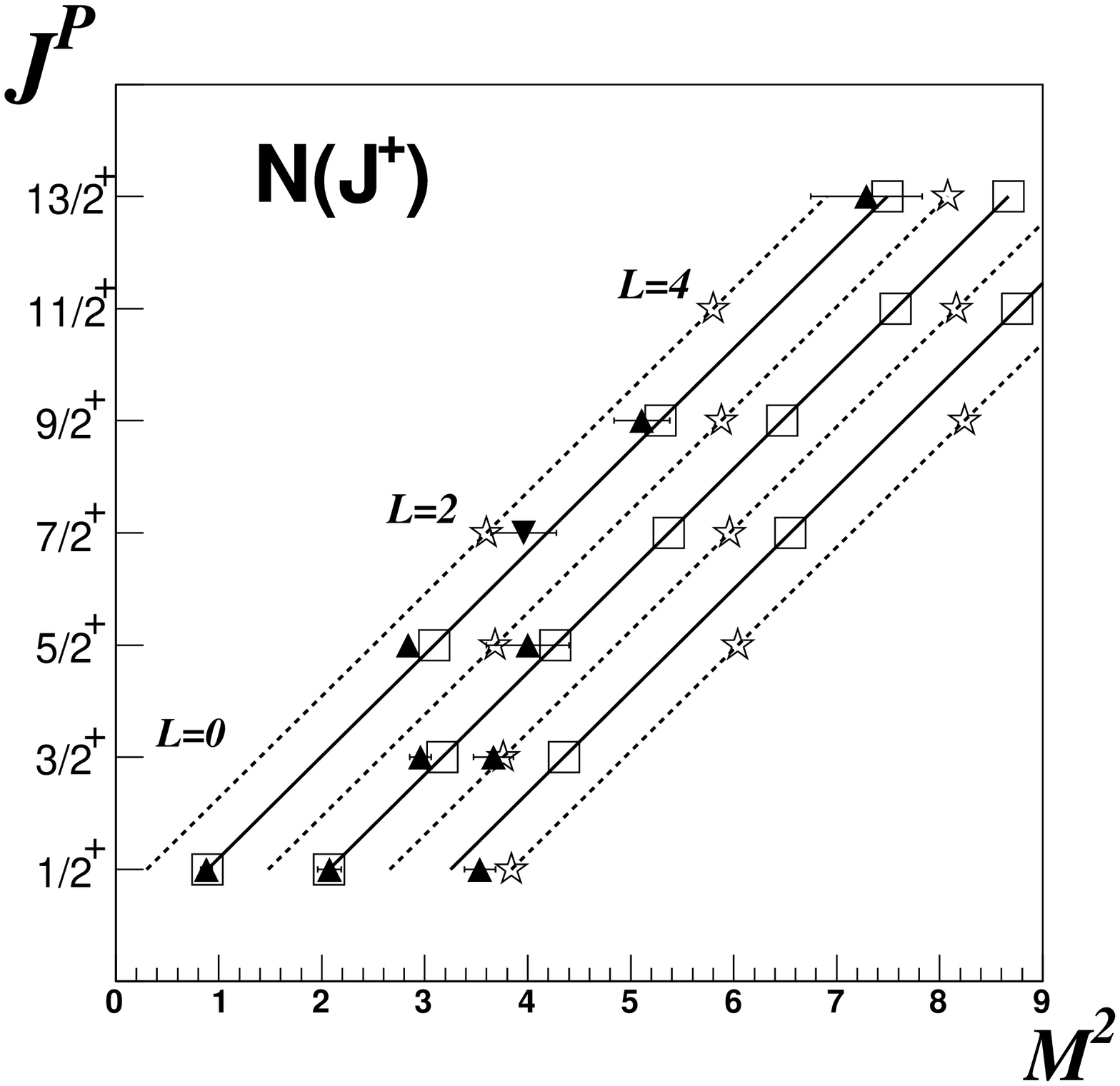,width=70mm}}
\caption{ ($J,M^2$)
planes for   baryons at
$M_{D^0_0} \neq M_{D^1_1}$ within the  $SU(6)$ constraints for wave
functions
  of the $L=0$ states -- the basic
states are shown only (notations are as in  Fig. \ref{J-M-G}).
 \label{J-M-G-1}}
\end{figure}

 For better presentation of the model, let us re-draw the ($J,M^2$)
 planes keeping
 the basic ($n=1$) states only -- they are shown in
 Fig. \ref{J-M-G-1}.

For $L=0$ we see two basic states: $N(1/2^+)$ and  $\Delta (3/2^+)$.

In the $I=3/2$ sector we have five states for $L=1$ and six states for
each $L$ at $L>1$.

For ($L=1, \, I=1/2$) states we have five basic states with
$J^P=\frac 12^-,\; \frac 32^-,\quad J^P=\frac 12^-,\; \frac 32^-,\;
\frac 52^-,\quad J^P=\frac 12^-,\; \frac 32^-$,
 while for the states with
$L>1$ we have $J^P=(L+\frac 12)^P,\; (L-\frac 12)^P,\quad J^P=(L+\frac
32)^P,\;(L+\frac 12)^P,\; (L-\frac 12)^P,\; (L-\frac 32)^P,\quad
J^P=(L+\frac 12)^P,\; (L-\frac 12)^P$.

The states belonging to the same
 trajectory have $\delta J^\pm=2^\pm$ and
 $M^2_{(J+2)^\pm}-M^2_{J^\pm}\simeq 2$ GeV$^2$. One pair of states with $S=\frac 12$
 and $I=\frac 12$ is formed by the $D_0^0$ diquark -- we cannot assert
 definitely what pair is built by $D_0^0$,  heavier or lighter
 one.

 \subsection{The setting of  $N(J^+)$ states at $M_{D^0_0} \neq M_{D^1_1}$ and
 $L\geq 2$}

 In this sector the trajectories of Fig. \ref{J-M-G} give us the
 following states at  $L\geq 2$:
\begin{equation}\label{GMG-1}
{\renewcommand{\arraystretch}{0,5}
\begin{tabular}{l|l|l|llll}
$L=2$&$S=\frac12$&     &    &$N(\frac32^+)$&$N(\frac52^+)$&   \\
     &     &$n=1$&    &$(1720\pm 30)$&$(1685\pm 5)$&  \\
     &     &$n=2$&    &$\sim 2040^{(*)}$&$(2000\pm 100)$&  \\
     &     &$n=3$&    &$\sim 2300^{(*)}$&$\sim 2300^{(*)}$&  \\
     &     &$n=4$&    &$\sim 2530^{(*)}$&$\sim 2530^{(*)}$&  \\
&$S=\frac12$&     &    &$N(\frac32^+)$&$N(\frac52^+)$&   \\
     &     &$n=1$&    &$\sim 2040^{(*)}$&$\sim 2040$&  \\
     &     &$n=2$&    &$\sim 2300^{(*)}$&$\sim 2300^{(*)}$&  \\
     &     &$n=3$&    &$\sim 2530^{(*)}$&$\sim 2530^{(*)}$&  \\
     &     &$n=4$&    &$\sim 2740$&$\sim 2740^{(*)}$&  \\
&$S=\frac32$&     &$N(\frac12^+)$&$N(\frac32^+)$&$N(\frac52^+)$&$N(\frac72^+)$     \\
     &     &$n=1$&$(1880\pm 40)$&$(1915\pm 50)$&$\sim 1900$&$(1990\pm 80)$  \\
     &     &$n=2$&$\sim 2180$&$\sim 2180$&$\sim 2170$&$\sim 2170$  \\
     &     &$n=3$&$\sim 2420$&$\sim 2420$&$\sim 2420^{(*)}$&$\sim 2410^{(*)}$  \\
     &     &$n=4$&$\sim 2640$&$\sim 2640$&$\sim 2640^{(*)}$&$\sim 2630^{(*)}$  \\
\hline
$L=4$&$S=\frac12$&     &    &$N(\frac72^+)$&$N(\frac92^+)$&   \\
     &     &$n=1$&    &$\sim 2290$&$(2260\pm 60)$&  \\
     &     &$n=2$&    &$\sim 2520$&$\sim 2520^{(*)}$&  \\
     &     &$n=3$&    &$\sim 2740^{(*)}$&$\sim 2730^{(*)}$&  \\
     &     &$n=4$&    &$\sim 2930^{(*)}$&$\sim 2930^{(*)}$&  \\
&$S=\frac12$&     &    &$N(\frac72^+)$&$N(\frac92^+)$&   \\
     &     &$n=1$&    &$\sim 2520$&$\sim 2520^{(*)}$&  \\
     &     &$n=2$&    &$\sim 2740^{(*)}$&$\sim 2730^{(*)}$&  \\
     &     &$n=3$&    &$\sim 2930^{(*)}$&$\sim 2930^{(*)}$&  \\
  %   &     &$n=4$&    &$\sim 3120$&$\sim 3120$&  \\
&$S=\frac32$&     &$N(\frac52^+)$&$N(\frac72^+)$&$N(\frac92^+)$&$N(\frac{11}2^+)$     \\
     &     &$n=1$&$\sim 2420^{(*)}$&$\sim 2410^{(*)}$&$\sim 2410$&$\sim 2400$  \\
     &     &$n=2$&$\sim 2640^{(*)}$&$\sim 2630^{(*)}$&$\sim 2630$&$\sim 2630$  \\
     &     &$n=3$&$\sim 2840^{(*)}$&$\sim 2840^{(*)}$&$\sim 2830^{(*)}$&$\sim 2830^{(*)}$  \\
   %  &     &$n=4$&$\sim 3030$&$\sim 3030$&$\sim 3020$&$\sim 3020$  \\
\hline
$L=6$&$S=\frac12$&     &    &$N(\frac{11}2^+)$&$N(\frac{13}2^+)$&   \\
     &     &$n=1$&    &$\sim 2730$&$(2700\pm 100)$&  \\
     &     &$n=2$&    &$\sim 2930$&$\sim 2920$&  \\
   %  &     &$n=3$&    &$\sim 3110$&$\sim 3110$&  \\
   %  &     &$n=4$&    &$\sim 3290$&$\sim 3280$&  \\
&$S=\frac12$&     &    &$N(\frac{11}2^+)$&$N(\frac{13}2^+)$&   \\
     &     &$n=1$&    &$\sim 2930$&$\sim 2920$&  \\
     &     &$n=2$&    &$\sim 3110$&$\sim 3110$&  \\
    % &     &$n=3$&    &$\sim 3290$&$\sim 3280$&  \\
   %  &     &$n=4$&    &$\sim 3450$&$\sim 3450$&  \\
&$S=\frac32$&     &$N(\frac92^+)$&$N(\frac{11}2^+)$&$N(\frac{13}2^+)$&$N(\frac{15}2^+)$     \\
     &     &$n=1$&$\sim 2830^{(*)}$&$\sim 2830^{(*)}$&$\sim 2830$&$\sim 2820$  \\
    &     &$n=2$&$\sim 3020^{(*)}$&$\sim 3020^{(*)}$&$\sim 3020$&$\sim 3010$  \\
  %   &     &$n=3$&$\sim 3200$&$\sim 3200$&$\sim 3200$&$\sim 3200$  \\
  %   &     &$n=4$&$\sim 3370$&$\sim 3370$&$\sim 3370$&$\sim 3370$
%\hline
\end{tabular}
}
\end{equation}
The symbol $^{(*)}$ means that in this mass region we should have two
poles.

We see reasonable agreement of our predictions with data.

  \subsection{The setting of the  $N(J^-)$ states at $M_{D^0_0} \neq
  M_{D^1_1}$}

 In the $N(J^-)$ sector the lightest states have $L=1$, and we see that
 these states are in agreement with model predictions. But let us
 stress that the scheme requires a series of radial excitation states
 at $J^P= \frac 12^+,\frac 32^+,\frac 52^+$ at $M\simeq 2010$ MeV.
\begin{equation}\label{GMG-2}
{\renewcommand{\arraystretch}{0,5}
\begin{tabular}{l|l|l|llll}
$L=1$&$S=\frac12$&     &    &$N(\frac12^-)$&$N(\frac32^-)$&   \\
     &     &$n=1$&    &$(1530\pm 30)$&$(1524\pm 5)$&  \\
     &     &$n=2$&    &$(1905\pm 60)$&$(1875\pm 25)$&  \\
     &     &$n=3$&    &$(2180\pm 80)^{(*)}$&$(2160\pm 40)^{(*)}$&  \\
     &     &$n=4$&    &$\sim 2390^{(*)}$&$\sim 2390^{(*)}$&  \\
&$S=\frac12$&     &    &$N(\frac12^-)$&$N(\frac32^-)$&   \\
     &     &$n=1$&    &$\sim 1870$&$\sim 1860$&  \\
     &     &$n=2$&    &$\sim 2140^{(*)}$&$\sim 2140^{(*)}$&  \\
     &     &$n=3$&    &$\sim 2390^{(*)}$&$\sim 2390^{(*)}$&  \\
     &     &$n=4$&    &$\sim 2610^{(*)}$&$\sim 2610^{(*)}$&  \\
&$S=\frac32$&     &$   $&$N(\frac12^-)$&$N(\frac32^-)$&$N(\frac52^-)$     \\
     &     &$n=1$&$         $&$(1705\pm 30)$&$(1740\pm 20)$&$(1670\pm 20)$  \\
     &     &$n=2$&$         $&$\sim 2010$&$\sim 2000$&$\sim 2000$  \\
     &     &$n=3$&$         $&$\sim 2270$&$\sim 2270^{(*)}$&$\sim 2260^{(*)}$  \\
     &     &$n=4$&$         $&$\sim 2500$&$\sim 2500^{(*)}$&$\sim 2500^{(*)}$  \\
\hline
$L=3$&$S=\frac12$&     &    &$N(\frac52^-)$&$N(\frac72^-)$&   \\
     &     &$n=1$&    &$(2150\pm 80)$&$(2170\pm 50)$&  \\
     &     &$n=2$&    &$\sim 2380^{(*)}$&$\sim 2380^{(*)}$&  \\
     &     &$n=3$&    &$\sim 2610^{(*)}$&$\sim 2600^{(*)}$&  \\
     &     &$n=4$&    &$\sim 2810^{(*)}$&$\sim 2810^{(*)}$&  \\
&$S=\frac12$&     &    &$N(\frac52^-)$&$N(\frac72^-)$&   \\
     &     &$n=1$&    &$\sim 2380^{(*)}$&$\sim 2380^{(*)}$&  \\
     &     &$n=2$&    &$\sim 2610^{(*)}$&$\sim 2600^{(*)}$&  \\
     &     &$n=3$&    &$\sim 2810^{(*)}$&$\sim 2810^{(*)}$&  \\
  %   &     &$n=4$&    &$\sim 3000$&$\sim 3000$&  \\
&$S=\frac32$&     &$N(\frac32^-)$&$N(\frac52^-)$&$N(\frac72^-)$&$N(\frac92^-)$     \\
     &     &$n=1$&$\sim 2270^{(*)}$&$\sim 2260^{(*)}$&$\sim 2260$&$(2250\pm 50)$  \\
     &     &$n=2$&$\sim 2500^{(*)}$&$\sim 2500^{(*)}$&$\sim 2490$&$\sim 2490$  \\
     &     &$n=3$&$\sim 2720^{(*)}$&$\sim 2710^{(*)}$&$\sim 2710^{(*)}$&$\sim 2700^{(*)}$  \\
 %    &     &$n=4$&$\sim 2910$&$\sim 2910$&$\sim 2910$&$\sim 2900$  \\
\hline
$L=5$&$S=\frac12$&     &    &$N(\frac92^-)$&$N(\frac{11}2^-)$&   \\
     &     &$n=1$&    &$\sim 2600$&$\sim 2600$&  \\
     &     &$n=2$&    &$\sim 2810^{(*)}$&$\sim 2800^{(*)}$&  \\
    &     &$n=3$&    &$\sim 3000^{(*)}$&$\sim 3000^{(*)}$&  \\
 %    &     &$n=4$&    &$\sim 3180$&$\sim 3180$&  \\
&$S=\frac12$&     &    &$N(\frac92^-)$&$N(\frac{11}2^-)$&   \\
     &     &$n=1$&    &$\sim 2810^{(*)}$&$\sim 2800^{(*)}$&  \\
     &     &$n=2$&    &$\sim 3000^{(*)}$&$\sim 3000^{(*)}$&  \\
   %  &     &$n=3$&    &$\sim 3180$&$\sim 3180$&  \\
 %    &     &$n=4$&    &$\sim 3350$&$\sim 3350$&  \\
&$S=\frac32$&     &$N(\frac72^-)$&$N(\frac92^-)$&$N(\frac{11}2^-)$&$N(\frac{13}2^-)$     \\
     &     &$n=1$&$\sim 2710^{(*)}$&$\sim 2700^{(*)}$&$\sim 2700$&$\sim 2700$  \\
    &     &$n=2$&$\sim 2910^{(*)}$&$\sim 2900^{(*)}$&$\sim 2900$&$\sim 2900$  \\
    % &     &$n=3$&$\sim 3090$&$\sim 3090$&$\sim 3090$&$\sim 3080$
   %  &     &$n=4$&$\sim 3270$&$\sim 3270$&$\sim 3260$&$\sim 3260$
%\hline
\end{tabular}
}
\end{equation}
\newpage

 \subsection{The setting of   $\Delta(J^+)$ states at $M_{D^0_0} \neq
 M_{D^1_1}$ and $L\geq 2$}

 Considering the equation (\ref{GMG-3}), we should remember that
 the ($L=0$) states  are excluded from
 the suggested classification -- they are given in section 3.
\begin{equation} \label{GMG-3}
{\renewcommand{\arraystretch}{0,5}
\begin{tabular}{l|l|l|llll}
$L=2$&$S=\frac12$&     &    &$\Delta(\frac32^+)$&$\Delta(\frac52^+)$&   \\
     &     &$n=1$&    &$\sim 1760$&$\sim 1760$&  \\
     &     &$n=2$&    &$\sim 2060$&$\sim 2050$&  \\
     &     &$n=3$&    &$\sim 2310$&$\sim 2310$&  \\
     &     &$n=4$&    &$\sim 2540$&$\sim 2540$&  \\
&$S=\frac32$&     &$\Delta(\frac12^+)$&$\Delta(\frac32^+)$&$\Delta(\frac52^+)$&$\Delta(\frac72^+)$     \\
     &     &$n=1$&$(1895\pm 25)$&$(1990\pm 35)^{(*)}$&$(1890\pm 25)$&$(1895\pm 20)$  \\
     &     &$n=2$&$\sim 2190$&$\sim 2190^{(*)}$&$\sim 2180$&$\sim 2180$  \\
     &     &$n=3$&$\sim 2430$&$\sim 2430^{(*)}$&$\sim 2430^{(*)}$&$\sim 2420^{(*)}$  \\
     &     &$n=4$&$\sim 2650$&$\sim 2650^{(*)}$&$\sim 2650^{(*)}$&$\sim 2640^{(*)}$  \\
\hline
$L=4$&$S=\frac12$&     &    &$\Delta(\frac72^+)$&$\Delta(\frac92^+)$&   \\
     &     &$n=1$&    &$\sim 2300$&$\sim 2300$&  \\
     &     &$n=2$&    &$\sim 2530$&$\sim 2530$&  \\
     &     &$n=3$&    &$\sim 2750$&$\sim 2740$&  \\
     &     &$n=4$&    &$\sim 2940$&$\sim 2940$&  \\
&$S=\frac32$&     &$\Delta(\frac52^+)$&$\Delta(\frac72^+)$  &$\Delta(\frac92^+)$&$\Delta(\frac{11}2^+)$     \\
     &      &$n=1$&$\sim 2430^{(*)}$  &$(2390\pm 100)^{(*)}$&$(2400\pm 50)$&$(2420\pm 100)$  \\
     &      &$n=2$&$\sim 2650^{(*)}$  &$\sim 2640^{(*)}$    &$\sim 2640$&$\sim 2640$  \\
     &      &$n=3$&$\sim 2850^{(*)}$  &$\sim 2850^{(*)}$    &$\sim 2840^{(*)}$&$\sim 2840^{(*)}$  \\
     &      &$n=4$&$\sim 3040^{(*)}$  &$\sim 3040^{(*)}$    &$\sim 3030^{(*)}$&$\sim 3030^{(*)}$  \\
\hline
$L=6$&$S=\frac12$&     &    &$\Delta(\frac{11}2^+)$&$\Delta(\frac{13}2^+)$&   \\
     &     &$n=1$&    &$\sim 2740$&$\sim 2740$&  \\
     &     &$n=2$&    &$\sim 2940$&$\sim 2930$&  \\
    &     &$n=3$&    &$\sim 3120$&$\sim 3120$&  \\
    &     &$n=4$&    &$\sim 3300$&$\sim 3290$&  \\
&$S=\frac32$&     &$\Delta(\frac92^+)$&$\Delta(\frac{11}2^+)$&$\Delta(\frac{13}2^+)$&$\Delta(\frac{15}2^+)$     \\
     &     &$n=1$&$\sim 2840^{(*)}$&$\sim 2840^{(*)}$&$\sim 2840$&$(2920\pm 100)$  \\
     &     &$n=2$&$\sim 3030^{(*)}$&$\sim 3030^{(*)}$&$\sim 3030$&$\sim 3020$  \\
    &     &$n=3$&$\sim 3210^{(*)}$&$\sim 3210^{(*)}$&$\sim 3210$&$\sim 3200$  \\
    &     &$n=4$&$\sim 3380^{(*)}$&$\sim 3380^{(*)}$&$\sim 3380$&$\sim 3370$  \\
%\hline
\end{tabular}
}
\end{equation}

\subsection{The setting of  $\Delta(J^-)$ states at $M_{D^0_0} \neq
M_{D^1_1}$}

In the $\Delta(J^-)$ sector we face a problem with the $\frac 52^-$
state observed in the analysis \cite{Arndt:2006bf}
with mass 2233 MeV and quoted as $D_{35}(1930)$ \cite{PDG}. However, it
can be hardly compatible with other observations, which are
in the region of 1930 MeV. In addition, the result of
\cite{Arndt:2006bf} is compatible with the analysis of Manley
\cite{Manley:1992yb}, quoted by PDG as $D_{35}(2350)$. Thus,
we introduce the $\Delta(\frac 52^-)$ state with the mass 2210 MeV and
with the error which covers the results  of
\cite{Manley:1992yb,Arndt:2006bf}. Then, the mass of the $D_{35}(2350)$
state is taken as an average value over the results of
\cite{Cutkosky:1980rh,Vrana:1999nt,Hohler:1973ww}.

\begin{equation} \label{GMG-4}
{\renewcommand{\arraystretch}{0,5}
\begin{tabular}{l|l|l|llll}
$L=1$&$S=\frac12$&     &    &$\Delta(\frac12^-)$&$\Delta(\frac32^-)$&   \\
     &     &$n=1$&    &$(1650\pm 25)$&$(1640\pm 40)$&  \\
     &     &$n=2$&    &$(1900\pm 50)$&$(1990\pm 40)$&  \\
     &     &$n=3$&    &$(2150\pm 50)$&$\sim 2210$&  \\
     &     &$n=4$&    &$\sim 2460$&$\sim 2450$&  \\
&$S=\frac32$&     &$   $&$\Delta(\frac12^-)$&$\Delta(\frac32^-)$&$\Delta(\frac52^-)$     \\
     &     &$n=1$&$         $&$\sim 1800$&$\sim 1800$&$(1910\pm 80)$  \\
     &     &$n=2$&$         $&$\sim 2090$&$\sim 2080$&$\sim 2080$  \\
     &     &$n=3$&$         $&$\sim 2340$&$\sim 2340^{(*)}$&$\sim 2330^{(*)}$  \\
     &     &$n=4$&$         $&$\sim 2570$&$\sim 2560^{(*)}$&$\sim 2560^{(*)}$  \\
\hline
$L=3$&$S=\frac12$&     &    &$\Delta(\frac52^-)$&$\Delta(\frac72^-)$&   \\
     &     &$n=1$&    &$\sim 2210$&$(2240\pm 60)$&  \\
     &     &$n=2$&    &$\sim 2450$&$\sim 2440$&  \\
     &     &$n=3$&    &$\sim 2670$&$\sim 2660$&  \\
&$S=\frac32$&     &$\Delta(\frac32^-)$&$\Delta(\frac52^-)$&$\Delta(\frac72^-)$&$\Delta(\frac92^-)$     \\
     &     &$n=1$&$\sim 2340^{(*)}$&$(2350\pm 50)^{(*)}$&$\sim 2330$&$\sim 2320$  \\
     &     &$n=2$&$\sim 2560^{(*)}$&$\sim 2560^{(*)}$&$\sim 2560$&$\sim 2550$  \\
     &     &$n=3$&$\sim 2770^{(*)}$&$\sim 2770^{(*)}$&$\sim 2770^{(*)}$&$\sim 2760^{(*)}$  \\
\hline
$L=5$&$S=\frac12$&     &    &$\Delta(\frac92^-)$&$\Delta(\frac{11}2^-)$&   \\
     &     &$n=1$&    &$(2633\pm 30)$&$\sim 2660$&  \\
     &     &$n=2$&    &$\sim 2860$&$\sim 2860$&  \\
     &     &$n=3$&    &$\sim 3050$&$\sim 3050$&  \\
&$S=\frac32$&     &$\Delta(\frac72^-)$&$\Delta(\frac92^-)$&$\Delta(\frac{11}2^-)$&$\Delta(\frac{13}2^-)$     \\
     &     &$n=1$&$\sim 2770^{(*)}$&$\sim 2760^{(*)}$&$\sim 2760$&$(2750\pm 100)$  \\
     &     &$n=2$&$\sim 2960^{(*)}$&$\sim 2960^{(*)}$&$\sim 2950$&$\sim 2950$  \\
     &     &$n=3$&$\sim 3140^{(*)}$&$\sim 3140^{(*)}$&$\sim 3140$&$\sim 3140$  \\
%\hline
\end{tabular}
}
\end{equation}

\subsection{ Overlapping of    baryon resonances}

In equations
(\ref{GMG-1}),  (\ref{GMG-2}), (\ref{GMG-3}), (\ref{GMG-4}), the
overlapping resonances are denoted by the symbol $(^*)$ -- the model
predicts a lot of such states. Decay processes lead to a mixing of
the overlapping states (owing  to the transition
$baryon(1)\to hadrons\to baryon(2)$). It results in  a specific
phenomenon, that is,
when several resonances overlap, one of them accumulates the widths of
neighbouring resonances and transforms into the broad state,
see \cite{book3,ufn} and references therein.

This phenomenon had been observed in \cite{km1900,km} for meson
scalar--isoscalar states and applied to the interpretation
   of
the broad state $f_0(1200-1600)$,  being
 a descendant of a pure glueball \cite{glueball,ZPhys}.

In meson physics this phenomenon can play an important role,
in particular, for exotic states which are beyond the $q\bar q$
systematics. Indeed, being among  $q\bar q$ resonances, the exotic
state creates a group of overlapping resonances. The exotic state,
 after accumulating the
"excess" of widths, turns into a broad one. This broad resonance
should be accompanied by narrow states which are the descendants of
states from which the widths have been borrowed. In this way, the
existence of a broad resonance accompanied by narrow ones may be a
signature of exotics.  This possibility, in context of searching for
exotic states, was discussed in \cite{exotic}.

In the considered case of quark--diquarks baryons (equations
(\ref{GMG-1}),  (\ref{GMG-2}), (\ref{GMG-3}), (\ref{GMG-4})), we deal
mainly with two overlapping states: it means that we should observe one
comparatively narrow resonance together with the second one which  is
comparatively broad. Experimental observation of the corresponding
two-pole singularities in  partial amplitudes looks as rather intricate
problem.

\subsubsection{The setting of baryons with  $M_{D^0_0} \neq M_{D^1_1}$ on
$(n,M^2)$ planes}

\begin{figure}[h]
%Fig. 6
\centerline{\epsfig{file=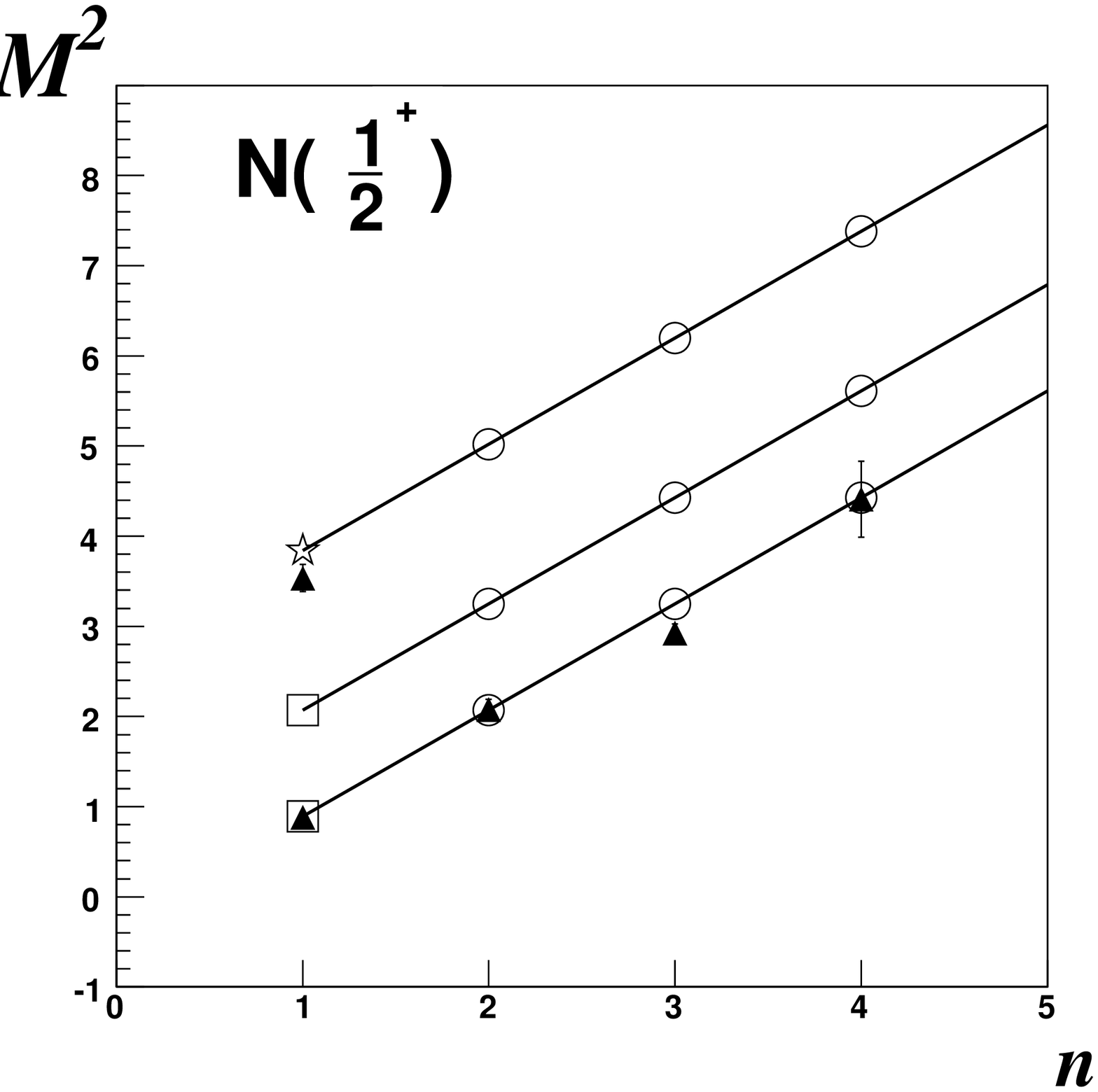,width=55mm}
            \epsfig{file=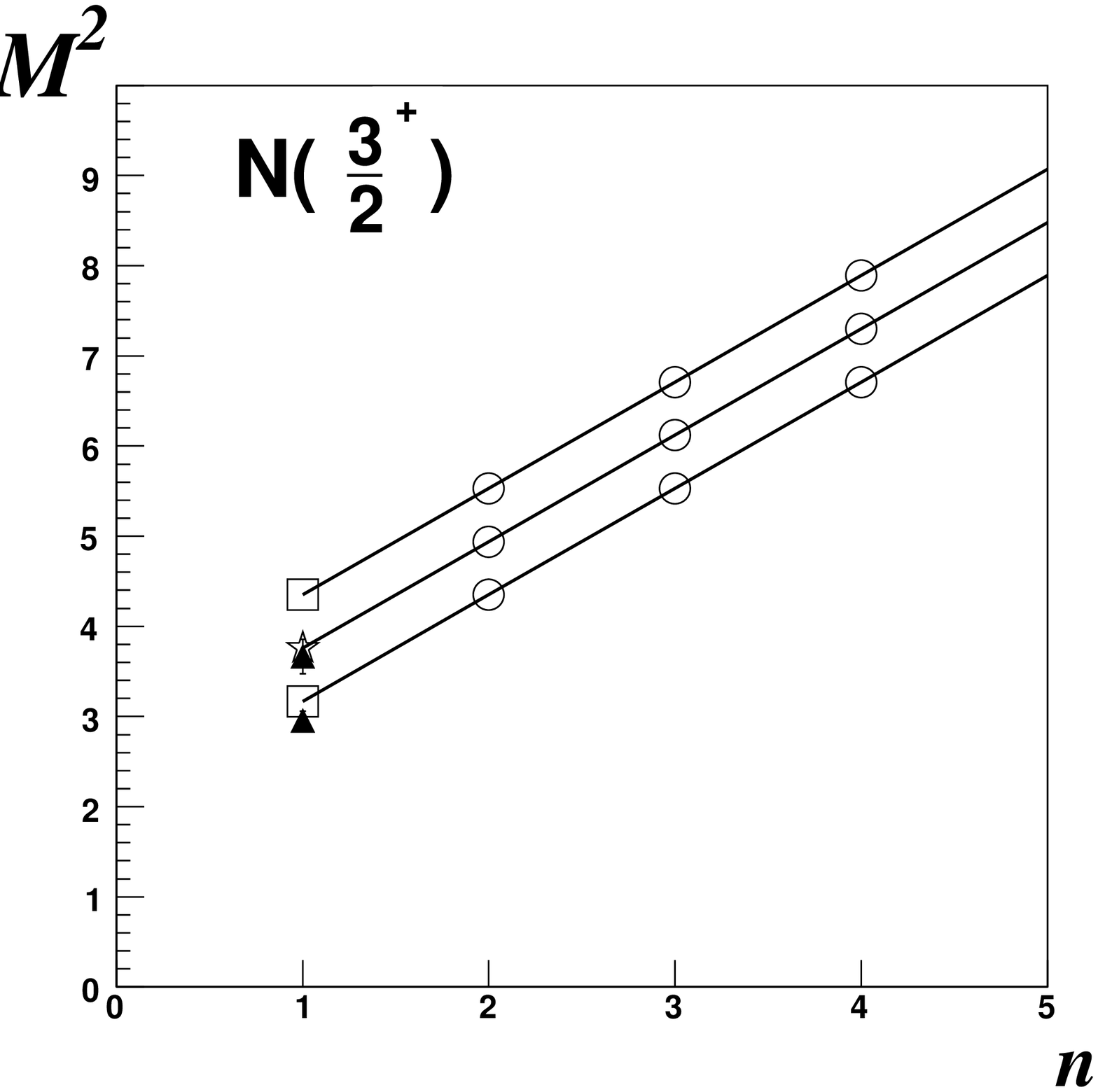,width=55mm}}
\vspace{-0.4cm}
\centerline{\epsfig{file=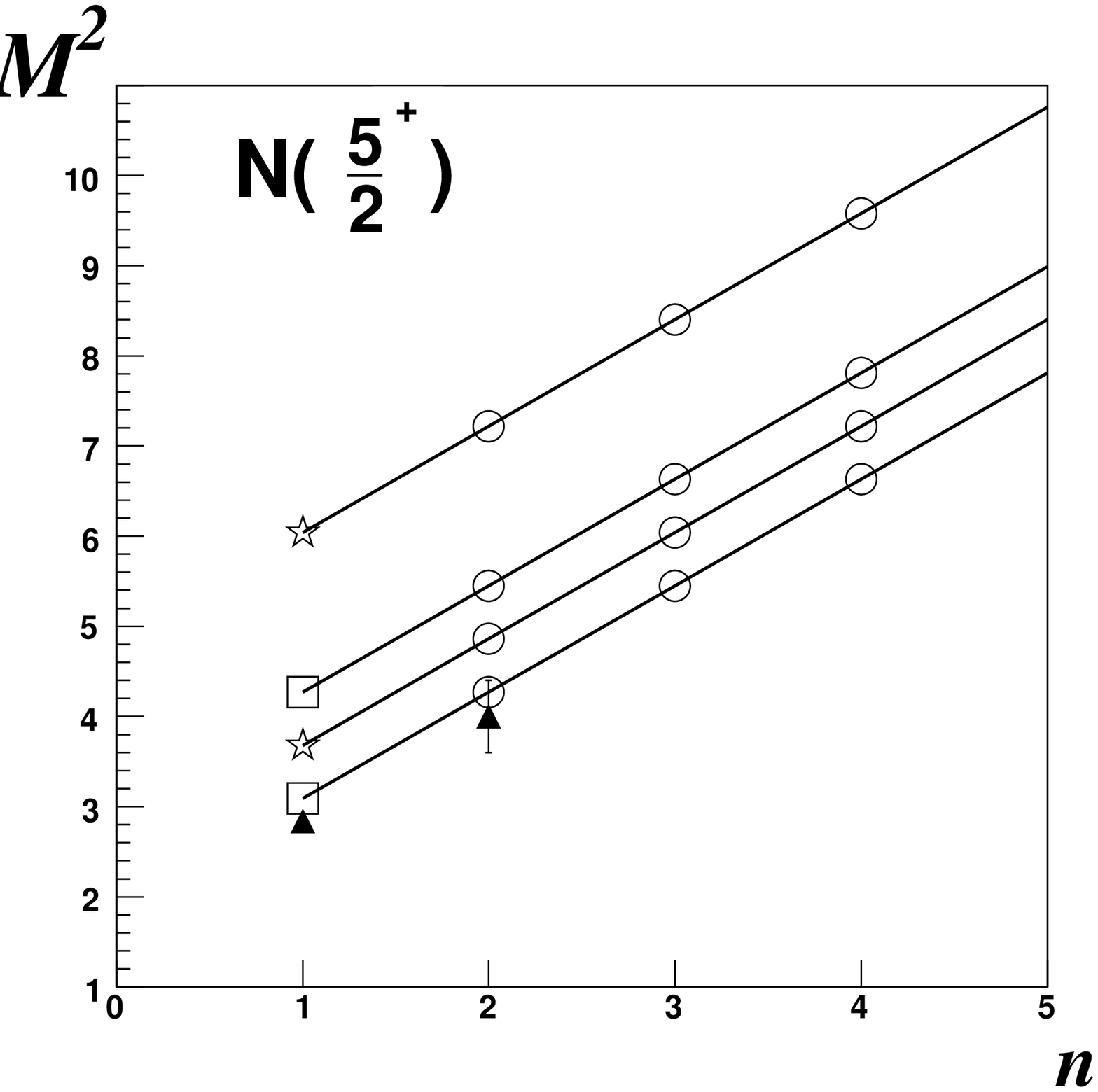,width=55mm}
            \epsfig{file=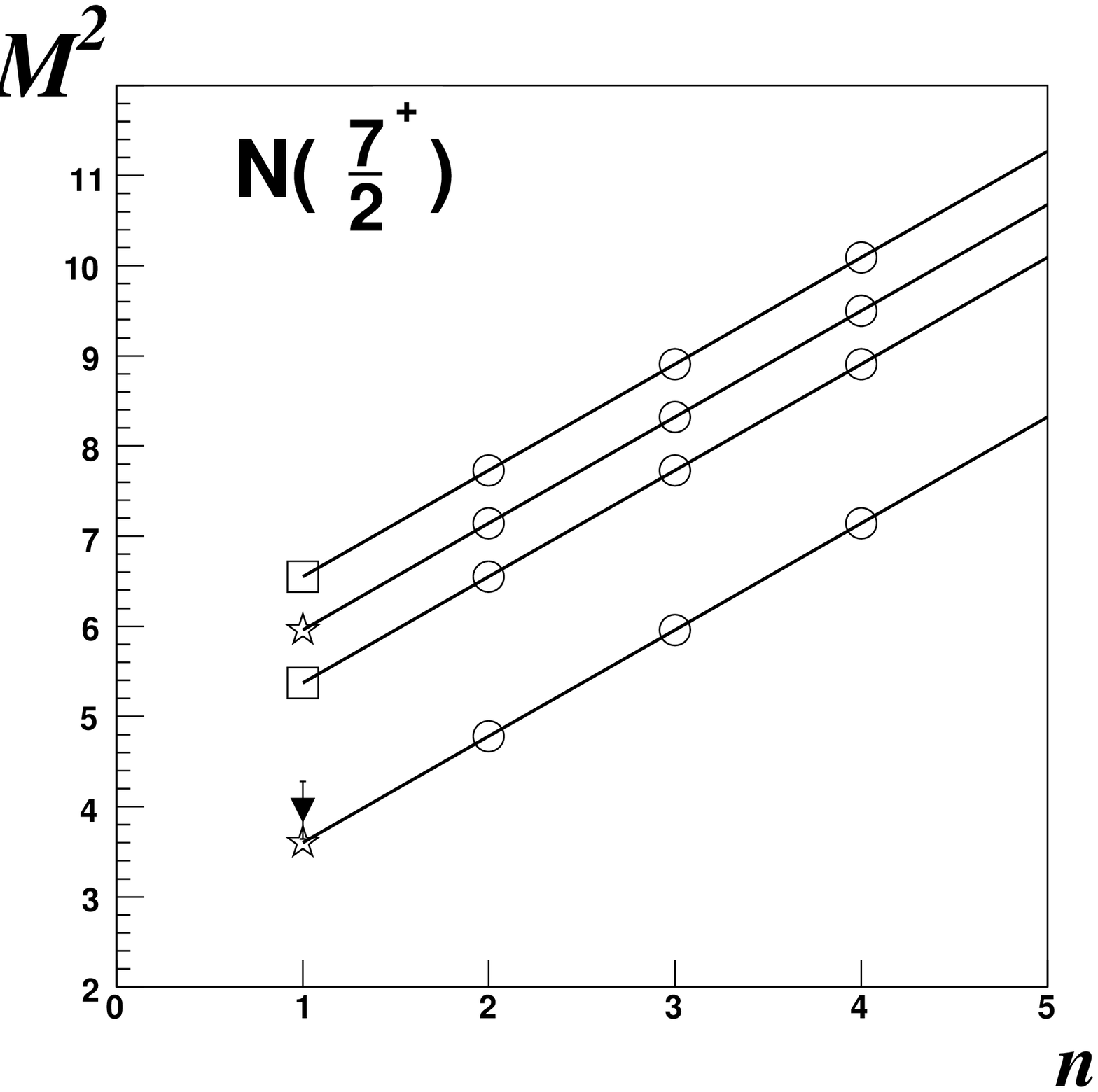,width=55mm}}
\vspace{-0.4cm}
\centerline{\epsfig{file=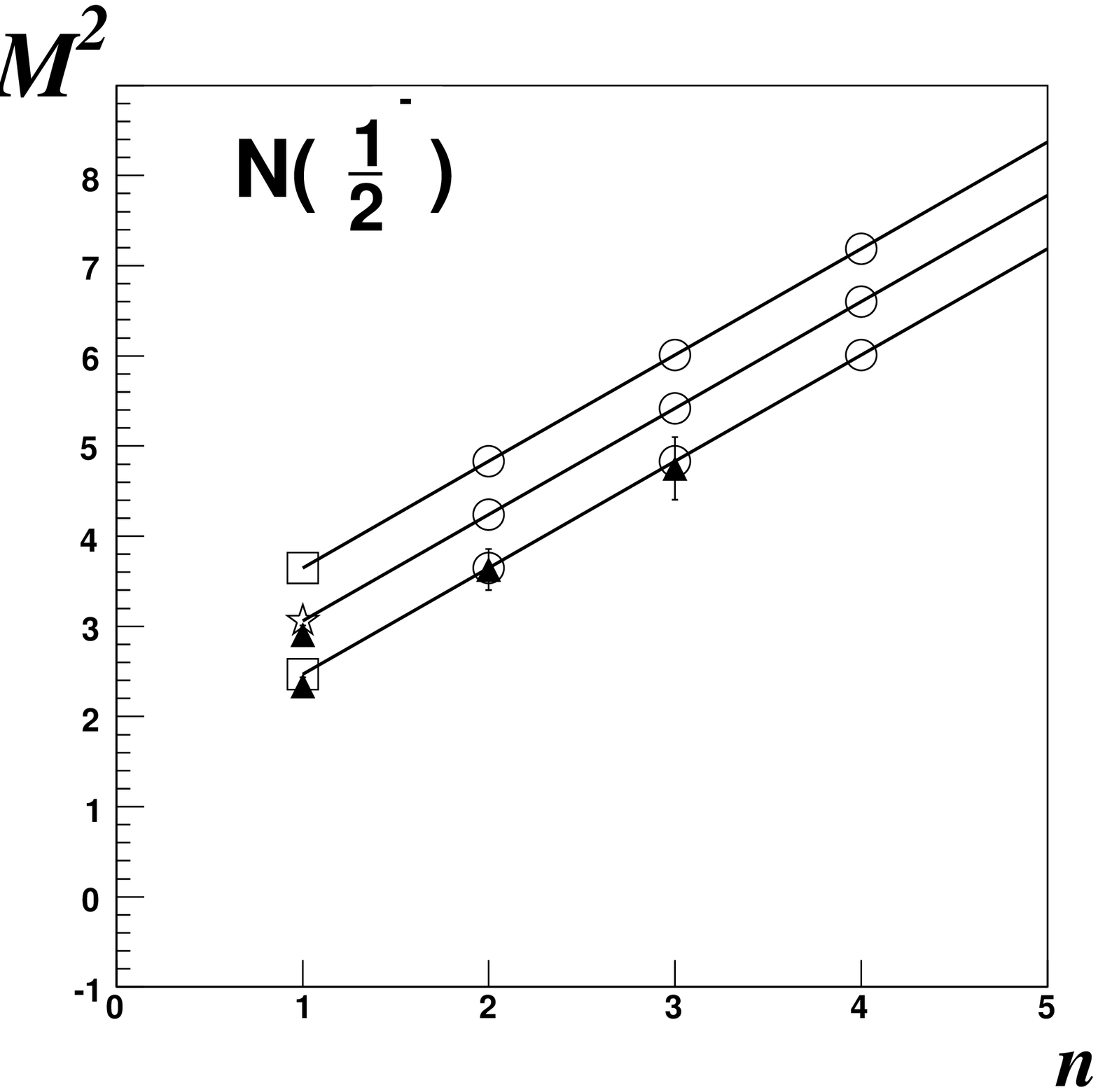,width=55mm}
            \epsfig{file=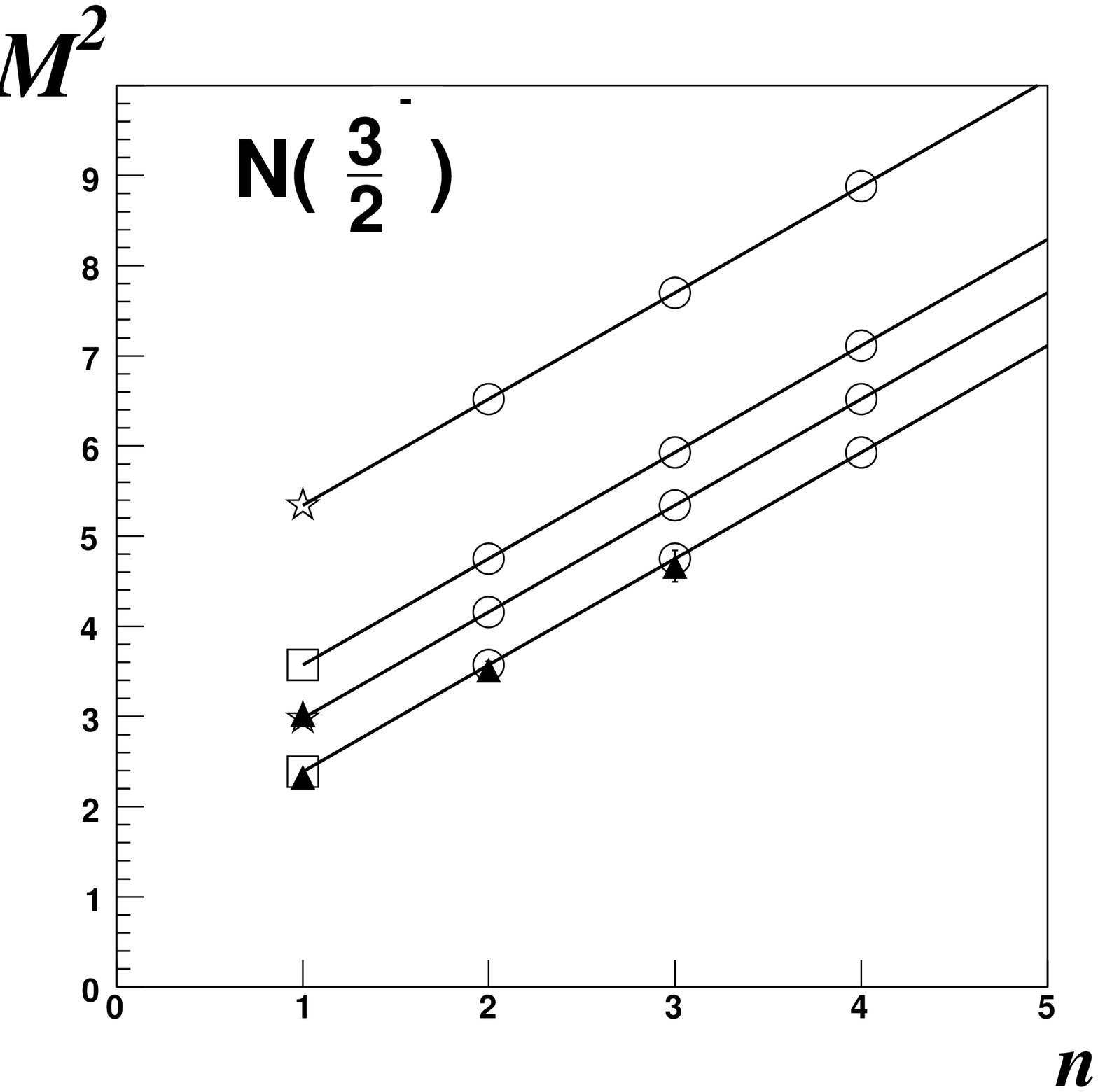,width=55mm}}
\vspace{-0.4cm}
\centerline{\epsfig{file=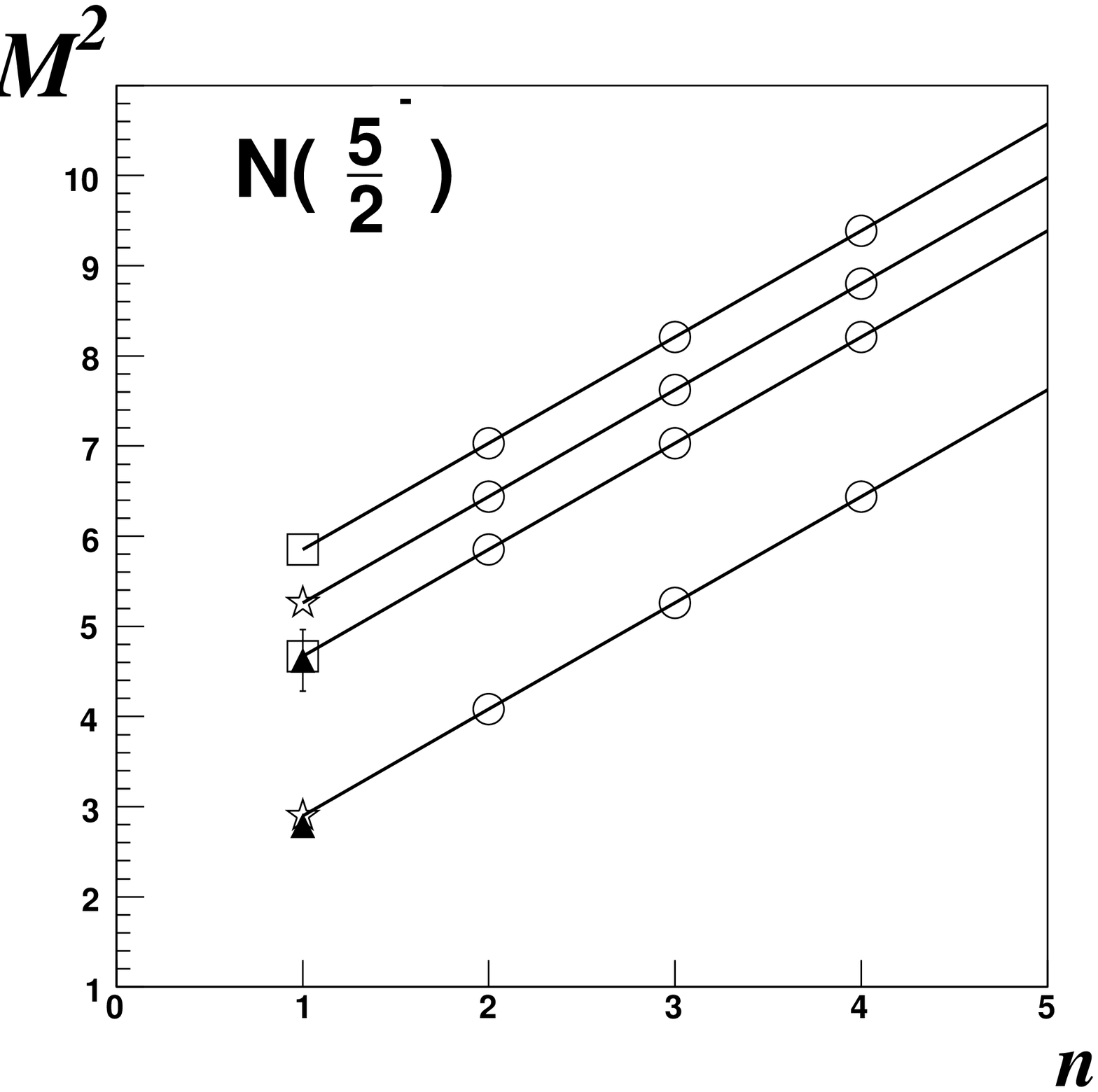,width=55mm}
            \epsfig{file=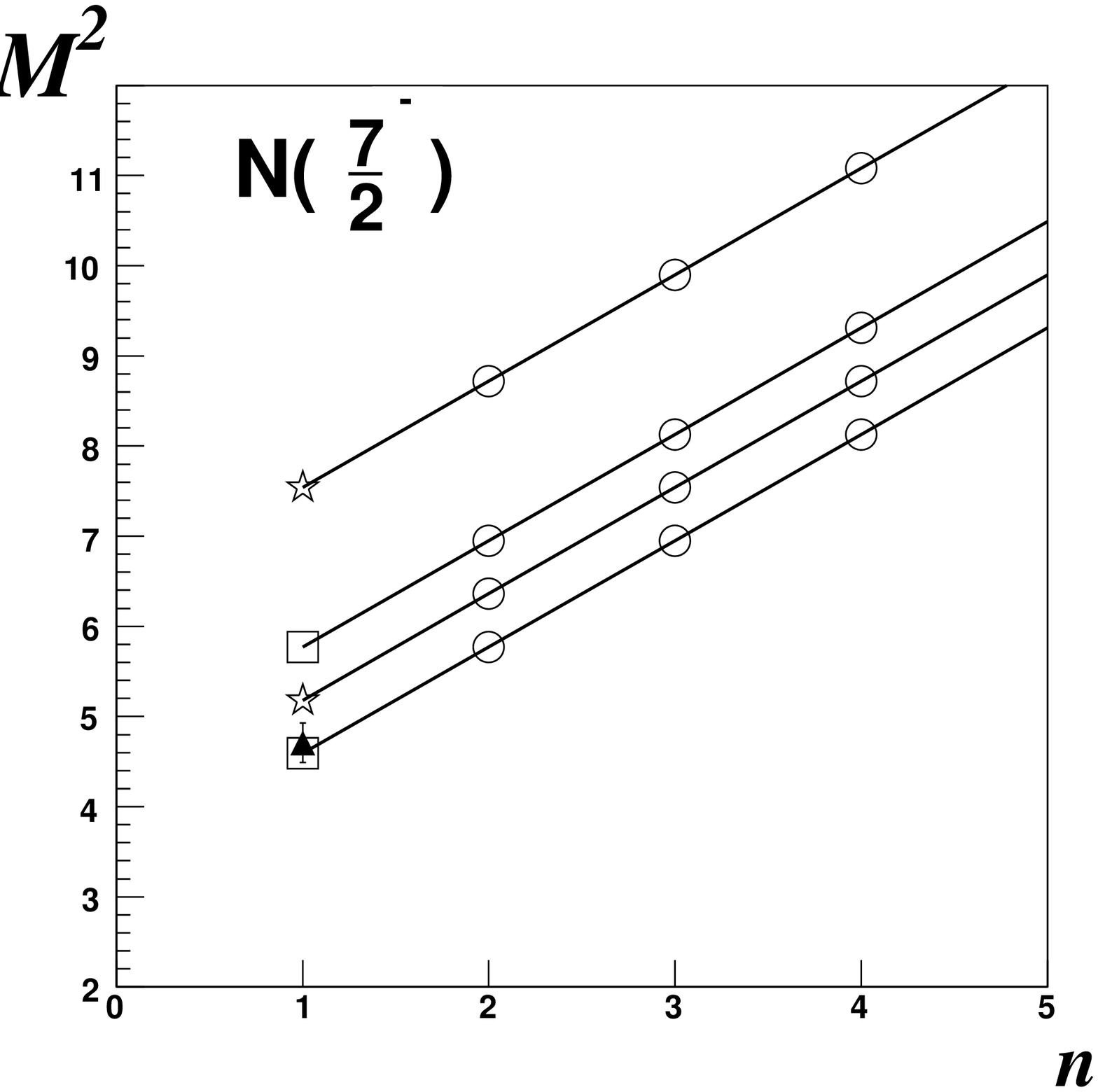,width=55mm}}
\caption{ $(n,M^2)$ planes for $N(J^{\pm})$ states,
 $M_{D^0_0} \neq M_{D^1_1}$.\label{NM1}}
\end{figure}

\begin{figure}[h]
%Fig. 7
\centerline{\epsfig{file=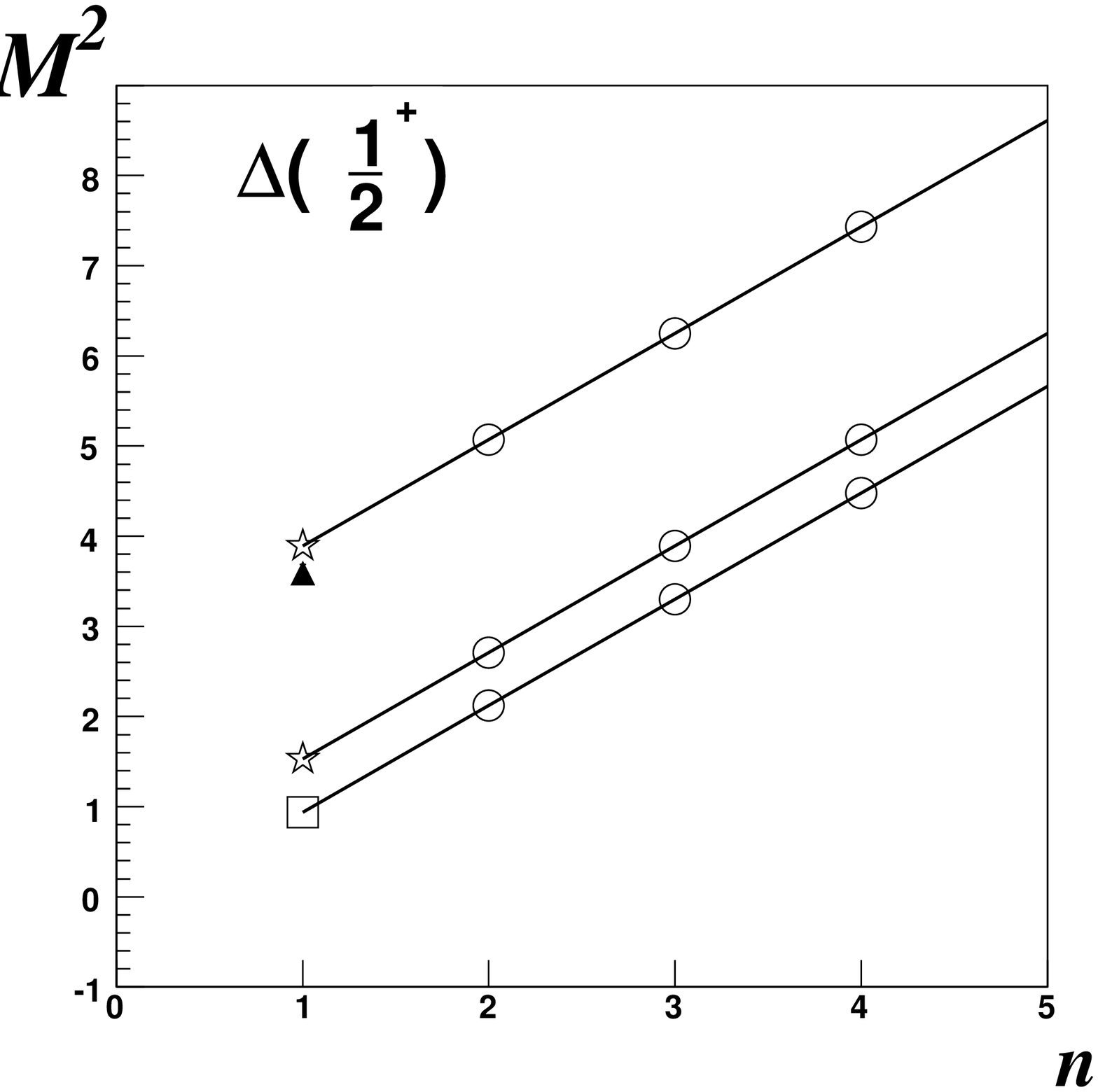,width=55mm}
            \epsfig{file=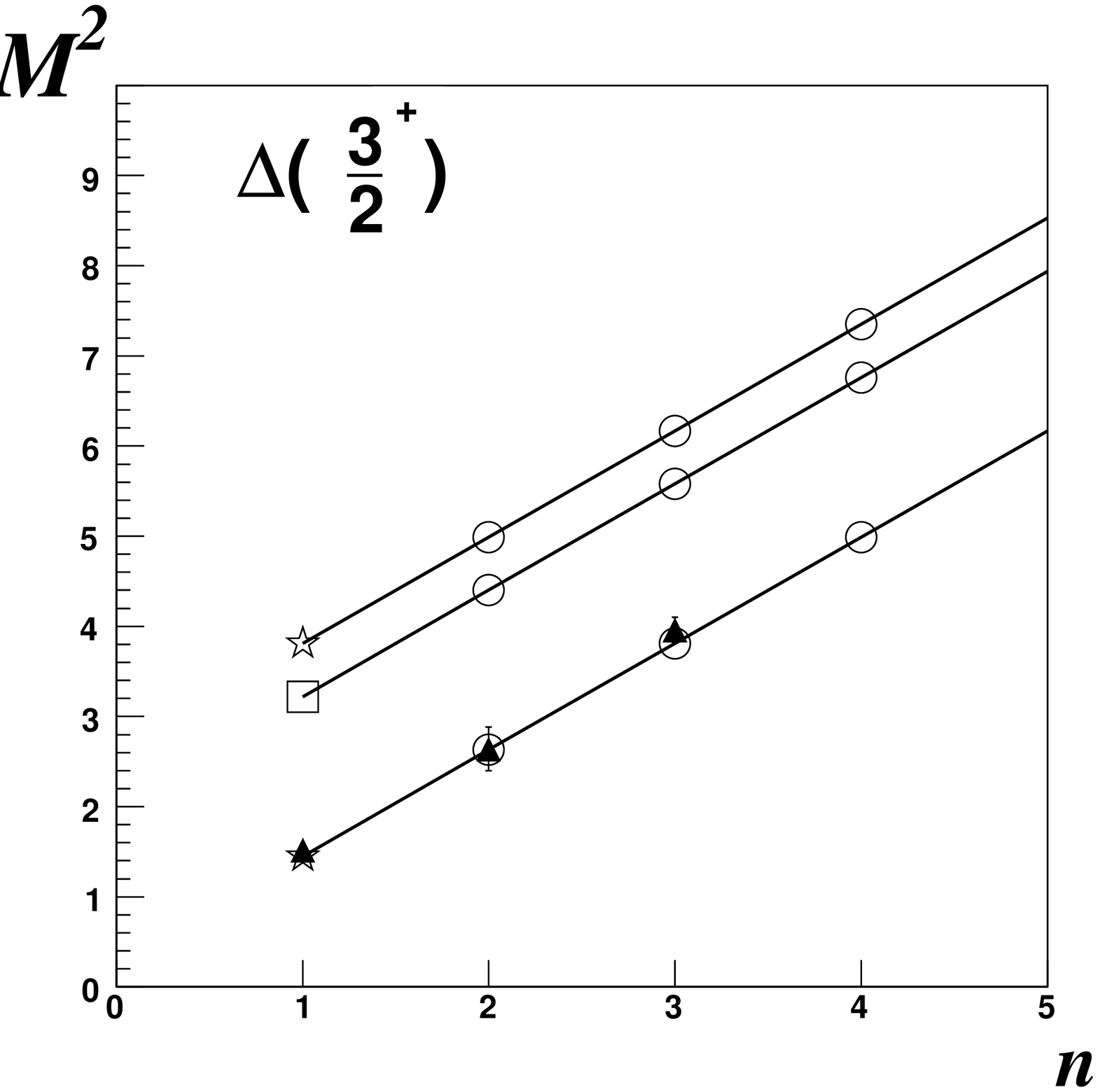,width=55mm}}
\vspace{-0.4cm}
\centerline{\epsfig{file=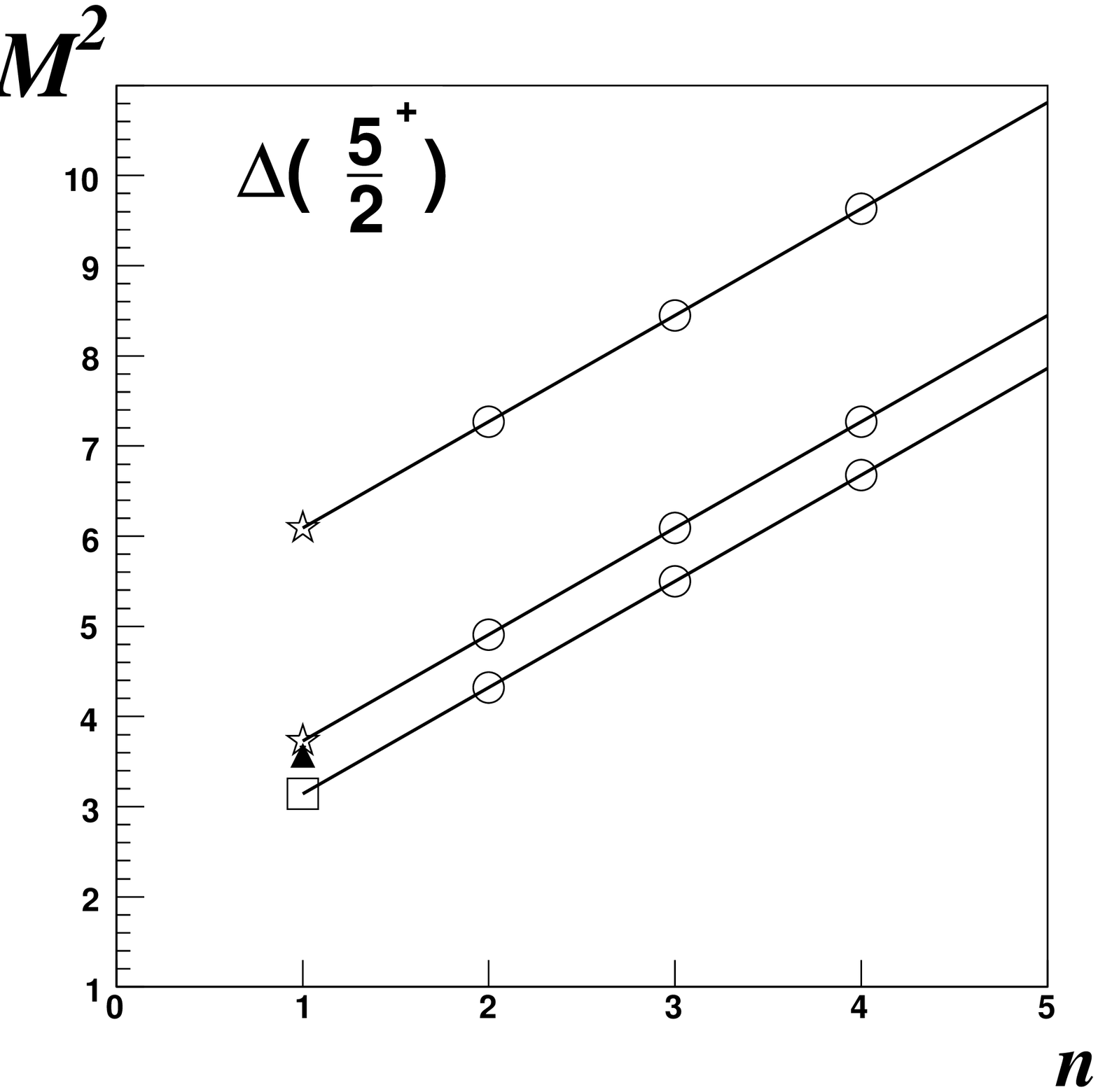,width=55mm}
            \epsfig{file=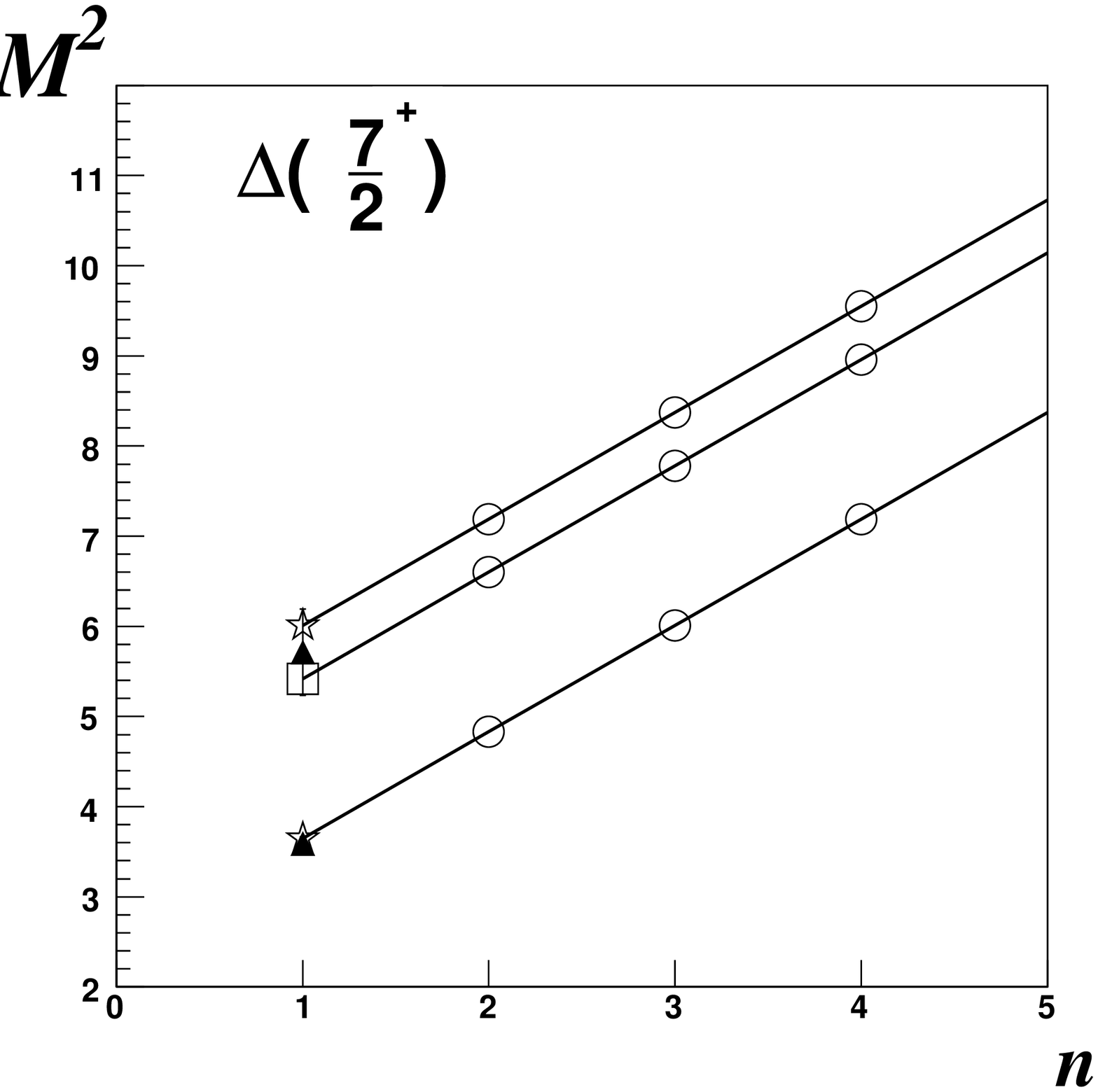,width=55mm}}
\vspace{-0.4cm}
\centerline{\epsfig{file=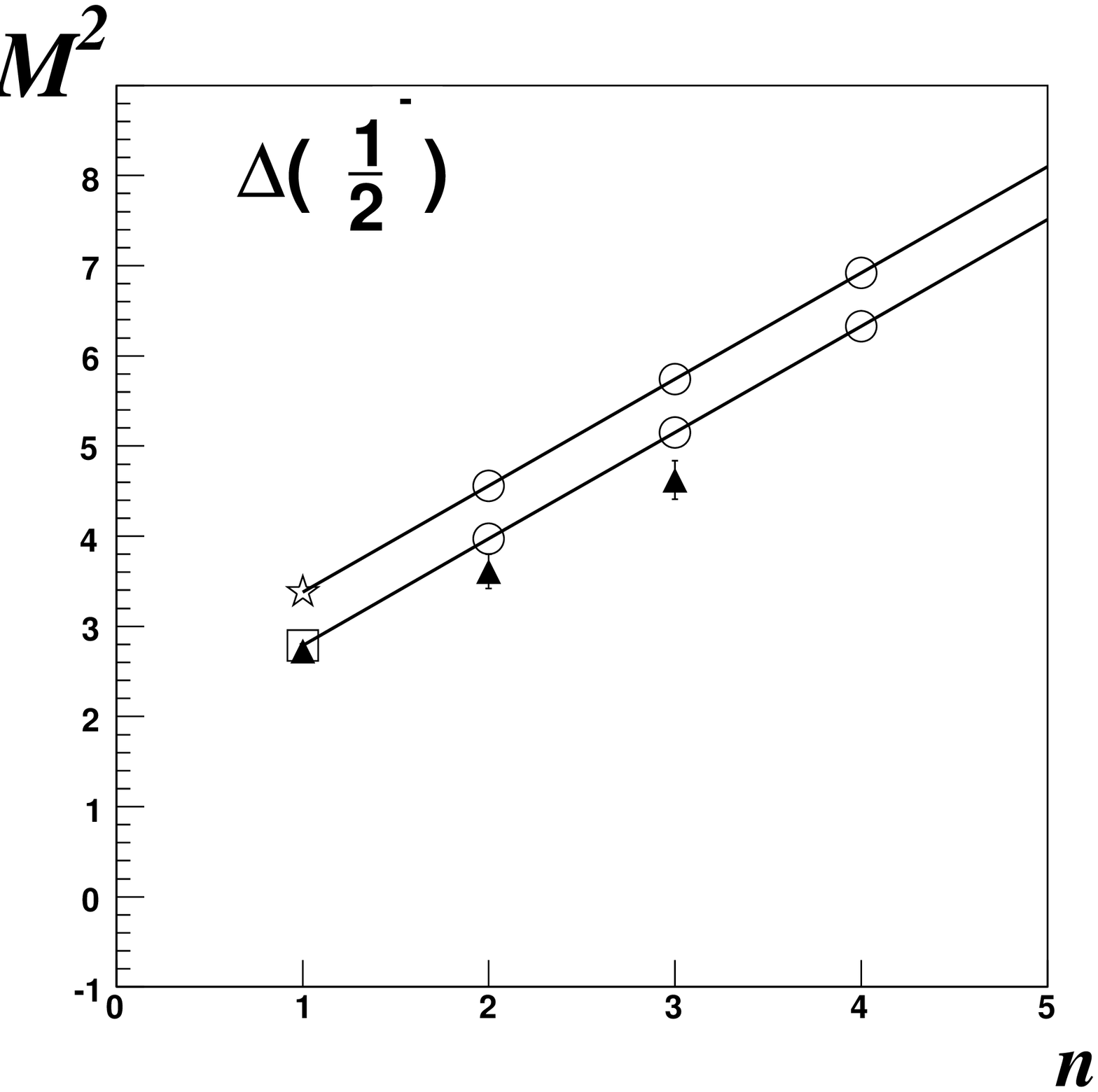,width=55mm}
            \epsfig{file=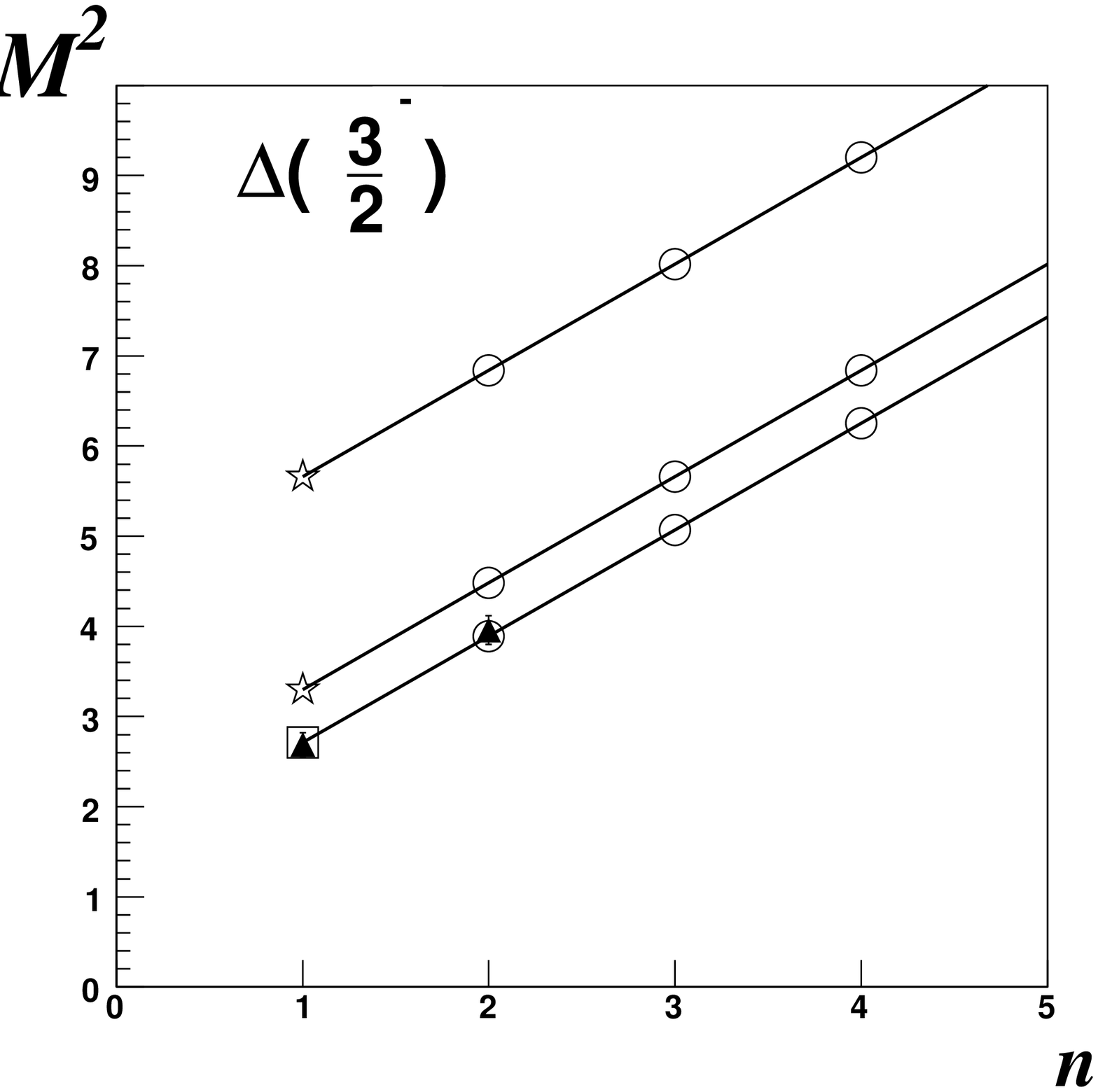,width=55mm}}
\vspace{-0.4cm}
\centerline{\epsfig{file=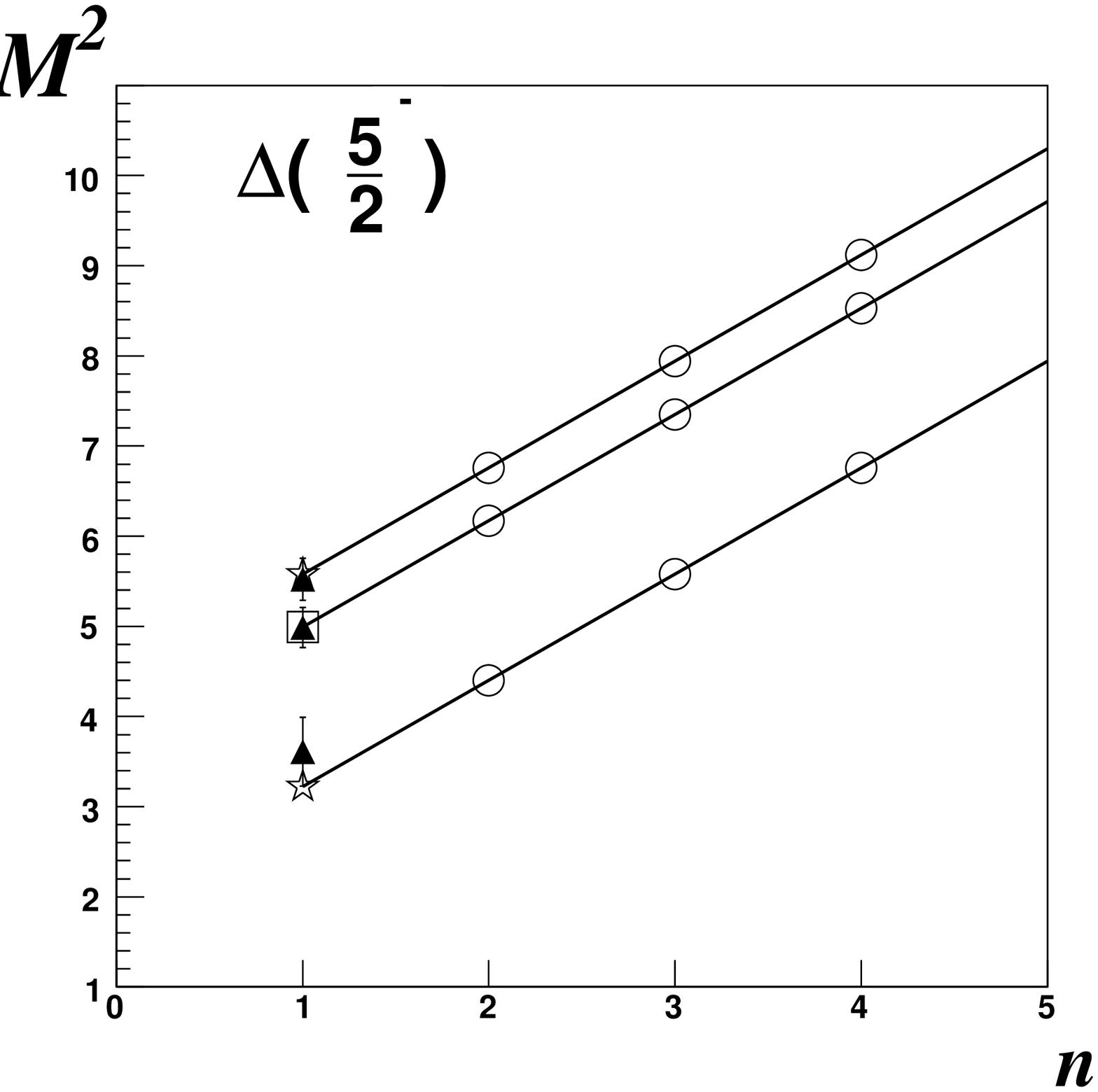,width=55mm}
            \epsfig{file=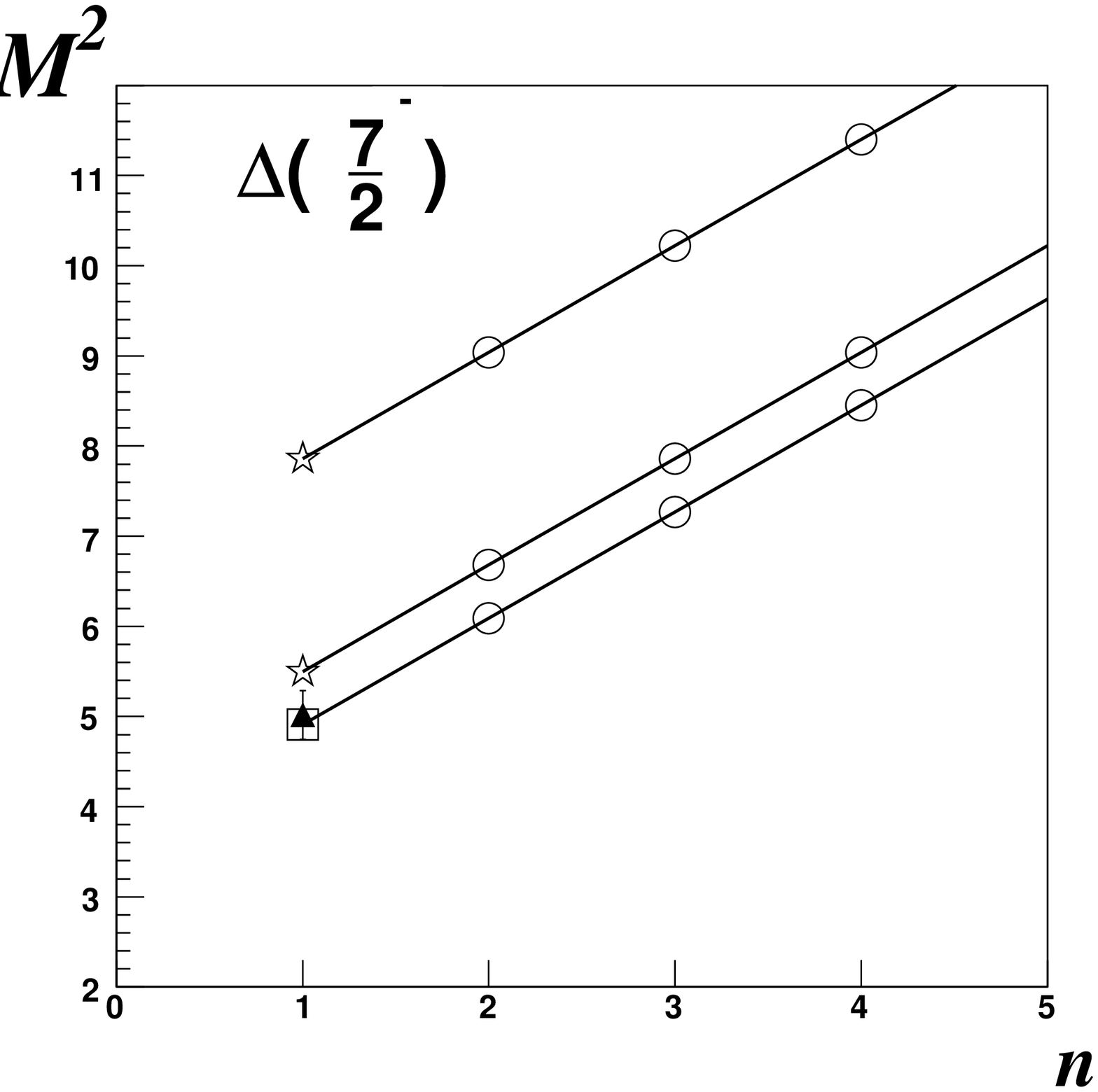,width=55mm}}
\caption
{ $(n,M^2)$ planes for $\Delta(J^{\pm})$ states,
 $M_{D^0_0} \neq M_{D^1_1}$. \label{NM2}}
\end{figure}

Equations  (\ref{GMG-1}),
 (\ref{GMG-2}), (\ref{GMG-3}), (\ref{GMG-4}) allow us to present the
 setting of baryons ($M_{D^0_0} \neq M_{D^1_1}$) on $(n,M^2)$ planes -- they
 are shown in Figs. \ref{NM1} and \ref{NM2}.

 We have three trajectories on the $(n,M^2)$ plot for $N(\frac12 ^+)$
 with the starting states shown in Fig. \ref{J-M-G-1} in the $L=0$
 group.
  The low trajectory is filled in by the states of equation (49).

 In the plot for $N(\frac32 ^+)$ we also have  three trajectories, with
 the starting $L=2$ states, see Fig. \ref{J-M-G-1}. Likewise,
  all other $(n,M^2)$ plots are constructed: the starting states are
  those  shown in Fig. \ref{J-M-G-1}.

 We have a lot of predicted radial excitation states, though not many
 observed ones -- the matter is that in  case of overlapping
 resonances the broad state is concealed under narrow one. As is well
 known, the mixing states repulse from one another. The mixing of
 overlapping states, due to the transition into real hadrons
 $baryon(1)\to real\; hadrons\to baryon(2)$ also leads to the
 repulsing of  resonance poles in the complex-$M$ plane: one is moving
 to real $M$ axis (i.e. reducing the width), another is diving
 deaper into  complex-$M$ plane (i.e. increasing the width) -- for
 more detail
see \cite{book3}, sections 3.4.2 and 3.4.3. To separate two
overlapping poles, one needs to carry out the measurements of decays into
different channels -- different resonances have, as a rule, different
partial widths, so the total width of the "two-pole resonance" should
depend on the  reaction type.

Radially excited states are seen in (see Figs. \ref{NM1} and
\ref{NM2}), namely,
 \\ $N(\frac 12^+)$ sector (four states on the
lowest trajectory), \\ $N(\frac 52^+)$ sector (two states on the lowest
trajectory),
\\ $N(\frac 12^-)$ sector (three states on the lowest
trajectory),
\\ $N(\frac 32^-)$ sector (three states on the lowest
trajectory),
\\ $\Delta(\frac 32^+)$ sector (three states on the lowest
trajectory),
\\ $\Delta(\frac 12^-)$ sector (three states on the lowest
trajectory),
\\ $\Delta(\frac 32^-)$ sector (two states on the lowest
trajectory),
\\ $\Delta(\frac 52^-)$ sector (two states on the lowest
trajectory).

However, in Figs. \ref{NM1} and \ref{NM2} we do not mark specially the
resonances which are "shadowed" by observed ones. The slopes of all
trajectories in Figs. \ref{NM1} and \ref{NM2} coincide with each other
and with those of meson trajectories (see \cite{book3,syst}).

\newpage

\clearpage

\clearpage

\section{A variant with $M_{D^0_0} = M_{D^1_1}$ and overlapping $qD^0_0(S=1/2)$
and $qD^1_1(S=1/2)$ states }

Here we consider the case with further decrease of states which can
be easily seen. We suppose that $M_{D^0_0} = M_{D^1_1}$ and the states
$qD^0_0(S=1/2)$ and $qD^1_1(S=1/2)$ overlap. So, for a naive observer,
who does not perform an analysis of double pole structure, the
number of states with $S=1/2$ decreases twice.

\begin{figure}[h]
%Fig. 8
\centerline{\epsfig{file=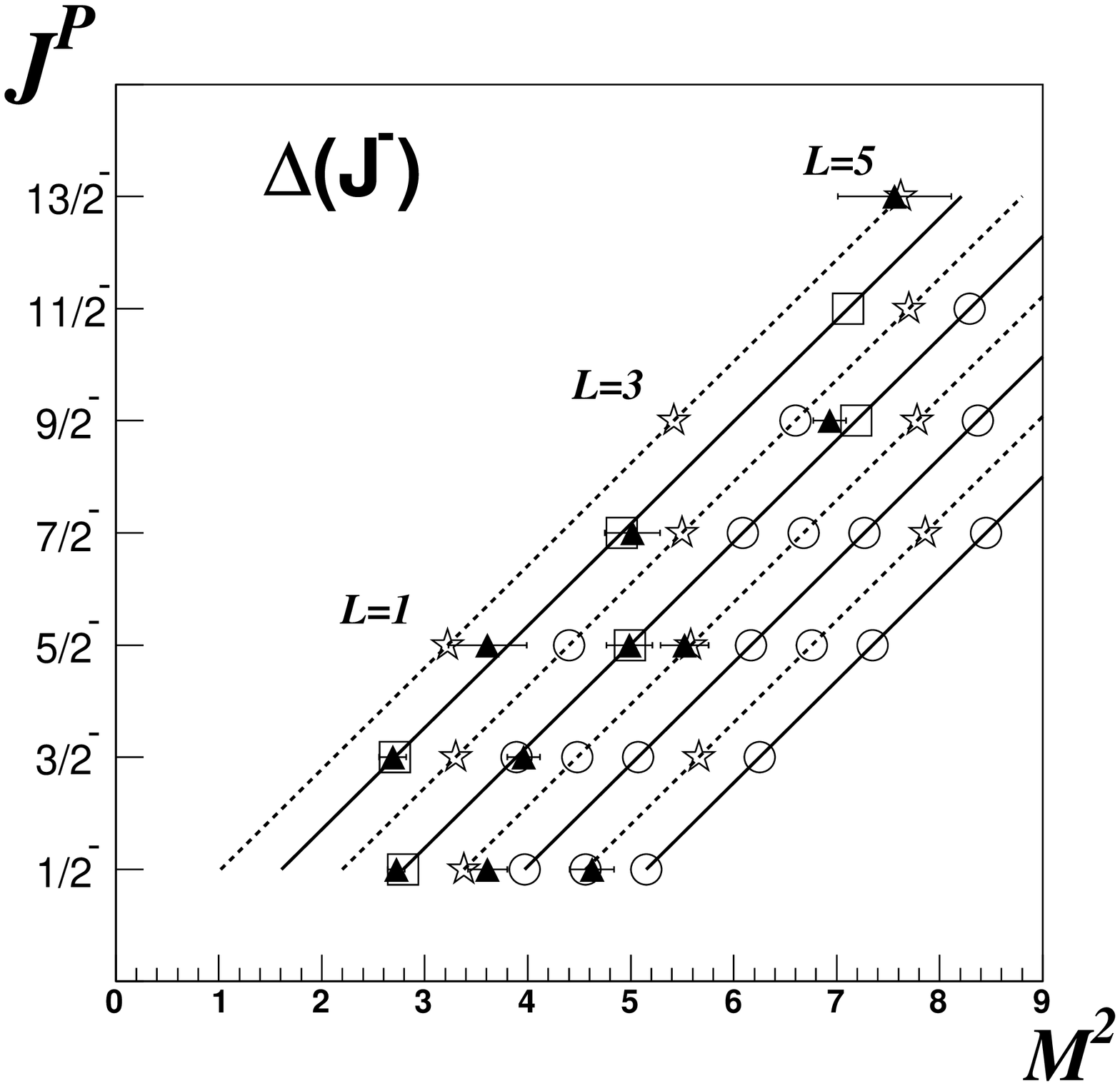,width=70mm}
             \epsfig{file=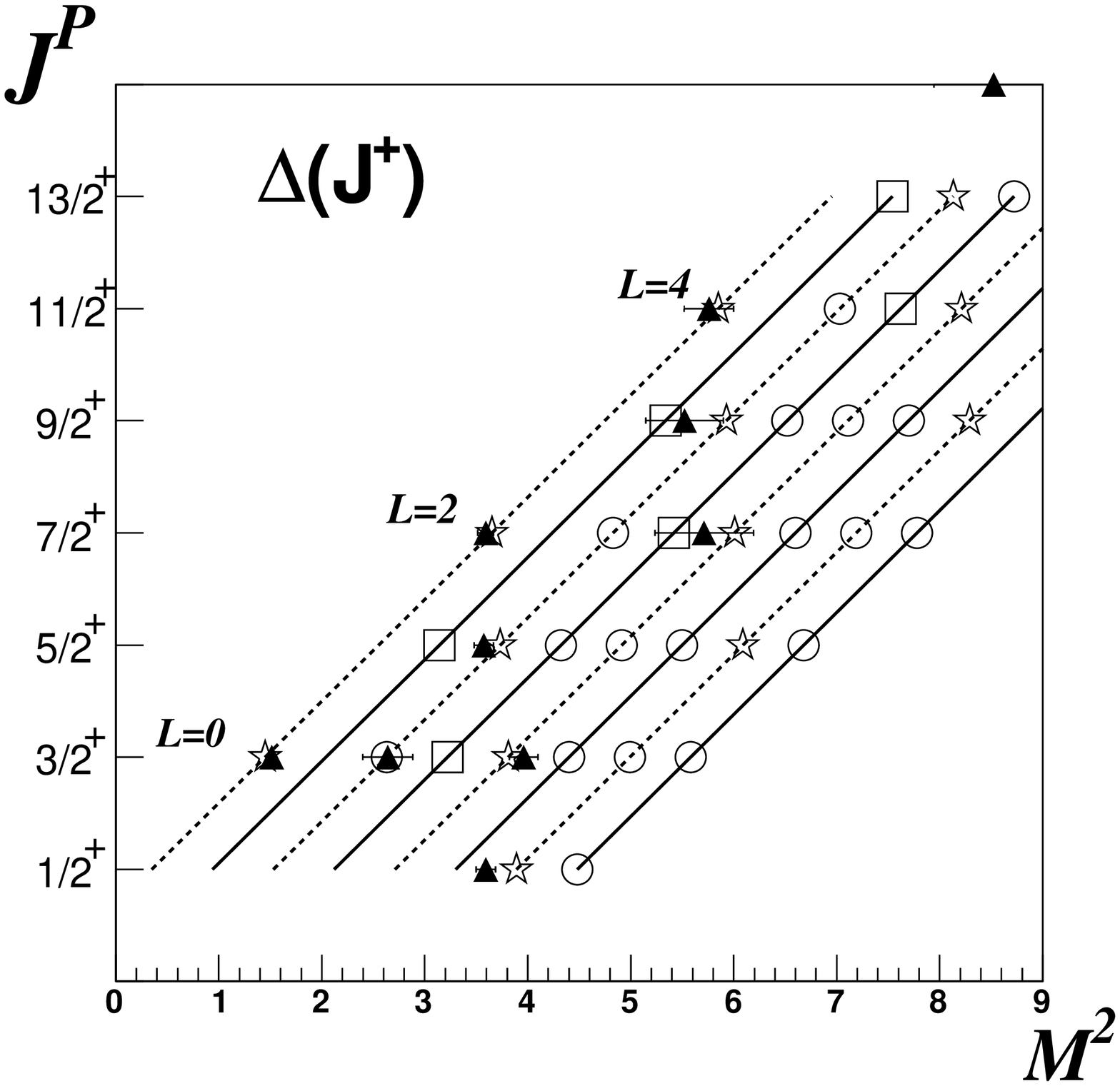,width=70mm}}
\centerline{\epsfig{file=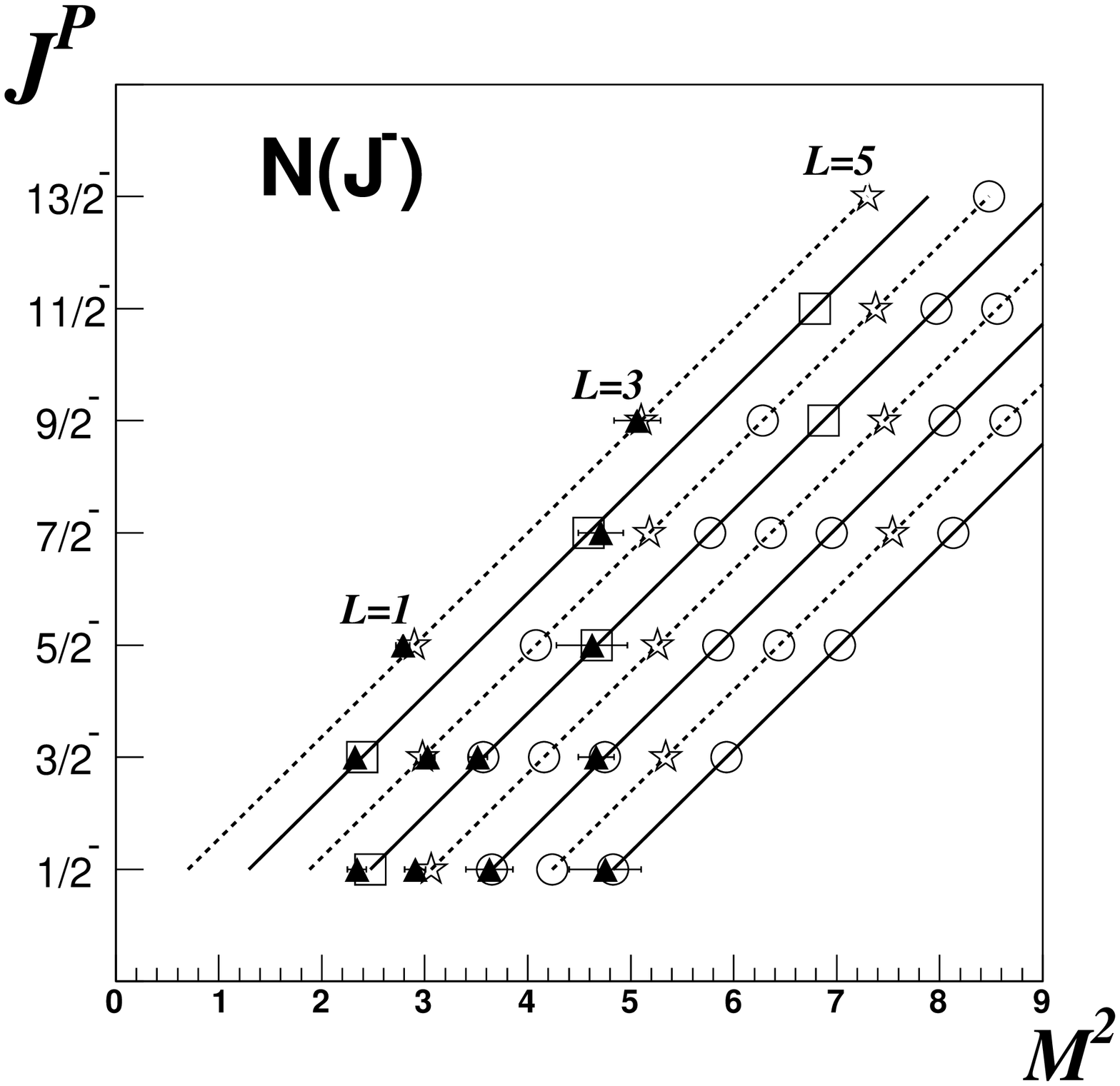,width=70mm}
             \epsfig{file=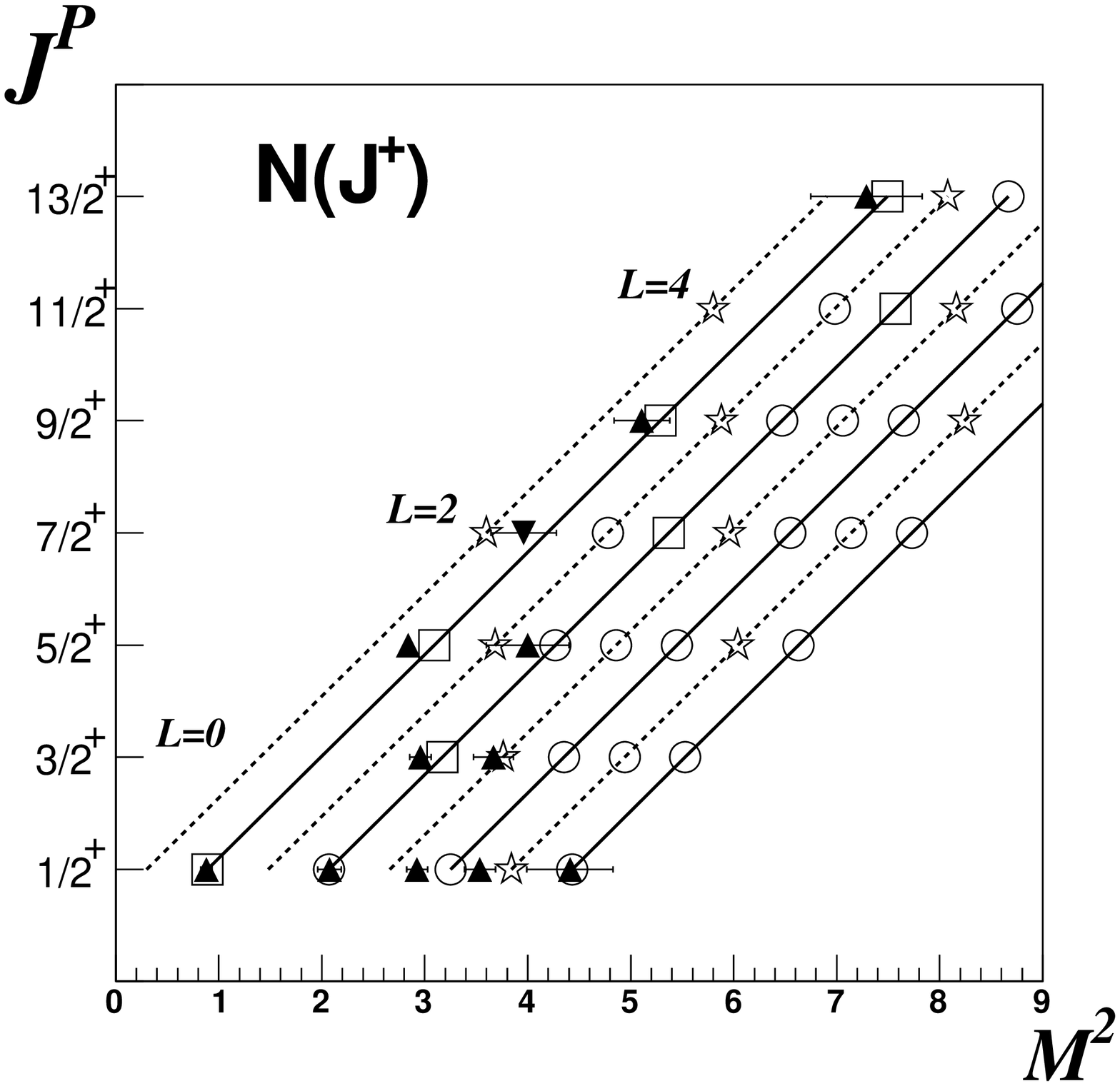,width=70mm}}
\caption{ Baryon setting  on
$(J^P,M^2)$ planes in the model with overlapping $qD^0_0(S=1/2)$ and
$qD^1_1(S=1/2)$ states (notations are as in Fig. \ref{J-M-G}).
}\label{J-M-eq-M}
\end{figure}

\begin{figure}[h]
%Fig. 9
\centerline{\epsfig{file=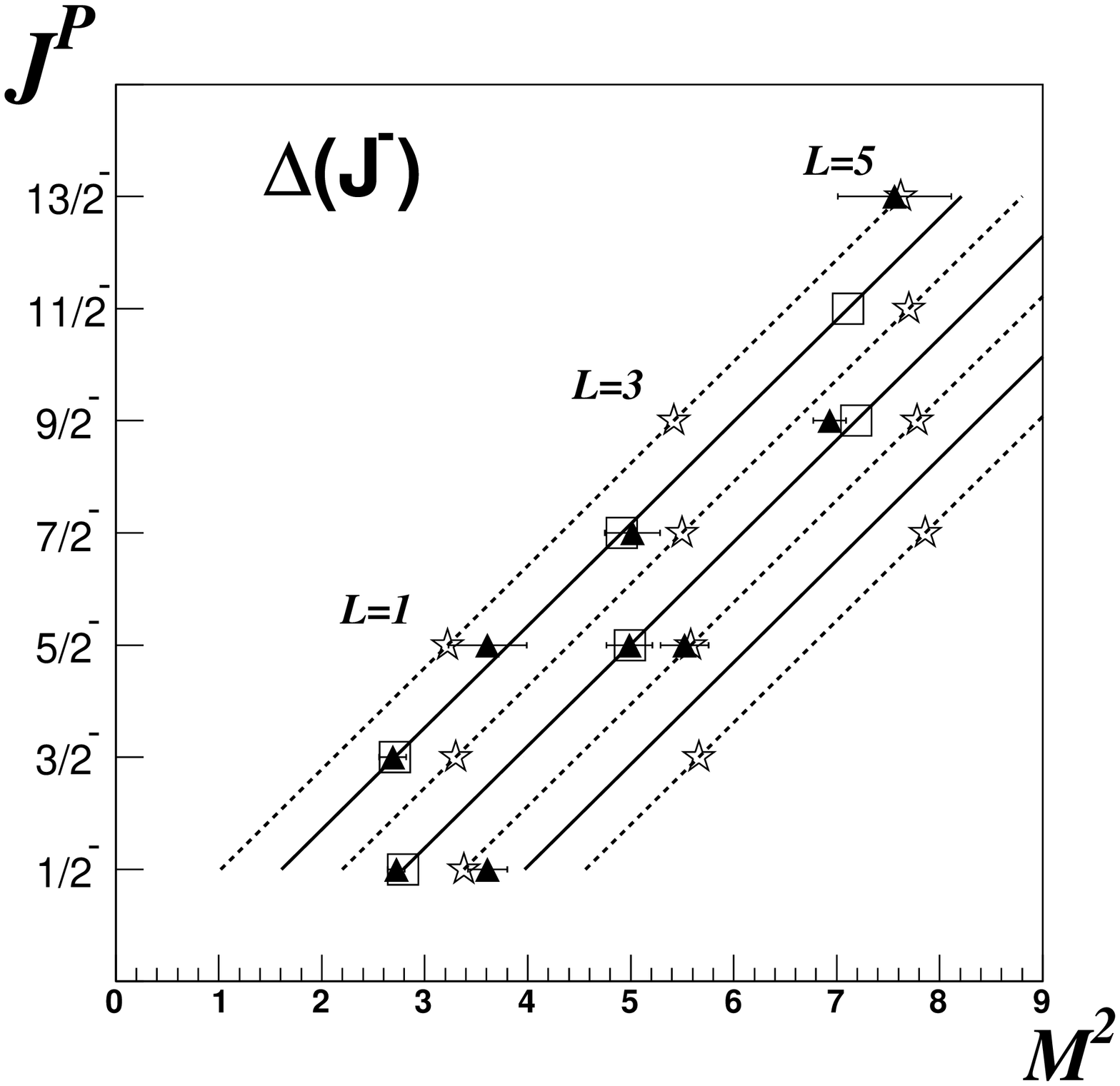,width=70mm}
             \epsfig{file=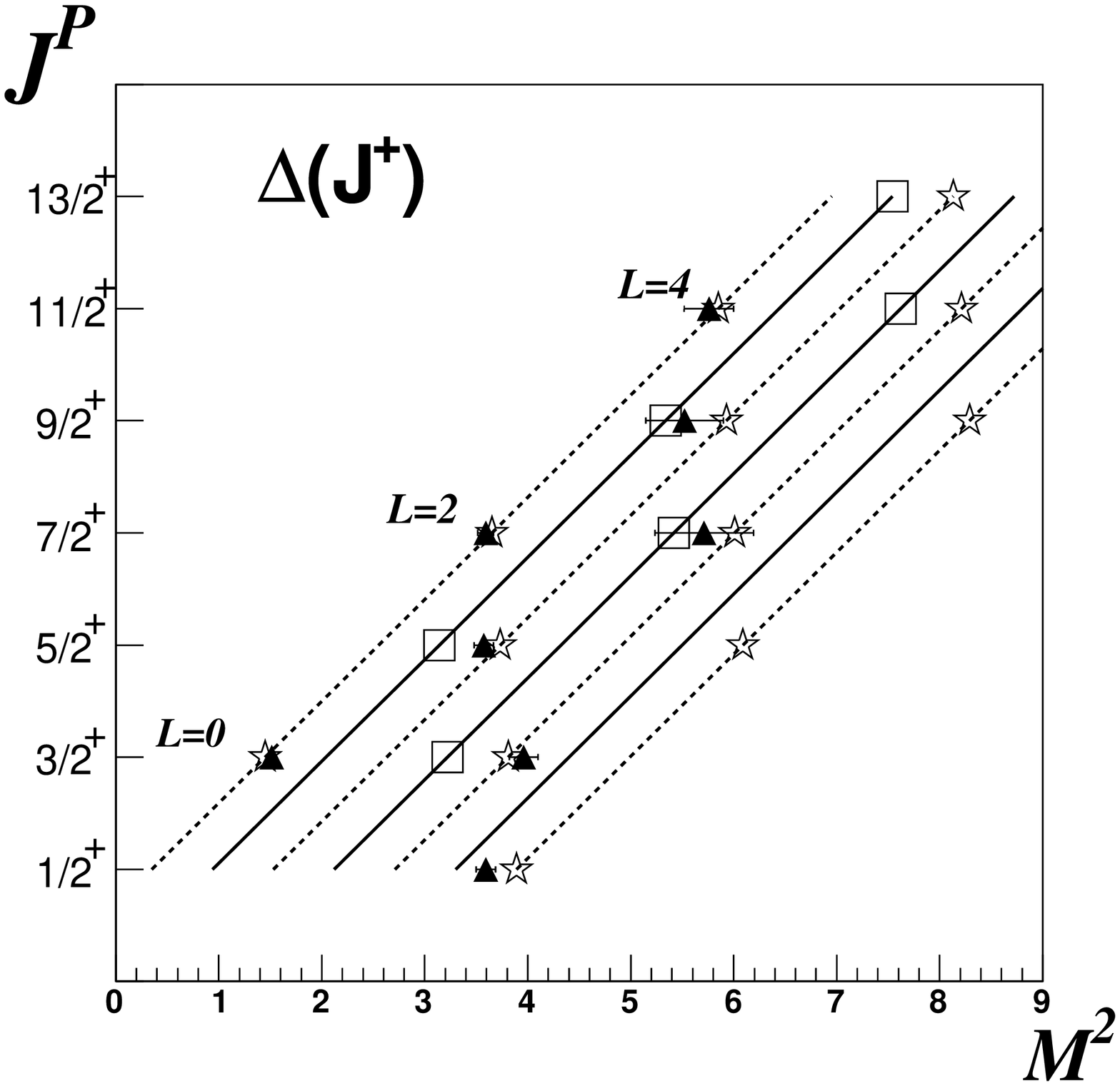,width=70mm}}
\centerline{\epsfig{file=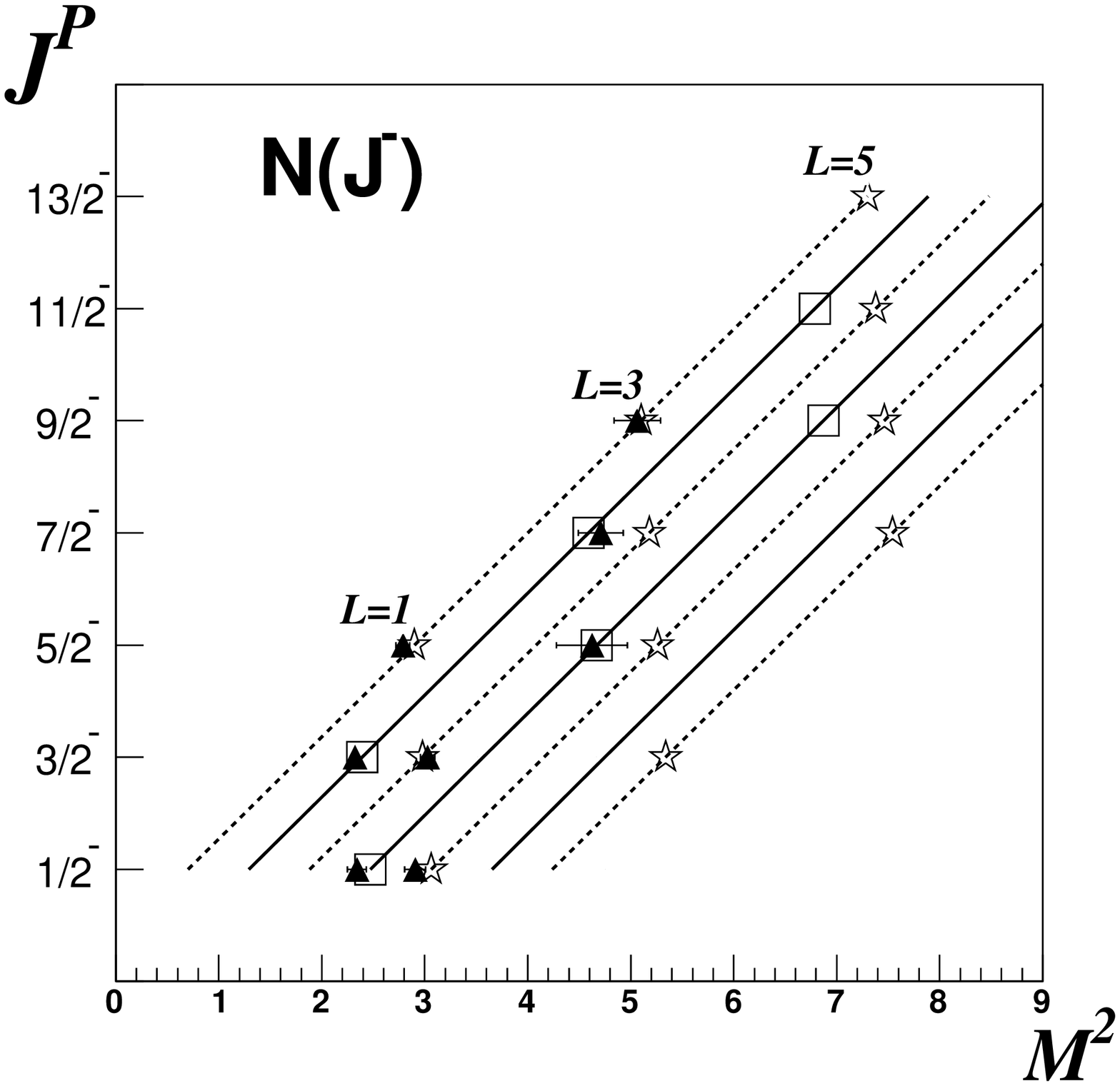,width=70mm}
             \epsfig{file=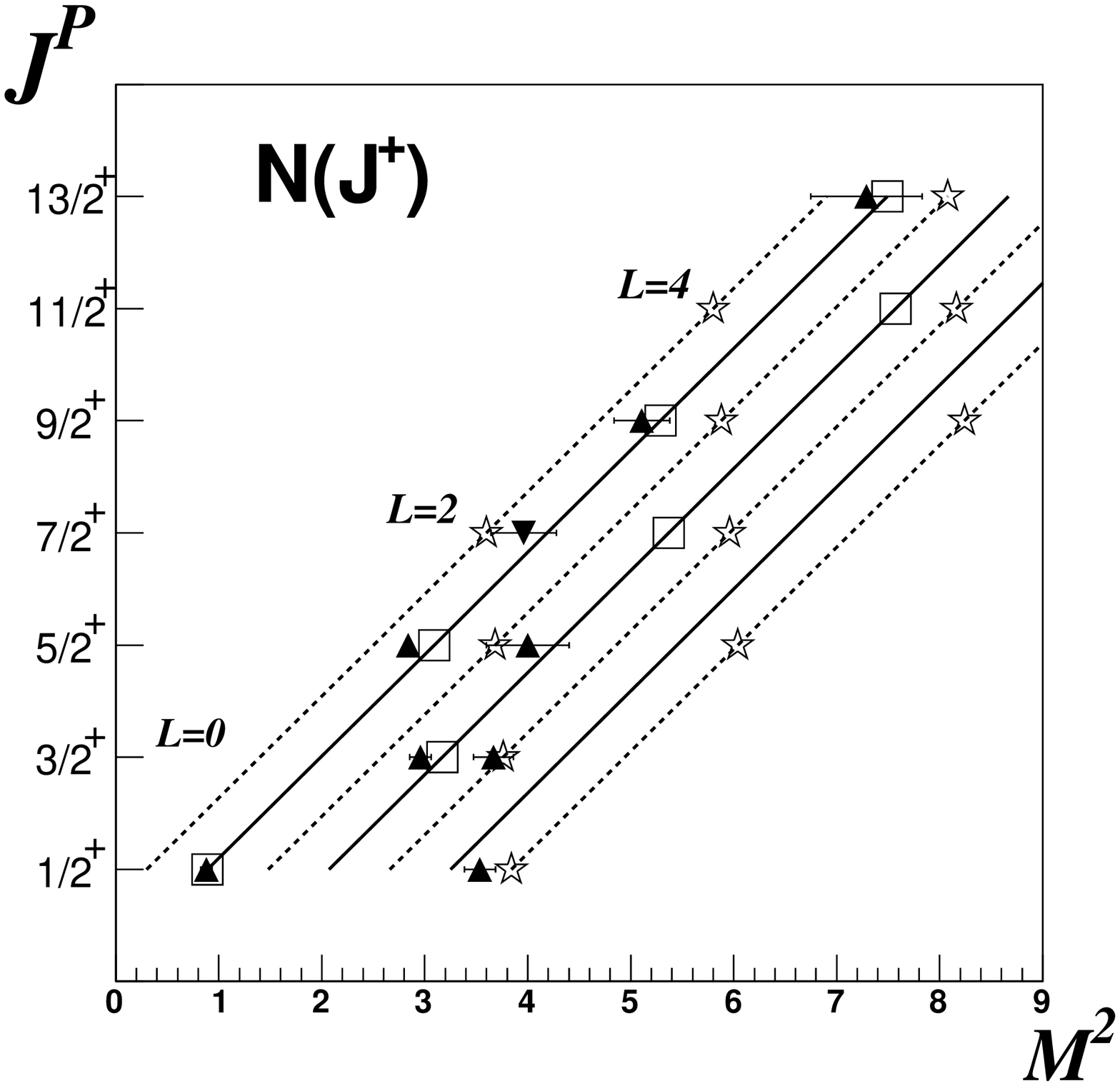,width=70mm}}
\caption{ Setting of  basic baryons ($n=1$)  on
$(J^P,M^2)$ planes in the model with overlapping $qD^0_0(S=1/2)$ and
$qD^1_1(S=1/2)$ states (notations are as in Fig). \ref{J-M-G}.
}\label{J-M-eq-M1}
\end{figure}

In Fig. \ref{J-M-eq-M}  we show $(J,M^2)$ plots as they look like
for  "naive observers", while Fig.
\ref{J-M-eq-M1} demonstrates the $(J,M^2)$ plots for ground states
($n=1$) only.

\clearpage

Figures \ref{J-M-eq-M}, \ref{J-M-eq-M1} present us more compact
scheme than that given in Figs. \ref{J-M-G}, \ref{J-M-G-1}.

\section{Conclusion}

We have systematized all baryon states in the framework of the
hypothesis of their quark--diquark structure. We cannot say whether
such a systematization is unambiguous, so we discuss possible
versions of setting baryons upon multiplets. To carry out a more
definite systematization,  additional efforts are needed in both
experiment and phenomenological comprehension of data.

Concerning the experiment, it is necessary:\\
(i) To investigate in  details the $\Delta$ spectrum in the
region around 1700 MeV. Here one should search for the $D_{15}$
and/or $F_{15}$ states. The double pole structures
should be searched for, first, in the regions $N_{\frac 12^+}(1400)$ and
$\Delta_{\frac 32^+}(1600)$.
\\
 (ii) To increase the interval of
available energies in order to get a possibility to investigate
resonances up to  the masses  3.0--3.5 GeV.\\
(iii) To measure
various types of reactions in order to analyze them simultaneously.

As to the phenomenology and theory, it is necessary
to continue the $K$-matrix analysis, the first results of which
were obtained in \cite{Horn:2007pp,Thoma:2007bm}, in order to
cover a larger mass interval and the most possible number of reactions.
One should take into account the expected overlapping of resonances.
Namely, the standard procedure should be elaborated for singling out
the amplitude poles in the complex-$M$ plane in case when one pole
is under another.

We thank L.G. Dakhno and E. Klempt for helpful discussions.
 This paper was
supported by the grants RFBR  No 07-02-01196-a and RSGSS-3628.2008.2.

 \end{document}